\documentclass[12pt]{article}
\usepackage{jheppub}

\usepackage{mathtools,amsmath,amssymb,graphicx,hyperref}

\usepackage{ifpdf}
\newcommand{\alt}[2]{\texorpdfstring{#1}{#2}}

\usepackage{framed}
\usepackage[most]{tcolorbox}
\usepackage{paracol}
\usepackage{mleftright}
\mleftright

\newcommand{\assign}{:=}
\newcommand{\mathd}{\mathop{}\!\mathrm{d}}
\newcommand{\noeq}{\phantom{=}}

\DeclareMathOperator{\diag}{diag}
\DeclareMathOperator{\sgn}{sgn}
\DeclareMathOperator{\re}{Re}
\DeclareMathOperator{\im}{Im}
\DeclareMathOperator{\sinc}{sinc}

\preprint{IFT-UAM/CSIC-26-80}

\title{Strongly-coupled black holes (are not) at weak coupling}

\author[a]{Ben Heidenreich}
\author[b]{and Matteo Lotito}

\affiliation[a]{Amherst Center for Fundamental Interactions,\\Department of Physics, University of Massachusetts, Amherst, MA 01003, USA}
\affiliation[b]{Instituto de F\'isica Te\'orica IFT-UAM/CSIC,
  C/ Nicol\'as Cabrera 13-15, Campus de Cantoblanco, 28049 Madrid, Spain}

\emailAdd{bheidenreich@umass.edu}
\emailAdd{matteo.lotito@ift.csic.es}

\abstract{
  We prove sharp bounds on the moduli space excursion that is possible outside
  the event horizon of a spherically symmetric charged black hole. In regions
  where the black hole charge $Q (\phi)$ is no less than its asymptotic value,
  we prove that the total moduli space excursion is less than $\frac{\pi}{2
  \sqrt{k_N}}$ where $k_N$ is the rationalized Newton constant. We also show
  that the moduli space excursion in a specified direction in which $Q (\phi)$
  is non-decreasing satisfies the same bound. Applying our bound to
  electrically charged black holes, we rule out a potential loophole in a
  recent proof [\href{https://arxiv.org/abs/2401.14449}{2401.14449}] of the
  Weak Gravity Conjecture in perturbative bosonic string theory.
}

\begin{document}

\maketitle

\section{Introduction} \label{sec:intro}

A remarkable property of charged black holes is that they explore the scalar
field space of the theory in which they live. For nearly extremal black holes
in theories with exact moduli, this exploration is extreme: near the black
hole horizon, the moduli always approach a critical point of the
moduli-dependent canonical charge of the black hole $Q (\phi)$, no matter how
far away in field space this critical point is. This is the famous attractor
mechanism \cite{Ferrara:1995ih,Cvetic:1995bj,Strominger:1996kf,Ferrara:1996dd,Ferrara:1996um,Ferrara:1997tw,
Goldstein:2005hq}.

The reason for this field-space exploration is relatively simple: the energy
stored in the electric field of the black hole decreases when the magnitude of
the charge function $Q (\phi)$ decreases. For a nearly extremal black hole
there is in some sense an arbitarily amount of ``space'' near the horizon for
the black hole to relax into its lowest energy configuration, so it seeks out
the lowest available value of the charge function $Q (\phi)$.\footnote{If the
boundary conditions at infinity are fine-tuned, the black hole may get
``stuck'' at a saddle point of $Q (\phi)$, which is why the moduli do not
always approach a local \emph{minimum} of $\phi$ near the horizon of an
extremal black hole.}

The sensitivity of nearly extremal black holes to physics in arbitrarily
distant parts of the moduli space is both a blessing and a curse. On the one
hand, such black holes could hypothetically act as experimental probes of
distant parts of the moduli space~\cite{Delgado:2022dkz, Delgado:2025crl}.\footnote{Note,
however, that the presence of a non-zero potential reduces the moduli space
excursion of the black hole, especially when the black hole is very large~\cite{Delgado:2025crl}.} On
the other hand, the landscape of quantum gravity theories (QGTs) is vast and
not fully explored, so the fact that black holes might be sensitive to
arbitarily distant parts of it makes it difficult to compute their properties
reliably. In particular, even if $Q (\phi)$ has a nearby local minimum, there
might be a deeper minimum elsewhere in the moduli space---a ``hidden
attractor''---that is energetically preferred. While we can construct black
hole solutions that flow towards the known attractor using local knowledge in
the moduli space, there may be other solutions flowing towards the hidden
attractor, and these solutions could even be lighter than the ones we know
about.

In this paper, we prove rigorous bounds on the circumstances in which a
spherically symmetric black hole can reach such a hidden attractor. In
particular, we specify several conditions under which local knowledge of the
moduli space (up to a field distance of order the Planck scale) can be enough
to rule out black hole solutions that move outside this known region of moduli
space. On the one hand, we prove that the total moduli space distance that the
black hole solution can traverse in regions where the charge function $Q
(\phi)$ equals or exceeds its initial value has a sharp upper bound $\Delta
\phi < \frac{\pi}{2 \sqrt{k_N}}$, where $k_N \assign \frac{d - 3}{d - 2}
\kappa_d^2$ is the rationalized Newton constant. We also prove that the
component of the moduli space distance along a direction in which $Q (\phi)$
is non-decreasing is bounded by the same amount, assuming certain additional
conditions are met. The former bound (the ``charge threshold bound'') is
discussed in \S\ref{sec:basinbound} and the latter (the
``directional derivative bound'') is discussed in
\S\ref{sec:distFuncBound}.

Our bounds are conceptually connected to more general considerations about
trans-Planckian spatial variations of scalar fields in QGTs
\cite{Nicolis:2008wh,Klaewer:2016kiy,Dolan:2017vmn,Draper:2019zbb,Draper:2019utz}. However, while these works have focused on the \emph{consequences} of trans-Planckian scalar field variations---such as requiring an exponentially large spatial volume to avoid gravitational collapse---our bounds simply \emph{rule out} such variations in the limited circumstances where they apply.

Applying our bounds, we show that in a perturbative string vacuum electrically
charged (and some types of dyonically charged) black holes cannot explore the
strongly coupled (i.e., $g_s \gtrsim 1$) regions of moduli space,\footnote{Magnetically charged black holes, on the other hand, \emph{do} have strongly-coupled cores. This is not in conflict with our bounds, since the charge function $Q^2(\phi)$ for such black holes does not satisfy the appropriate constraints, e.g., it decreases as $g_s$ increases.} and in fact
for $g_s \ll 1$ we have parametric control over all such solutions. This fills
in a hypothetical hole in the proof \cite{Heidenreich:2024dmr,HLSuperstring} that the Weak Gravity Conjecture~\cite{Arkani-Hamed:2006emk,Palti:2020mwc,Harlow:2022ich} is satisfied in perturbative
string theory. Despite the fact that the charge function $Q (\phi)$ increases
with $g_s$, one might worry that the lightest black holes of a given charge
have strongly coupled cores. However, we will show that this is impossible, at
least for spherically symmetric black holes.\footnote{We thank the referee of
\cite{Heidenreich:2024dmr} for motivating us to definitively rule out this
possibility.}

The outline of our paper is as follows. In \S\ref{sec:background}
we lay out some basic facts about spherically symmetric charged black holes
that we will use in our argument. In \S\ref{sec:1dmodulispace} we
derive bounds on black holes exploring 1d moduli spaces, and provide geometric
interpretations of these bounds using Maupertuis's principle. In
\S\ref{sec:multidimModuliSpace} we discuss multi-dimensional
moduli spaces and derive our two main bounds. In
\S\ref{sec:applications} we apply our bounds to electrically and
dyonically charged black holes in perturbative string theory. We conclude in
\S\ref{sec:conclusions}. Appendix \ref{app:Maupertuis} reviews
Maupertuis's principle, Appendix \ref{app:4ddyonicQ2bounds} derives lower
bounds on the dyonic charge function in toroidal compactifications of
heterotic string theory, and Appendix \ref{app:wormhole} explores the connection between our bounds and properties of Euclidean wormholes.

\section{Background and notation}\label{sec:background}

In this paper, we limit our attention to spherically symmetric charged black
hole solutions in Minkowski space for simplicity. To study such solutions it
is sufficient to consider the two-derivative effective action
\begin{equation}
  S = \int d^d x \sqrt{- g}  \biggl[ \frac{1}{2 \kappa_d^2} R - \frac{1}{2}
  G_{i j} (\phi) \nabla \phi^i \cdot \nabla \phi^j - \frac{1}{2}
  \mathfrak{f}_{a b} (\phi) F_2^a \cdot F_2^b \biggr] \,,
  \label{eqn:actionansatz}
\end{equation}
as shown, e.g., in \cite{Heidenreich:2020upe}, where $\phi^i$ are the neutral moduli with metric on moduli space $G_{i j}
(\phi)$ and $F_2^a = \mathd A_1^a$ are the abelian fields.\footnote{In the
case of a non-abelian gauge symmetry, we truncate to the Cartan subalgebra.}
Making a suitable gauge choice, these solutions take the form:
\begin{equation}
\begin{aligned}
  \mathd s^2 &= - e^{2 \psi (r)} f (r) \mathd t^2 + e^{- \frac{2}{d - 3} \psi
  (r)}  \biggl[ \frac{\mathd r^2}{f (r)} + r^2 \mathd \Omega_{d - 2}^2 \biggr]\,,
  \qquad f (r) = 1 - \frac{r_h^{d - 3}}{r^{d - 3}}\,, \\
  F_2^a &= \frac{\mathfrak{f}^{a b} (\phi (r)) Q_b}{V_{d - 2}}  \frac{e^{2
  \psi (r)}}{r^{d - 2}} \mathd t \wedge \mathd r, 
\end{aligned}
\end{equation}
where $r_h \geqslant 0$ is a constant that vanishes when the black hole is
quasiextremal (i.e., has zero temperature and/or horizon area), $V_{d - 2} = 2
\pi^{\frac{d - 1}{2}} / \Gamma \bigl( \frac{d - 1}{2} \bigr)$ is the volume
of a unit $S^{d - 2}$ sphere, $\mathfrak{f}^{a b} (\phi)$ is the inverse of
$\mathfrak{f}_{a b} (\phi)$, and $\psi (r), \phi^i (r)$ solve the ODEs:
\begin{subequations} \label{eqn:BHeqnsz}
\begin{align}
  k_N^{- 1} \frac{\mathd}{\mathd z} [f \dot{\psi}] &=\mathfrak{f}^{a b} (\phi)
  Q_a Q_b e^{2 \psi}, \\ 
  \frac{\mathd}{\mathd z}  [f \dot{\phi}^i] + f
  \Gamma^i_{\; j k} (\phi)  \dot{\phi}^j  \dot{\phi}^k &=
  \frac{1}{2} G^{i j} (\phi) \mathfrak{f}_{, j}^{a b} (\phi) Q_a Q_b e^{2
  \psi}, \\
  k_N^{- 1} \dot{\psi}  (f \dot{\psi} + \dot{f}) + f G_{i j}
  (\phi) \dot{\phi}^i \dot{\phi}^j &=\mathfrak{f}^{a b} (\phi) Q_a Q_b e^{2
  \psi}, \end{align}
\end{subequations}
where $z \assign \frac{1}{(d - 3) V_{d - 2} r^{d - 3}}$, dots denote $z$
derivatives, $\Gamma^i_{\; j k} (\phi)$ is the Levi-Civita
connection for the metric $G_{i j} (\phi)$ (whose inverse is $G^{i j} (\phi)$)
and $k_N \assign \frac{d - 3}{d - 2} \kappa_d^2$ is the rationalized Newton
constant appearing in the gravitational force law $F_{\text{grav}} = -
\frac{k_N m m'}{V_{d - 2} r^{d - 2}}$. Note that in terms of $z$, $f (z) = 1 -
z / z_h$ where $z_h \assign \frac{1}{(d - 3) V_{d - 2} r_h^{d - 3}}$.

Spherically symmetric black hole solutions are solutions of
\eqref{eqn:BHeqnsz} that remain smooth at $r = r_h$. Assuming non-zero charge,
$Q_a \neq 0$, one can show that this occurs for a subextremal ($r_h > 0$)
solution if and only if \cite{Harlow:2022ich}:\footnote{For a neutral
black hole, $\psi (r) = 0$ so $\dot{\psi} = 0$ everywhere.}
\begin{equation}
  \dot{\psi} < 0 \qquad \text{for all} \qquad z < z_h .
  \label{eqn:horizoncond}
\end{equation}
The ADM mass of such solutions is $M_{\text{BH}} = k_N^{- 1} \bigl[ -
\dot{\psi}_{\infty} + \frac{1}{2 z_h} \bigr]$, where the ``$\infty$''
subscript denotes a quantity evaluated at $r = \infty$, i.e., at $z = 0$.
Notice that $M_{\text{BH}} > 0$ as a consequence of \eqref{eqn:horizoncond}.

In the quasiextremal case, $r_h = 0$ $(z_h = \infty)$, \eqref{eqn:horizoncond}
is a necessary but insufficient condition. To have a smooth horizon in this
case, we also require that the horizon area
\begin{equation}
  A_h = \lim_{r \rightarrow 0} V_{d - 2} r^{d - 2} e^{- \frac{d - 2}{d - 3}
  \psi},
\end{equation}
is non-zero,\footnote{Note that it is impossible for $A_h$ to be infinite. In
particular, combining the first and third equations in \eqref{eqn:BHeqnsz},
one obtains $\ddot{\psi} = \dot{\psi}^2 + k_N  \dot{s}^2$ where $\dot{s}^2
\assign G_{i j}  \dot{\phi}^i  \dot{\phi}^j \geqslant 0$. Therefore $v \assign
\frac{d - 3}{d - 2}  \frac{\mathd \log A}{\mathd \log z}$ satisfies
$\frac{\mathd v}{\mathd \log z} \leqslant - v (v + 1)$, so $A (z)$ is bounded
as $z \rightarrow \infty$.} and that the moduli take finite values $\phi^i_h$
at the horizon.

The condition \eqref{eqn:horizoncond} will be essential for placing
constraints on the regions of moduli space that the black hole solution can
reach. As an aside, we note that it is a consequence of the dominant energy
condition. For a spherically symmetric solution of Einstein gravity coupled to
an arbitary stress energy tensor $T_{\mu \nu}$, one finds \cite{Etheredge:2022rfl}:
\begin{equation}
  \frac{\mathd}{\mathd z}  (f \dot{\psi} - \Sigma) = - k_N e^{2 \psi} A^2 
  (T^t_t + T^r_r), \qquad \dot{\psi} [f \dot{\psi} + \dot{f}] - \dot{\Sigma} = -
  2 k_N e^{2 \psi} A^2 T_r^r, \label{eqn:hdcEOMs}
\end{equation}
where $A (r) \assign V_{d - 2} r^{d - 2} e^{- \frac{d - 2}{d - 3} \psi}$ is
the $r$-dependent area of the $S^{d - 2}$ and $\Sigma \assign \frac{1 - f}{z}$.
We fix $f (z_h) = 0$ and $\dot{f} (z_h) = - \frac{1}{z_h}$ for $z_h > 0$ by a
gauge choice, with $f (z) > 0$ for $z < z_h$.\footnote{Here $f (z)$ is an
\emph{a priori} undetermined function. If $T^{\hat{\theta}}_{\hat{\theta}}
= - T_r^r$ for a direction $\hat{\theta}$ along the $S^{d - 2}$ (as follows
spherical symmetry with action \eqref{eqn:actionansatz}) then one of the
Einstein equations can be integrated to fix $f (z) = 1 - z / z_h$.} Combining
the two equations \eqref{eqn:hdcEOMs}, one finds:
\begin{equation}
  f \ddot{\psi} = k_N e^{2 \psi} A^2  (- T^t_t + T_r^r) + f \dot{\psi}^2 .
\end{equation}
If $T_{\mu \nu}$ satisfies the null energy condition then $T^r_r \geqslant
T^t_t$, and therefore $\ddot{\psi} \geqslant 0$ for $z < z_h$. Moreover,
evaluating \eqref{eqn:hdcEOMs} at $z = z_h$, one finds:
\begin{equation}
  \dot{\psi} (z_h) = 2 z_h k_N e^{2 \psi_h} A^2_h \lim_{z \rightarrow z_h}
  T^t_t, \qquad \text{where} \qquad \lim_{z \rightarrow z_h}  (T^t_t - T_r^r)
  = 0 .
\end{equation}
If $T_{\mu \nu}$ satisfies the weak energy condition then $T_{t t} \geqslant
0$ for $z < z_h$, hence $T_t^t = \frac{1}{g_{t t}} T_{t t} \leqslant 0$ for $z
< z_h$. This implies that $\lim_{z \rightarrow z_h} T^t_t \leqslant 0$, so
$\dot{\psi} (z_h) \leqslant 0$. Moreover, if $\dot{\psi} (z_h) = 0$ then
$T_t^t$ and $T_r^r$ both vanish at the horizon, so if $T_{\mu \nu}$ satisfies
the dominant energy condition then we must have $T_{\mu \nu} = 0$ at the
horizon, which is incompatible with the electric field generated by a charged
black hole. Thus $\dot{\psi} (z_h) < 0$, and since $\ddot{\psi} \geqslant 0$
for $z < z_h$ we conclude that for a charged black hole
\begin{equation}
  \dot{\psi} (z) < 0 \qquad \text{for all} \qquad z < z_h,
  \label{eqn:horizoncondRederive}
\end{equation}
as a consequence of the dominant energy condition, reproducing
\eqref{eqn:horizoncond}.\footnote{In the quasiextremal case, this condition
follows more simply from the requirement that $A (z_h)$ is finite together
with $\ddot{\psi} \geqslant 0$, see
[\href{https://arxiv.org/abs/2201.08380}{2201.08380}] Appendix A, so one need
only assume the null energy condition.}

For our purposes, it will be convenient to rewrite the equations
\eqref{eqn:BHeqnsz}, \eqref{eqn:horizoncond} slightly. We define the new
independent variable:
\begin{equation}
  \tau \assign - z_h \log f \qquad \Rightarrow \qquad \mathd \tau =
  \frac{\mathd z}{f} .
\end{equation}
Note that $\tau = 0$ at spatial infinity and $\tau \rightarrow + \infty$ as we
approach the horizon. Moreover, for small $z$ (near infinity):
\begin{equation}
  \tau = - z_h \log \biggl( 1 - \frac{z}{z_h} \biggr) = - z_h  \biggl[ -
  \frac{z}{z_h} - \frac{1}{2}  \biggl( \frac{z}{z_h} \biggr)^2 - \cdots \biggr]
  \simeq z + \frac{z^2}{2 z_h} + \ldots,
\end{equation}
so $\tau \simeq z$ far from the horizon, and in fact $\tau = z$ exactly in the
quasiextremal limit $z_h \rightarrow \infty$. In terms of $\tau$, the black
hole equations become:
\begin{subequations} \label{eqn:BHeqnstau} 
\begin{align}
  k_N^{- 1} \chi'' &= e^{2 \chi} Q^2 (\phi), \label{eqn:BHeqns-chi} \\
 \phi^{i\prime\prime} + \Gamma^i_{\; j k} \phi^{j\prime} \phi^{k\prime}&=
  \frac{1}{2} e^{2 \chi} G^{i j} Q^2_{, j}, \label{eqn:BHeqns-phi}  \\
  k_N^{- 1} {\chi'}^2 + G_{i j} \phi^{i\prime} \phi^{j\prime} &= e^{2 \chi} Q^2 (\phi)
  + k_N M_0^2, \label{eqn:BHeqns-cons} 
\end{align}
\end{subequations} 
where
\begin{equation}
  (\ldots)' \assign \frac{\mathd (\ldots)}{\mathd \tau}, \quad Q^2 (\phi)
  \equiv \mathfrak{f}^{a b} (\phi) Q_a Q_b, \quad \chi \assign \psi -
  \frac{1}{2 z_h} \tau, \quad M_0 \assign \frac{1}{2 k_N z_h} .
\end{equation}
Note that $e^{2 \chi} = f e^{2 \psi}$. The horizon condition
\eqref{eqn:horizoncond} now becomes:
\begin{equation}
  \chi' < - k_N M_0 \qquad \text{for} \qquad \tau < \infty,
  \label{eqn:horizonregularity}
\end{equation}
and ADM mass is $M_{\text{BH}} = - k_N^{- 1} \chi'_{\infty}$.

Note that the first two black hole equations \eqref{eqn:BHeqnstau} are the
Euler-Lagrange equations for the effective action:
\begin{equation}
  S = \int \biggl[ \frac{1}{2 k_N}  (\chi')^2 + \frac{1}{2} G_{i j} (\phi)
  {\phi^i}' {\phi^j}' + \frac{1}{2} e^{2 \chi} Q^2 (\phi) \biggr] \mathd \tau\,,
  \label{eqn:effaction}
\end{equation}
whereas the third equation is the condition that the associated conserved
energy is equal to:
\begin{equation}
  E \assign \frac{1}{2 k_N}  (\chi')^2 + \frac{1}{2} G_{i j} (\phi) {\phi^i}'
  {\phi^j}' - \frac{1}{2} e^{2 \chi} Q^2 (\phi) = \frac{1}{2} k_N M_0^2 .
  \label{eqn:conservedE}
\end{equation}

The quasiextremal case $M_0 = 0$ has unique properties. To have a smooth
horizon in this case, we require that $\tau e^{\chi} \rightarrow \tau_0$ and
$\phi^i \rightarrow \phi_h^i$ for a constant $\tau_0 > 0$ and a point
$\phi_h^i$ inside the moduli space as $\tau \rightarrow \infty$. Defining
$\hat{\chi} \assign \chi + \log \frac{\tau}{\tau_0}$ and $u \assign - \log
\tau$, the black hole equations \eqref{eqn:BHeqnstau} become:
\begin{subequations}\label{eqn:BHeqnsu}
\begin{align}
  \frac{\mathd^2 \hat{\chi}}{\mathd u^2} &= - \frac{\mathd \hat{\chi}}{\mathd
  u} + k_N \tau^2_0 e^{2 \hat{\chi}} Q^2 (\phi) - 1, \\
  \frac{\mathd^2 \phi^i}{\mathd u^2} + \Gamma^i_{\; j k} 
  \frac{\mathd \phi^j}{\mathd u}  \frac{\mathd \phi^k}{\mathd u} &= -
  \frac{\mathd \phi^i}{\mathd u} + \frac{1}{2} \tau_0^2 e^{2 \hat{\chi}} G^{i
  j} Q^2_{, j} (\phi), \\
  \biggl( \frac{\mathd \hat{\chi}}{\mathd u} \biggr)^2 + k_N G_{i j} 
  \frac{\mathd \phi^i}{\mathd u}  \frac{\mathd \phi^j}{\mathd u} &=  - 2
  \frac{\mathd \hat{\chi}}{\mathd u} + k_N \tau_0^2 e^{2 \hat{\chi}} Q^2
  (\phi) - 1 .
\end{align}
\end{subequations}
Requiring $\hat{\chi} \rightarrow 0$ and $\phi^i \rightarrow \phi^i_h$ as $u
\rightarrow - \infty$, we obtain the conditions:
\begin{equation}
  Q^2 (\phi_h) = \frac{1}{k_N \tau^2_0}, \qquad Q^2_{, j} (\phi_h) = 0 .
\end{equation}
Thus, the moduli approach a critical point of $Q^2 (\phi)$ at the horizon,
where the horizon area:
\begin{equation}
  A_h = \lim_{\tau \rightarrow \infty} \frac{1}{V_{d - 2}^{\frac{1}{d - 3}} 
  ((d - 3) \tau e^{\chi})^{\frac{d - 2}{d - 3}}} = \frac{1}{V_{d -
  2}^{\frac{1}{d - 3}}}  \biggl[ \frac{\sqrt{k_N} Q (\phi_h)}{d - 3}
  \biggr]^{\frac{d - 2}{d - 3}},
\end{equation}
is fixed by the value of $Q^2 (\phi)$ at the critical point. This is the
attractor mechanism \cite{Ferrara:1995ih,Cvetic:1995bj,Strominger:1996kf,Ferrara:1996dd,Ferrara:1996um,Ferrara:1997tw,
Goldstein:2005hq}.

\begin{figure}\centering
  \includegraphics[width=0.8\textwidth]{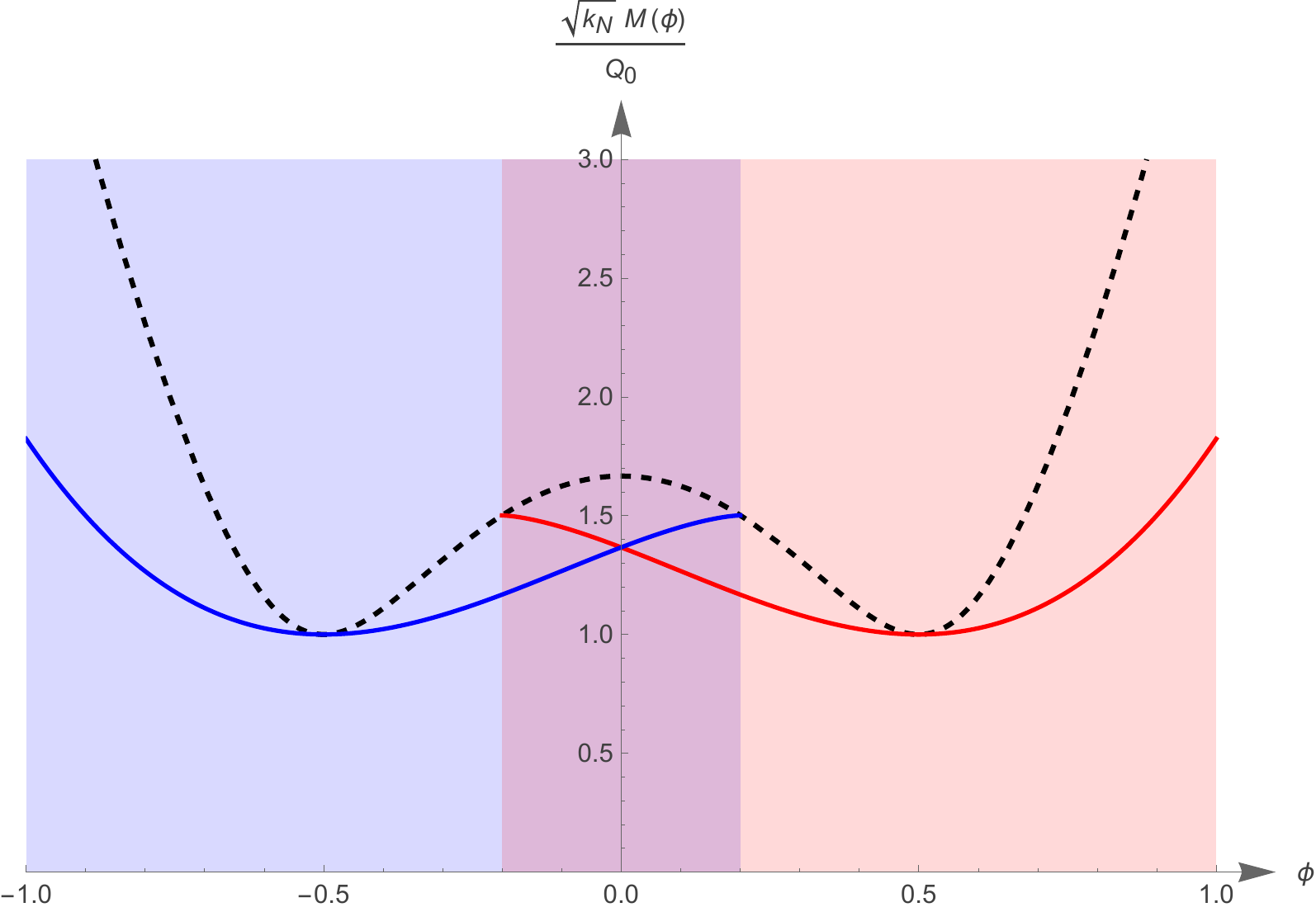}
  \caption{The fake superpotentials $M_{\pm} (\phi)$ for the attractor points
  $\phi_h^{\pm} = \pm \sqrt{\lambda} \phi_0$ of the theory
  \eqref{eqn:Q2example} with $\lambda = 4 / 3$ and $\phi_0 = \sqrt{3} / 4$.
  The blue (red) region on the left (right) is the attractor basin for
  $\phi_h^- = - \sqrt{\lambda} \phi_0$ ($\phi_h^+ = \sqrt{\lambda} \phi_0$),
  where the purple region in the middle denotes where the two basins
  overlap.\label{fig:twobasins}}
\end{figure}

To understand whether such solutions exist, note that the first two equations
in \eqref{eqn:BHeqnsu} are those of a Newtonian particle with position $
(\hat{\chi}, \phi^i)$ moving in a space with metric $G_{I J} = \diag
(k_N^{- 1}, G_{i j})$, subject to linear drag as well as the force induced by
the potential
\begin{equation}
  V_{\text{eff}} (\hat{\chi}, \phi^i) = k_N^{- 1}  \hat{\chi} - \frac{1}{2}
  \tau_0^2 e^{2 \hat{\chi}} Q^2 .
\end{equation}
If the critical point $\phi^i = \phi^i_h$ is a local minimum of $Q^2 (\phi)$,
then the particle starts off at rest at $u = - \infty$ at a local maximum of
$V_{\text{eff}} (\hat{\chi}, \phi^i)$, and can roll off in any direction,
generating a family of smooth, quasiextremal solutions to
\eqref{eqn:BHeqnstau} that can reach any point in some neighborhood
$\mathfrak{B}$ of moduli space surrounding $\phi_h^i$.\footnote{Note that, as
in the example shown in figure \ref{fig:twobasins}, the region $\mathcal{B}$
need not be the \emph{entire} moduli space.} Since the equations
\eqref{eqn:BHeqnstau} are invariant under $\tau \rightarrow \tau +
\text{const}$, we can set $\tau = 0$ at any of these points, and thereby
produce a quasiextremal black hole with any desired asymptotic value of the
moduli $\phi_{\infty}^i \in \mathfrak{B}$ but the same horizon value
$\phi_h^i$. Minima of $Q^2 (\phi)$ are thus \emph{attractor points}, where
quasiextremal solutions that reach a given attractor $\phi_h^i$ exist within
an \emph{attractor basin} $\mathfrak{B} \ni \phi_h^i$ surrounding that
point.

On the other hand, if $\phi^i = \phi^i_h$ is a saddle point or a local maximum
of $Q^2 (\phi)$, then the particle starts off at rest at a saddle point of
$V_{\text{eff}} (\hat{\chi}, \phi^i)$, and there are only some directions in
$\hat{\chi}$-$\phi^i$ space in which it can roll. The space of such solutions
does not fill the nearby moduli space, but for certain types of saddle points
there can still be a region $\mathfrak{B} \ni \phi_h^i$ of moduli space within
which quasiextremal solutions reaching $\phi_h^i$ exist; in these cases, the
attractor point $\phi_h^i$ lies on the boundary of the attractor basin
$\mathfrak{B}$. For other saddle points, such as those with a non-degenerate
(and non-positive) Hessian matrix $\partial_i \partial_j Q^2
|_{\phi_h}$, quasiextremal solutions only exist for $\phi_{\infty}^i$ along
some locus in the moduli space of codimension at least one, and thus do not
exist for generic $\phi_{\infty}^i$. We will not refer to such critical points
as ``attractor points,'' even though $\phi^i$ can flow to them for non-generic
values of $\phi_{\infty}^i$.

Now let $\phi_h^i$ be an attractor point of $Q^2 (\phi)$, which we will take
to be a local minimum for simplicity. One convenient way to construct
quasiextremal solutions flowing to this point is as follows. First, we look
for a function $M (\phi)$---sometimes called a ``fake superpotential'' \cite{Ceresole:2007wx,Andrianopoli:2007gt,Andrianopoli:2009je,Bossard:2009we,Andrianopoli:2010bj,Bellucci:2010aq,Trigiante:2012eb}---that solves the first-order PDE
\begin{equation}
  k_N M^2 (\phi) + G^{i j} (\phi) \partial_i M (\phi) \partial_j M (\phi) =
  Q^2 (\phi) . \label{eqn:Mpde}
\end{equation}
Given such an $M (\phi)$, any solution to the gradient flow equations:
\begin{equation}
  \chi' (\tau) = - e^{\chi (\tau)} k_N M (\phi (\tau)), \qquad {\phi^i}'
  (\tau) = - e^{\chi (\tau)} G^{i j} (\phi (\tau)) \partial_j M (\phi (\tau)),
  \label{eqn:BHfirstorder}
\end{equation}
is a solution to \eqref{eqn:BHeqnstau} with $M_0 = 0$, where the mass of this
solution is $M_{\text{BH}} = M (\phi_{\infty})$. Note that there are generally
many solutions to \eqref{eqn:Mpde}, at least in local patches of the moduli
space, but if we require that $M (\phi)$ also has a local minimum at $\phi^i =
\phi_h^i$ then the solution to \eqref{eqn:Mpde} is uniquely determined in some
neighborhood $\mathfrak{R} \ni \phi_h^i$, and for $\phi^i_{\infty}$ within a
(potentially smaller) neighborhood $\mathfrak{B}' \subseteq \mathfrak{R}$, the
gradient flow \eqref{eqn:BHfirstorder} ends at $\phi^i = \phi^i_h$, producing
a smooth quasiextremal black hole solution.\footnote{Thus, the attractor basin
$\mathfrak{B}$ for $\phi_h^i$ must contain $\mathfrak{B}' \subseteq
\mathfrak{B}$, but \emph{a priori} it might be larger.}

To illustrate these points, consider, e.g., a theory with a 1d moduli space
with metric and charge function:
\begin{equation}
  \qquad G_{\phi \phi} = k_N^{- 1}, \qquad Q^2 (\phi) = Q_0^2  (1 + [(\phi /
  \phi_0)^2 - \lambda]^2), \label{eqn:Q2example}
\end{equation}
as in \cite{Harlow:2022ich}, appendix A. Then \eqref{eqn:Mpde} becomes:
\begin{equation}
  [M (\phi)]^2 + [M' (\phi)]^2 = k_N^{- 1} Q_0^2  (1 + [(\phi / \phi_0)^2 -
  \lambda]^2) . \label{eqn:fakeWex}
\end{equation}
There are two attractor points at $\phi_h^{\pm} = \pm \sqrt{\lambda} \phi_0$.
Rewriting \eqref{eqn:fakeWex}, we obtain:
\begin{equation}
  M' (\phi) = \pm \sqrt{k_N^{- 1} Q_0^2  (1 + [(\phi / \phi_0)^2 - \lambda]^2)
  - M^2},
\end{equation}
One can numerically integrate this equation starting from either attractor
point with the initial condition $M (\phi_h) = Q (\phi_h) / \sqrt{k_N}$, where
we choose the positive (negative) sign for $\phi > \phi_h$ ($\phi < \phi_h$)
in order to construct a function $M (\phi)$ with a local minimum at the chosen
attractor point. An example of this is shown in figure \ref{fig:twobasins}. As
this example illustrates, there can be more than one attractor basin and the
attractor basins can overlap.

\section{One-dimensional moduli spaces}\label{sec:1dmodulispace}

We now turn to the question of how far from a given vacuum $\phi_{\infty}^i$
an arbitrary spherically symmetric black hole can travel in the moduli
space. As discussed in \S\ref{sec:intro}, dilatonic black holes
demonstrate that this distance can be arbitrarily large. However, apart from
some special cases, we rarely know the EFT across the entire moduli
space.\footnote{For instance, since Calabi-Yau compactifications are connected
by a web of geometric transitions, to understand the entire moduli space of
such compactifications we would \emph{a priori} need to have a complete
list of Calabi-Yau threefolds (or, at least, a complete list of those
connected by geometric transitions), which is an unsolved problem.} Thus, this
becomes a pressing question if we want to understand the spectrum of black
holes in a given vacuum. To approach this question, notice that dilatonic
black holes exist for specific charge, of the form $Q^2 (\phi) \propto e^{-
\alpha \phi}$, that decrease along the flow. One can construct other examples
with $Q^2 (\phi)$ decreasing using the fake superpotential method discussed
above, but it is interesting to ask whether it is possible for $Q^2 (\phi)$ to
\emph{increase} as we approach the horizon. We will show that this is
strongly constrained and subject to sharp bounds. Thus, if $Q^2 (\phi)$ has
the right properties in a known region of moduli space, we can ensure that for
vacua within this region, black hole solutions do not venture out into unknown
regions of moduli space.

For simplicity, we begin by considering the case of a
one-dimensional moduli space, generalizing to higher dimensional moduli spaces
in~\S\ref{sec:multidimModuliSpace}.

\subsection{A bound on monotonically increasing \alt{$Q^2 (\phi)$}{Q²(phi)}}

\begin{figure}\centering
  \includegraphics[width=0.8\textwidth]{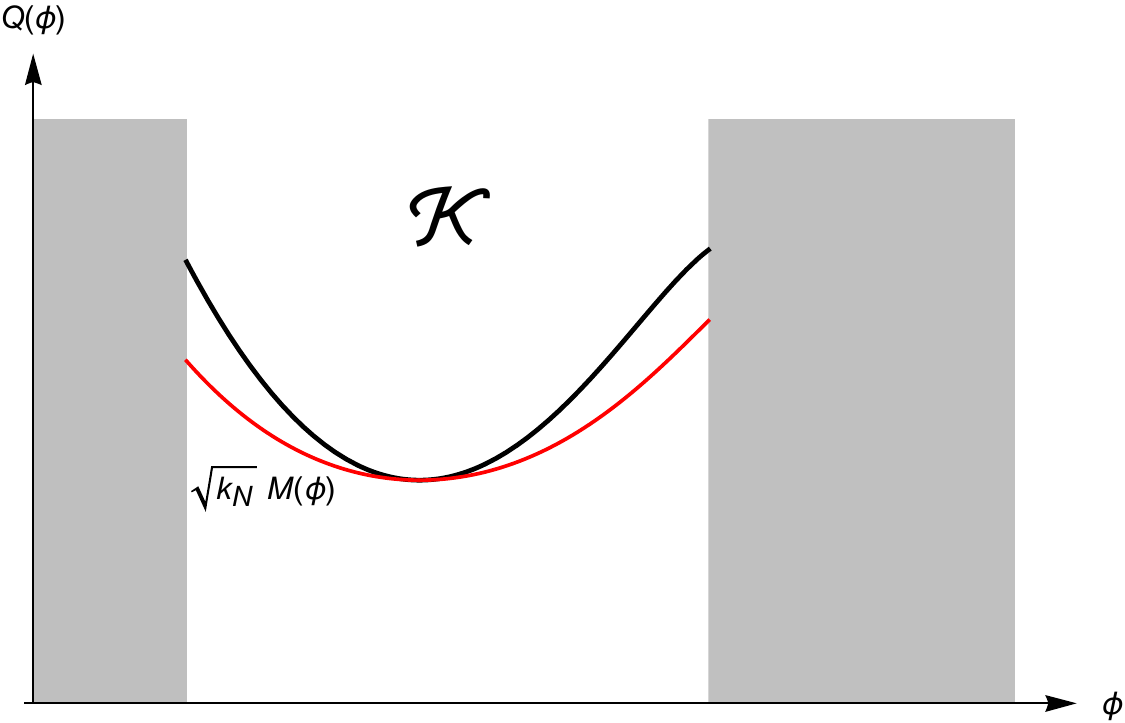}
  \caption{So long as $\phi_{\infty}$ lies within the known region
  $\mathcal{K}$ with $Q (\phi)$ increasing at the boundaries, we can find
  quasiextremal black hole solutions with $\phi (r)$ entirely within
  $\mathcal{K}$ using the fake superpotential method.\label{fig:knownregion}}
\end{figure}

For a one-dimensional moduli space the effective action \eqref{eqn:effaction}
can be written as:
\begin{equation}
  S = \int \biggl[ \frac{1}{2 k_N}  (\chi')^2 + \frac{1}{2}  (\phi')^2 +
  \frac{1}{2} e^{2 \chi} Q^2 (\phi) \biggr] \mathd \tau,
\end{equation}
after performing a field redefinition to set $G_{\phi \phi} = 1$. Suppose we
know the charge function $Q^2 (\phi)$ only in some local patch $\mathcal{K}$
(the \textbf{known region}) of the moduli space, with $Q^2 (\phi)$
increasing at the boundaries of $\mathcal{K}$. Provided $\mathcal{K}$ is
compact, $Q^2 (\phi)$ must then have one or more local minima within
$\mathcal{K}$, and for any $\phi_{\infty} \in \mathcal{K}$ there is a
quasiextremal solution flowing to one of these attractor points, where the
entire solution $\phi (r)$ lies within $\mathcal{K}$. We can find such
solutions using the fake superpotential method, as shown in figure
\ref{fig:knownregion}.

\begin{figure}\centering
  \includegraphics[width=0.8\textwidth]{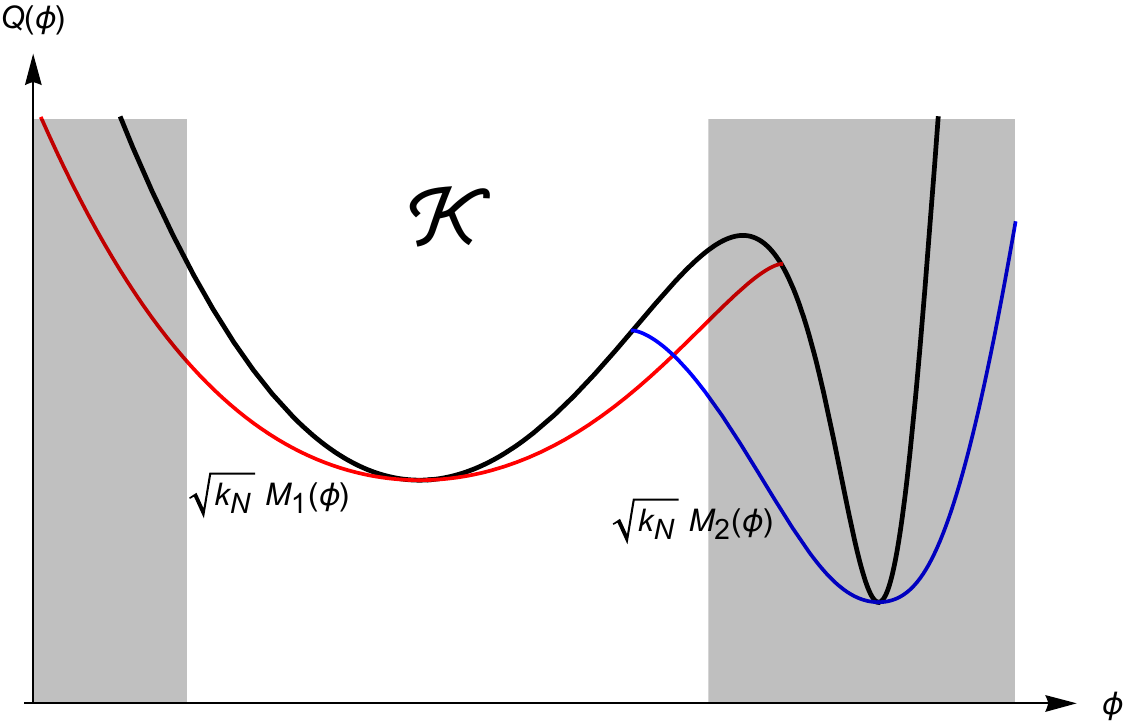}
  \caption{As shown here, $Q (\phi)$ may have ``hidden'' attractor points
  outside of $\mathcal{K}$, where the red (blue) curve shows the mass of
  quasiextremal solutions flowing to the visible (hidden) attractor point.
  Thus, there may be black hole solutions with $\phi_{\infty} \in \mathcal{K}$
  for which the trajectory $\phi (r)$ is not entirely within $\mathcal{K}$.
  These solutions can even be \emph{lighter} than the solutions with $\phi
  (r) \subseteq \mathcal{K}$, as illustrated by the fact that the blue curve
  is sometimes below the red curve.\label{fig:hiddenattractor}}
\end{figure}

The question we wish to answer is: are there other black hole solutions with
$\phi_{\infty} \in \mathcal{K}$ but with $\phi (r)$ moving outside of
$\mathcal{K}$? An example of this phenomenon is shown in figure
\ref{fig:hiddenattractor}. If these black holes exist, we cannot predict their
properties without a better knowledge of the EFT outside of $\mathcal{K}$.
They could be lighter than equal-charged black holes with $\phi (r) \in
\mathcal{K}$ or have other unique properties that are important for
understanding the QGT in question, but with our limited control of the EFT we
might not even know that they exist.

\begin{figure}\centering
  \includegraphics[width=0.8\textwidth]{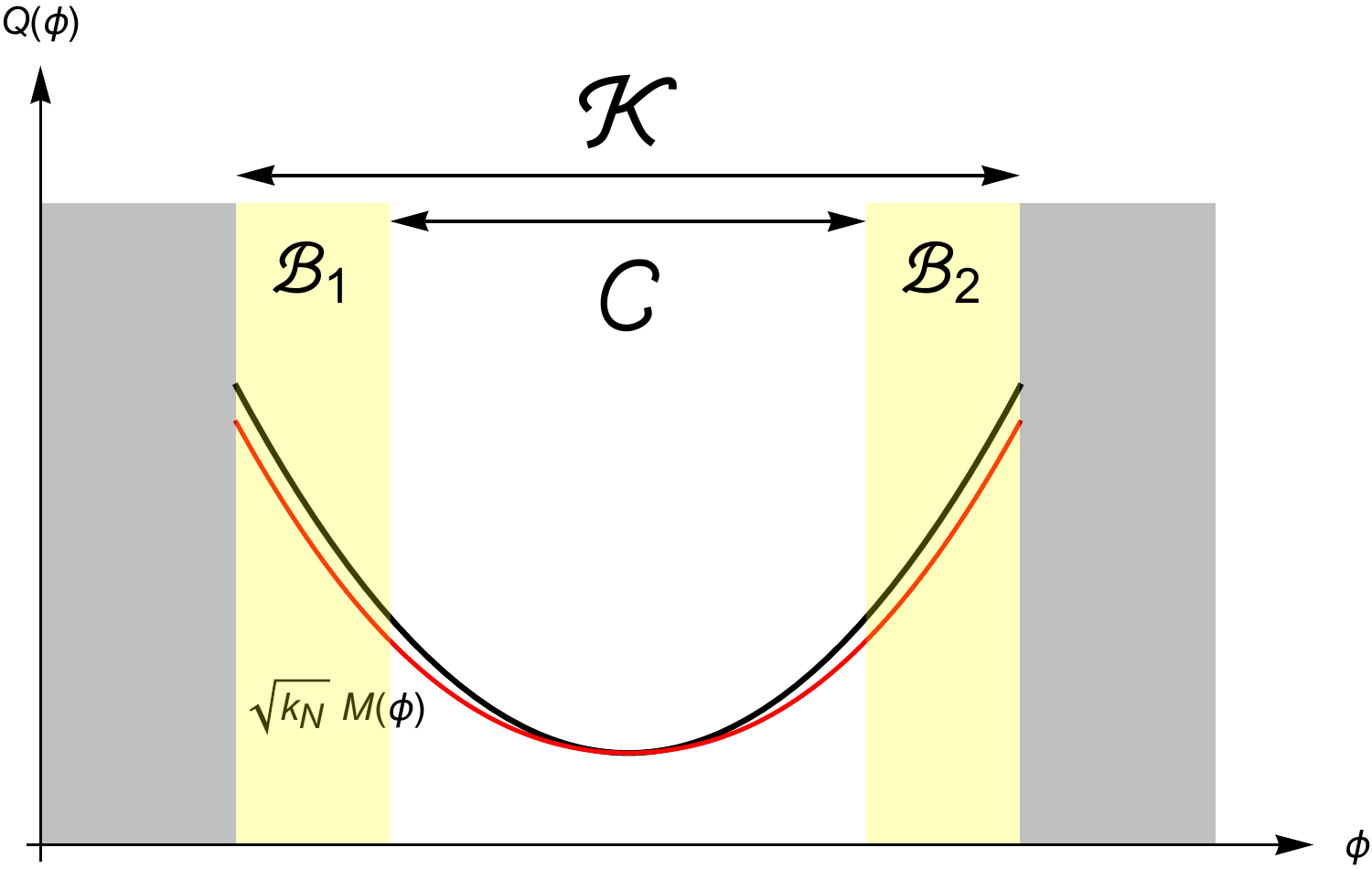}
  \caption{By requiring $Q^2 (\phi)$ to be non-decreasing within bands
  $\mathcal{B}_{\alpha}$ of thickness at least $\frac{\pi}{2 \sqrt{k_N}}$ at
  the edges of the known region $\mathcal{K}$, we can ensure that for any
  $\phi_{\infty}$ within a smaller ``controlled region''
  $\mathcal{C}=\mathcal{K} \setminus \bigcup_{\alpha} \mathcal{B}_{\alpha}$,
  the moduli trajectory $\phi (r)$ will remain entirely within the known
  region for any spherically symmetric black hole of this
  charge.\label{fig:testbands}}
\end{figure}

To answer this question, we leverage the assumption that $Q^2 (\phi)$ is
increasing at the boundaries of $\mathcal{K}$. More precisely, we assume that
it is \emph{non-decreasing} within a \textbf{test band}
$\mathcal{B}_{\alpha}$ of width $w (\mathcal{B}_{\alpha})$ near each of these
boundaries. We will show that as we approach the horizon, the distance $\Delta
\phi$ that the modulus can roll in a direction in which $Q^2 (\phi)$ is
non-decreasing is bounded by $\Delta \phi < \frac{\pi}{2 \sqrt{k_N}}$. Thus,
if $w (\mathcal{B}_{\alpha}) \geqslant \frac{\pi}{2 \sqrt{k_N}}$ for each test
band, then in a vacuum $\phi_{\infty}$ within the \textbf{controlled region}
$\mathcal{C} \assign \mathcal{K} \setminus \bigcup_{\alpha}
\mathcal{B}_{\alpha}$, it is not possible for a spherically symmetric black
hole of this charge to explore a part of moduli space outside of
$\mathcal{K}$, see figure \ref{fig:testbands}.

Now let us study this problem in detail. The black hole equations are:
\begin{equation}
  k_N^{- 1} \chi'' = e^{2 \chi} Q^2 (\phi), \qquad \phi'' = \frac{1}{2} e^{2
  \chi}  \frac{\partial Q^2}{\partial \phi}, \qquad k_N^{- 1} {\chi'}^2 {+
  \phi'}^2 = e^{2 \chi} Q^2 + k_N M_0^2 . \label{eqn:1dBHeqns}
\end{equation}
We are interested in solutions within the test bands, where $Q^2 (\phi)$ is
monotonic, e.g.,
\begin{equation}
  \frac{\partial}{\partial \phi} Q^2 \geqslant 0, \label{eqn:Qmonotone}
\end{equation}
within the band on the right side of $\mathcal{K}$. We focus on this band for
definiteness below. The band on the left side is analogous with $\phi
\rightarrow - \phi$.

Our argument proceeds roughly as follows:
\begin{enumerate}
  \item We define a ``clock'' $\xi (\tau)$ such that, given our assumptions,
  $\xi_{\min} \leqslant \xi (\tau) \leqslant \xi_{\max}$ and $\xi' (\tau)
  \geqslant 0$ (i.e., the clock has some minimum and maximum settings and
  moves only in one direction).
  
  \item We place a lower bound on the rate $\frac{\mathd \xi}{\mathd \phi}$ at
  which the clock advances per moduli space distance as we move through the
  test band.
  
  \item By integrating the inverse of this bound, we place on upper bound on
  the distance $\Delta \phi$ travelled in moduli space before the clock
  ``expires'' at $\xi = \xi_{\max}$:
  \begin{equation}
    \Delta \phi \leqslant \int_{\xi_{\min}}^{\xi_{\max}} \biggl[ \frac{\mathd
    \xi}{\mathd \phi} \biggr]^{- 1} \mathd \xi .
  \end{equation}
\end{enumerate}
The clock we will use is:
\begin{equation}
  \xi (\tau) \assign \frac{\sqrt{k_N} \phi'}{\sqrt{(\chi')^2 - k_N^2 M_0^2}} .
  \label{eqn:xidefn}
\end{equation}
Note that the smooth horizon condition $\chi' < - k_N M_0$ ensures that
expression inside the square root is positive, so that $\xi < \infty =
\xi_{\max}$. Since we are interested in solutions that are moving through the
test band, $\phi' > 0$ initially and so the initial setting of the clock is
positive, $\xi > \xi_{\min} = 0$. Moreover, we find:
\begin{align}
  \xi' &= \frac{\sqrt{k_N} \phi''}{\sqrt{(\chi')^2 - k_N^2 M_0^2}} -
  \frac{\sqrt{k_N} \phi' \chi' \chi''}{[(\chi')^2 - k_N^2 M_0^2]^{3 / 2}}
  \nonumber\\
  &= \left[ {\chi'}^2 + k_N {\phi'}^2 - k_N^2 M_0^2 \right] \left[
  \frac{\frac{1}{2 \sqrt{k_N}}  \frac{\partial}{\partial \phi} \log
  Q^2}{\sqrt{(\chi')^2 - k_N^2 M_0^2}} - \frac{\sqrt{k_N} \phi'
  \chi'}{[(\chi')^2 - k_N^2 M_0^2]^{3 / 2}} \right] \nonumber\\
  &= \sqrt{k_N} \phi'  [1 + \xi^2] \left[ \frac{1}{2 \xi \sqrt{k_N}} 
  \frac{\partial}{\partial \phi} \log Q^2 + \frac{1}{\sqrt{1 - \mu^2 /
  \mu_h^2}} \right],  \label{eqn:clockrate}
\end{align}
where in the second step we use the black hole equations rewritten as:
\begin{equation}
  \phi'' = \frac{1}{2 k_N}  \left[ {\chi'}^2 - k_N^2 M_0^2 + k_N {\phi'}^2
  \right]  \frac{\partial}{\partial \phi} \log Q^2, \qquad \chi'' {= \chi'}^2
  + k_N {\phi'}^2 - k_N^2 M_0^2,
\end{equation}
and in the last step we define:
\begin{equation}
  \mu \assign \frac{1}{- \chi'}, \qquad \mu_h \assign \frac{1}{k_N M_0},
\end{equation}
so that $0 < \mu < \mu_h$ by the smooth horizon condition $\chi' < - k_N M_0$.
Thus, if $\phi' > 0$ initially and the condition \eqref{eqn:Qmonotone} is
satisfied then the ``clock'' moves only in one direction $\xi' > 0$, which in
turn ensures that $\phi'$ remains positive thereafter.

Using \eqref{eqn:clockrate}, we can read off the rate at which the clock
advances per moduli space distance:
\begin{equation}
  \frac{\mathd \xi}{\mathd \phi} = \frac{\xi'}{\phi'} = \sqrt{k_N}  [1 +
  \xi^2] \left[ \frac{1}{2 \xi \sqrt{k_N}}  \frac{\partial}{\partial \phi}
  \log Q^2 + \frac{1}{\sqrt{1 - \mu^2 / \mu_h^2}} \right] \geqslant \sqrt{k_N}
  [1 + \xi^2], \label{eqn:ratebound}
\end{equation}
using the condition \eqref{eqn:Qmonotone}. Therefore,
\begin{equation}
  \Delta \phi = \int \mathd \phi = \int \left[ \frac{\mathd \xi}{\mathd \phi}
  \right]^{- 1} \mathd \xi < \int_0^{\infty} \frac{\mathd \xi}{\sqrt{k_N}  [1
  + \xi^2]} = \frac{\pi}{2 \sqrt{k_N}},
\end{equation}
where the bound is strict because the initial setting of the clock is strictly
positive, $\xi > 0$. Thus, if the test band is sufficiently thick, $w
(\mathcal{B}) \geqslant \frac{\pi}{2 \sqrt{k_N}}$, then it is impossible for a
black hole solution to cross through it beginning from inside $\mathcal{C}$!

\begin{figure}\centering
  \includegraphics[width=0.8\textwidth]{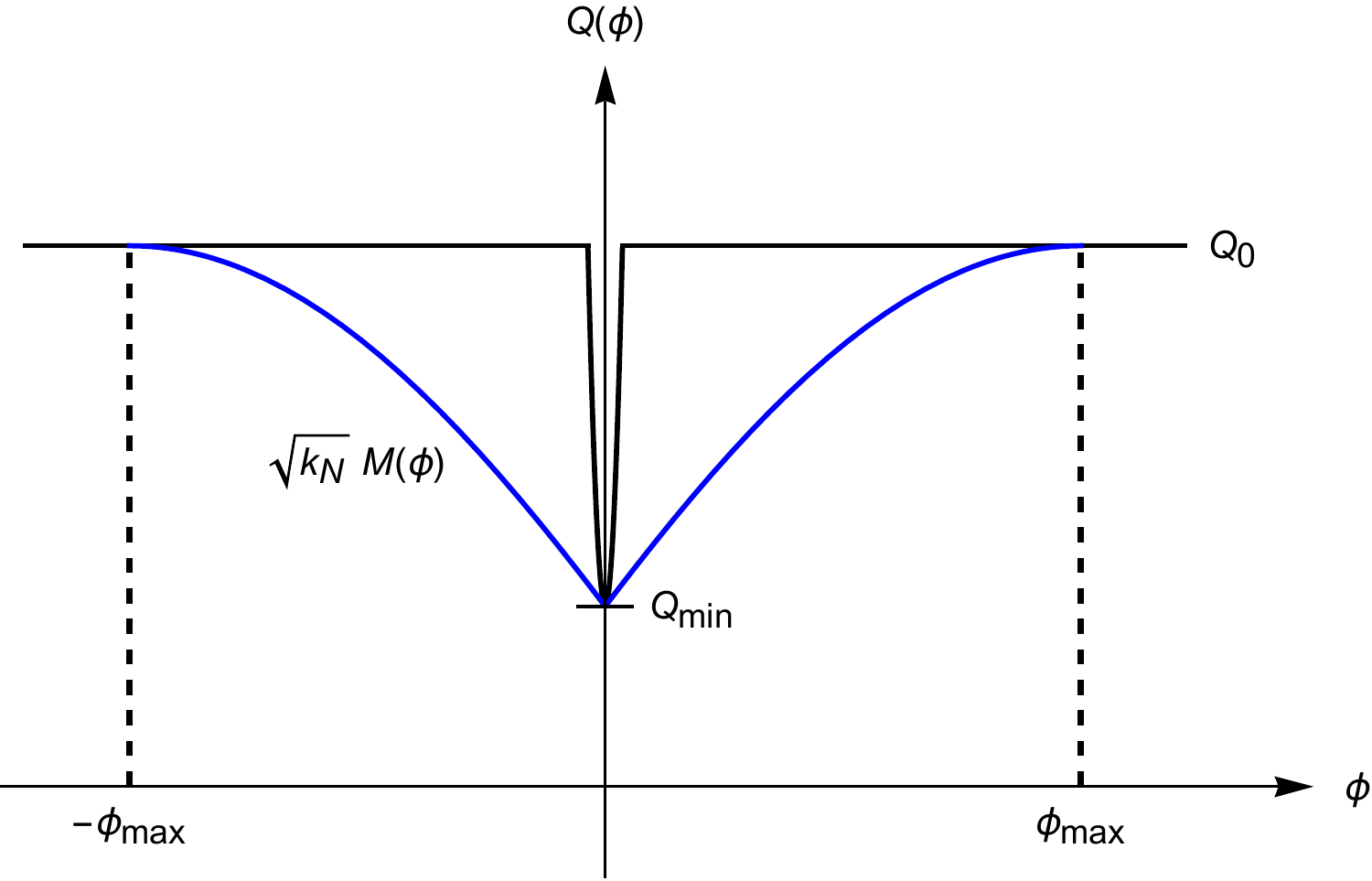}
  \caption{A charge function $Q (\phi)$ which is flat away from a sharply
  localized dip at $\phi = 0$. There is a finite-sized attractor basis $\phi
  \in [- \phi_{\max}, \phi_{\max}]$ around the attractor point at $\phi = 0$,
  where $\phi_{\max} \rightarrow \frac{\pi}{2 \sqrt{k_N}}$ in the limit
  $Q_{\min} / Q_0 \rightarrow 0$.\label{fig:saturatingexample}}
\end{figure}

Thus, we conclude that:
\begin{tcolorbox}[colback=yellow!20, colframe=black,  boxrule=0.8pt,  arc=0pt,  left=2mm,  right=2mm,  top=1mm,  bottom=1mm]
The distance $\Delta\phi$ that can be traversed with $Q^2$ monotonically increasing is bounded,
\begin{equation}
  \Delta\phi < \frac{\pi}{2 \sqrt{k_N}}\,. \label{eqn:monotonicBound}
\end{equation}
\end{tcolorbox}
\noindent
The preceding bound is a strict inequality, but it is not yet clear whether
this inequality is \emph{optimum}. We now show that there are valid black
hole solutions that come arbitrarily close to saturating this bound. For this
purpose, we consider a $Q^2 (\phi)$ that is constant, $Q^2 = Q_0^2$, except in
a very narrow region near $\phi = 0$, where it briefly dips down to
$Q_{\text{min}}^2$, see figure \ref{fig:saturatingexample}.

We can determine construct quasiextermal solutions by solving the fake
superpotential equation:
\begin{equation}
  k_N M^2 (\phi) + G^{i j} \partial_i M \partial_j M = Q^2 (\phi),
\end{equation}
with the boundary condition $M (0) = \frac{1}{\sqrt{k_N}} Q_{\text{min}}^2$ at
the attractor point $\phi = 0$. Away from $\phi = 0$, $Q^2 \rightarrow Q_0^2$
rapidly, and so the equation we have to solve is:
\begin{equation}
  k_N  [M (\phi)]^2 + [M' (\phi)]^2 = Q_0^2 \qquad \text{with the boundary
  condition} \qquad M (0) = \frac{1}{\sqrt{k_N}} Q_{\min} .
\end{equation}
The general solution to the ODE is:
\begin{equation}
  M (\phi) = \frac{Q_0}{\sqrt{k_N}} \sin \left[ \sqrt{k_N} (\phi - \phi_0)
  \right],
\end{equation}
for some constant $\phi_0$. Imposing the boundary condition, we find:
\begin{equation}
  M (\phi) = \frac{1}{\sqrt{k_N}}  \left[ \sqrt{Q_0^2 - Q_{\min}^2} \sin
  \left( \sqrt{k_N} | \phi | \right) + Q_{\min} \cos \left( \sqrt{k_N} \phi
  \right) \right],
\end{equation}
as plotted in the figure above. The attractor basin associated to the $\phi =
0$ attractor is the region in which an $M (\phi)$ gradient flow ends up at
$\phi = 0$. Thus, the boundaries of the attractor basin are the two critical
points of $M (\phi)$ adjacent to the one at $\phi = 0$. We find:
\begin{align}
  M' (\phi) &= \sqrt{Q_0^2 - Q_{\min}^2} \sgn (\phi) \cos \left(
  \sqrt{k_N} \phi \right) - Q_{\min} \sin \left( \sqrt{k_N} \phi \right) = 0
  \nonumber\\
  &\noeq \Rightarrow \quad \phi_{\text{crit}} = \pm \phi_{\max}, \quad
  \text{where} \quad \phi_{\max} \assign \frac{1}{\sqrt{k_N}} \arctan \sqrt{1
  - \frac{Q_{\min}^2}{Q_0^2}} . 
\end{align}

Since its argument is at most 1, the arctan is at most $\frac{\pi}{2}$, so the
attractor basin is no wider than $\frac{\pi}{2 \sqrt{k_N}}$, where the width
is maximized when we take $Q_{\min} \rightarrow 0$ (where we have already
assumed that the dip in $Q^2 (\phi)$ at $\phi = 0$ is arbitrarily narrow). In
fact, the width of this attractor basin can never reach $\frac{\pi}{2
\sqrt{k_N}}$, both because $Q_{\min} > 0$ is required to have a horizon of
finite area, and because the dip in $Q (\phi)$ cannot have zero width.

\begin{figure}\centering
  \includegraphics[width=0.8\textwidth]{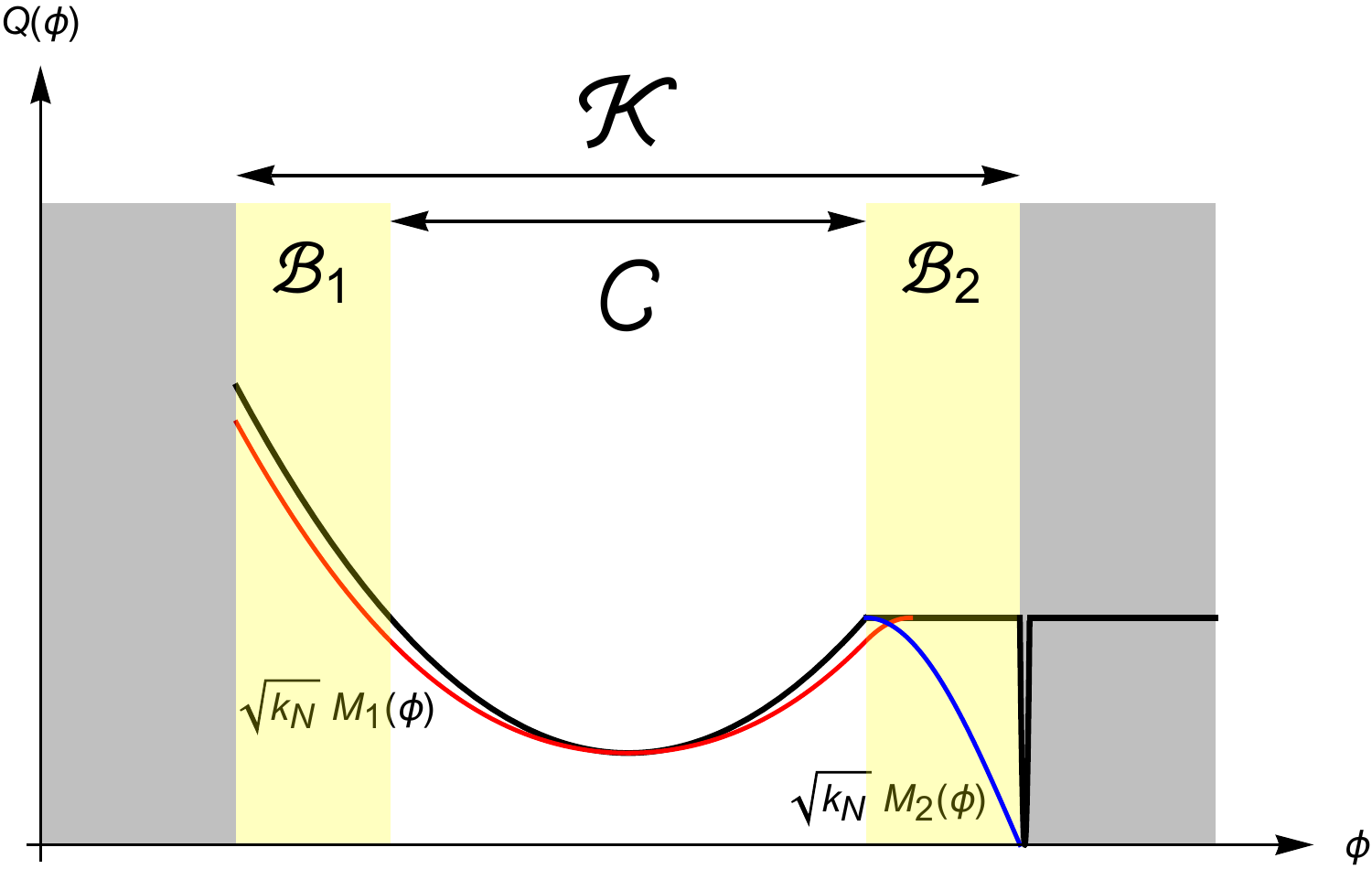}
  \caption{An example of a charge function with an attractor point that is
  ``hidden'' just outside of the known region $\mathcal{K}$ with an attractor
  basin that covers an arbitrarily large fraction of the test band
  $\mathfrak{B}_2$, thereby coming arbitrarily close to saturating the bound
  \eqref{eqn:monotonicBound}.\label{fig:saturatingexample2}}
\end{figure}

Using this mechanism, we can construct examples that come arbitrarily close to
saturating the bound \eqref{eqn:monotonicBound}, as shown in figure
\ref{fig:saturatingexample2}. Notice that, in order to do so, we must have
$Q^2 (\phi)$ be very close to \emph{flat} in the test band. If, on the
contrary, $Q^2 (\phi)$ increases more rapidly in this band, then we can set a
stronger bound, as follows. Suppose that
\begin{equation}
  \frac{\mathd}{\mathd \phi} \log Q^2 \geqslant 2 \alpha \sqrt{k_N},
  \label{eqn:QslopeCond}
\end{equation}
throughout $\mathcal{B}$, where $\alpha$ is some non-negative dimensionless
constant.\footnote{Note that in comparison to \cite{Heidenreich:2015nta},
$\alpha_{\text{there}} = \alpha_{\text{here}}  \sqrt{2 \frac{d - 3}{d - 2} }$.
In our current conventions, the extremality bound for a dilatonic black hole
can be written simply as $k_N M^2 \geqslant \frac{1}{1 + \alpha^2} Q^2$.} Then
\eqref{eqn:ratebound} becomes:
\begin{equation}
  \frac{\mathd \xi}{\mathd \phi} = \sqrt{k_N}  [1 + \xi^2] \left[ \frac{1}{2
  \xi \sqrt{k_N}}  \frac{\partial}{\partial \phi} \log Q^2 + \frac{1}{\sqrt{1
  - \mu^2 / \mu_h^2}} \right] \geqslant \sqrt{k_N}  [1 + \xi^2] [1 + \alpha /
  \xi],
\end{equation}
which implies that
\begin{equation}
  \Delta \phi < \int_0^{\infty} \frac{\mathd \xi}{\sqrt{k_N}  [1 + \xi^2] [1 +
  \alpha / \xi]} = \frac{\pi + 2 \alpha \log \alpha}{2 \sqrt{k_N}  (1 +
  \alpha^2)} .
\end{equation}
Note that the right-hand side is a decreasing function of $\alpha$, so for
$\alpha > 0$ this is strictly stronger than our previous bound $\Delta \phi <
\frac{\pi}{2 \sqrt{k_N}}$.

\subsection{Geometric interpretation}

We can reinterpret the above bound geometrically use Maupertuis's principle
(reviewed in Appendix \ref{app:Maupertuis}). Recall the conserved ``energy''
\eqref{eqn:conservedE}
\begin{equation}
  \frac{1}{2} k_N M_0^2 = \frac{1}{2 k_N} {\chi'}^2 + \frac{1}{2} G_{i j}
  {\phi^i}' {\phi^j}' - \frac{1}{2} e^{2 \chi} Q^2 (\phi) .
\end{equation}
According to Maupertuis's principle, the solutions of fixed $M_0$ are the
geodesics of the Maupertuis metric:
\begin{equation}
  \widetilde{\mathd s}^2 = [e^{2 \chi} Q^2 (\phi) + k_N M_0^2] \left[
  \frac{1}{k_N} \mathd \chi^2 + G_{i j} (\phi) \mathd \phi^i \mathd \phi^j
  \right] . \label{eqn:MaupertuisMetric}
\end{equation}
For convenience, we call this extended moduli space the \textbf{Maupertuis
space}. Thus, the problem of finding black hole solutions reduces to a
geodesic problem in Maupertuis space.

The quasiextremal case, $M_0 = 0$, is particularly interesting for several
reasons. Firstly, in this case we have:
\begin{align}
  S_{\text{geom}} &= \int \sqrt{e^{2 \chi} Q^2 (\phi) \left[ \frac{1}{k_N}
  (\chi')^2 + G_{i j} (\phi) {\phi^i}' {\phi^j}' \right]} \mathd \tau \nonumber\\&= \int
  e^{2 \chi} Q^2 (\phi) \mathd \tau = \int k_N^{- 1} \chi'' \mathd \tau = -
  k_N \chi' (0) = M_{\text{BH}}, \label{eqn:SgeomQuasiExt}
\end{align}
evaluated on a black hole solution, where we use $\chi' \rightarrow 0$ as
$\tau \rightarrow \infty$. Thus, the length of the geodesic is the mass of the
corresponding black hole solution. Moreover, the horizon condition is $\chi' <
0$, so the allowable geodesics are those that go from $\chi = 0$ to $\chi = -
\infty$ without ever reversing direction.

\begin{figure}\centering
  \includegraphics[width=0.6\textwidth]{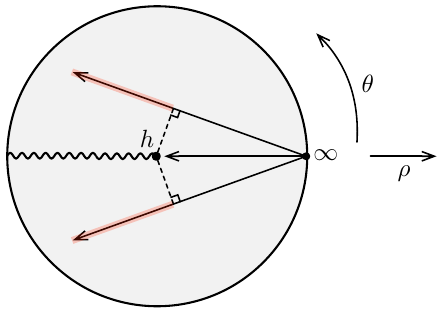}
  \caption{Geodesics on the Maupertuis disk originating at the point
  ``$\infty$'' representing the vacuum with
  $\theta_{\infty} = 0$. When $Q^2 (\theta)$ is constant, these geodesics are
  straight lines, exactly one of which passes through the point
  ``$h$'' representing the horizon. If $Q^2 (\theta)$ is not globally constant, then these lines may eventually bend when they reach a
  region with varying $Q^2$, but after they start moving outwards (highlighted in pink) it is no longer possible for them to reach the
  horizon point, no matter what $Q^2 (\theta)$ they encounter
  subsequently.\label{fig:constQ}}
\end{figure}

One can write the quasiextremal Maupertuis metric somewhat more simply in
terms of $\rho \equiv e^{\chi}$:
\begin{equation}
  \widetilde{\mathd s}^2 = \frac{1}{k_N} Q^2 (\phi)  [\mathd \rho^2 + \rho^2 k_N
  G_{i j} (\phi) \mathd \phi^i \mathd \phi^j] . \label{eqn:MaupertuisCone}
\end{equation}
Thus, the Maupertuis space is a cone over the moduli space, warped by $Q^2
(\phi)$. Now let us specialize to a 1d moduli space, writing $\mathd s^2 =
\frac{1}{k_N} \mathd \theta^2$, so that:
\begin{equation}
  \widetilde{\mathd s}^2 = \frac{1}{k_N} Q^2 (\phi)  [\mathd \rho^2 + \rho^2
  \mathd \theta] .
\end{equation}
If $Q^2$ were constant and $\theta$ were periodic with period $2 \pi$ then
this would be the flat metric on a disk of radius $Q / \sqrt{k_N}$. If,
instead, $\theta$ is not periodic then we get a branched cover of the disk.
This, however, has no effect on the geodesics, which are straight lines on the
disk provided that $Q^2$ remains constant, see figure \ref{fig:constQ}.

\begin{figure}\centering
  \includegraphics[width=0.6\textwidth]{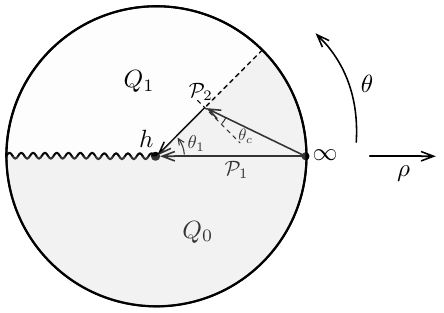}
  \caption{The effect of a nearby region where $Q(\theta)$
  \emph{decreases}. The path $\mathcal{P}_2$ can be shorter than the path
  $\mathcal{P}_1$, provided that $\theta_1 < \arccos \frac{Q_1}{Q_0}$. Thus, a
  hidden attractor is possible if the region where $Q(\theta)$ is smaller
  is close enough. Note that $Q_0, Q_1$ can be thought of as indices of
  refraction, where the angle $\theta_c = \arcsin \frac{Q_1}{Q_0}$ is the
  critical angle for total internal reflection.\label{fig:piecewiseConstQ}}
\end{figure}

Suppose for instance, that $Q^2 (\theta)$ is constant in some interval $[-
\varepsilon, \Delta \theta]$ including the point $\theta_{\infty} = 0$.
Clearly, one possible geodesic joining the infinity and horizon points is the
straight line $\theta = 0$. Let us ask whether it is possible to pick $Q^2
(\theta)$ in the region $\theta > \Delta \theta$ such that there is
\emph{another} geodesic that also joins $\infty$ to $h$ and is shorter.
This is indeed possible. For example, let $Q (\theta)$ be piecewise constant,
\begin{equation}
  Q (\theta) = \begin{cases}
    Q_0,  & \theta < \theta_1,\\
    Q_1,  & \theta > \theta_1,
  \end{cases} \qquad \text{for some $\theta_1 > 0$} .
\end{equation}
Then due to the warping in the Maupertuis metric, there may be a
shorter path to $h$ that diverts to the region $\theta > \theta_1$ before
moving to the origin, see figure \ref{fig:piecewiseConstQ}.

In particular, by a simple minimization problem, one finds that a shorter path
exists provided that
\begin{equation}
  \theta_1 < \gamma \equiv \arccos \left( \frac{Q_1}{Q_0} \right),
  \label{eqn:thetabound}
\end{equation}
which results in a lighter extremal black hole, of mass:
\begin{equation}
  M_{\text{light}} = \frac{Q_0}{\sqrt{k_N}} \cos (\gamma - \theta_1) <
  \frac{Q_0}{\sqrt{k_N}} .
\end{equation}
Note that since $Q_0 \geqslant 0$, such a path can only exist if $\theta_1 <
\frac{\pi}{2}$, which is precisely the bound \eqref{eqn:monotonicBound}. In
fact, this is just an optics problem, where $Q (\theta)$ plays the role of the
refractive index and the ray joining the point $\infty$ to the interface
strikes at the threshold angle for total internal reflection:
\begin{equation}
  \theta_c = \arcsin \left( \frac{Q_1}{Q_0} \right) .
\end{equation}
This provides a simple and intuitive explanation for the limit
\eqref{eqn:thetabound}.\footnote{Here we use $\theta_1 + ( \frac{\pi}{2}
+ \theta_c ) < \pi$, since the internal angles of a triangle add up to
$\pi$.}

More generally, $Q^2 (\theta)$ will vary continuously with $\theta$. As above,
we can think of this as a refractive index on the disk, which varies with the
angular coordinate. Whenever $Q^2 (\theta)$ is \emph{increasing}, this has
the effect of bending the trajectory towards the normal to a constant $Q$
surface, i.e., towards the direction $\vec{\nabla} Q$ in which $Q$ is
increasing. Since the trajectory must be inward to satisfy the horizon
condition $\chi' < 0$ and $\vec{\nabla} Q$ points in the angular direction,
this means that as the path moves into a region with larger $Q^2 (\theta)$, it
bends \emph{outward}, see figure \ref{fig:increasingQ}. This means that
the horizon condition will be violated \emph{sooner} (i.e., after a
smaller angular displacement $\Delta \theta$) then if $Q^2$ were constant:

\begin{figure}\centering
  \includegraphics[width=0.6\textwidth]{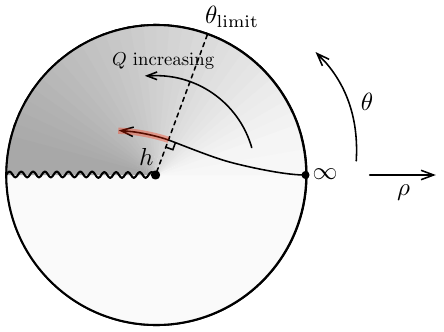}
  \caption{When $Q^2$ is increasing, it causes the path to bend towards
  $\hat{\theta}$, so that the horizon condition is violated more quickly than
  in figure \ref{fig:constQ}.\label{fig:increasingQ}}
\end{figure}

Thus, so long as $\frac{\mathd}{\mathd \theta} Q^2 \geqslant 0$ in the region
in question, we have geometrically proved the bound:
\begin{equation}
  \Delta \theta < \frac{\pi}{2} \qquad \text{i.e.,} \qquad \Delta \phi <
  \frac{\pi}{2 \sqrt{k_N}},
\end{equation}
for a quasiextremal black hole, which is the same as the bound
\eqref{eqn:monotonicBound}.

\begin{figure}\centering
  \includegraphics[width=0.65\textwidth]{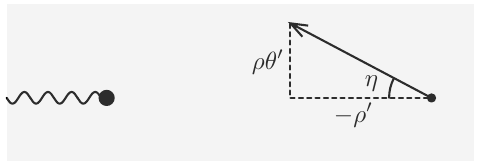}
  \caption{The angle and velocity of the trajectory on the Maupertuis
  disk.\label{fig:MaupertuisAngle}}
\end{figure}

To make the connection between this argument and our previous argument more
precise, consider the angle $\eta$ between the direction of the path on the
disk and the inwards $(- \hat{\rho})$ direction, see figure
\ref{fig:MaupertuisAngle}. We find
\begin{equation}
  \tan \eta = - \frac{\rho \theta'}{\rho'} = - \frac{\sqrt{k_N} \phi'}{\chi'}
  = \xi,
\end{equation}
where we refer to \eqref{eqn:xidefn}, noting that $\mu_h = \infty$ since we
are discussing the quasiextremal case. Note that the horizon condition $\chi'
< 0$ translates to $- \tfrac{\pi}{2} < \eta < \frac{\pi}{2}$.

We now compute:
\begin{align}
  \frac{\mathd \xi}{\mathd \theta} &= \frac{\xi'}{\theta'} = \frac{- \chi'
  \phi'' + \chi'' \phi'}{\phi' {\chi'}^2} = \frac{- \frac{1}{2} \chi' e^{2
  \chi}  \frac{\partial Q^2}{\partial \phi} + k_N e^{2 \chi} Q^2 \phi'}{\phi'
  {\chi'}^2} \nonumber\\
  &= \left[ 1 + \frac{k_N {\phi'}^2}{{\chi'}^2} \right]  \left[ 1 +
  \frac{1}{2}  \frac{- \chi'  \frac{\partial \log Q^2}{\partial
  \phi}}{k_N \phi'} \right] = [1 + \xi^2] \left[ 1 + \frac{1}{2 \xi} 
  \frac{\mathd}{\mathd \theta} \log Q^2  \right], 
\end{align}
using the black hole equations \eqref{eqn:1dBHeqns}. Thus,
\begin{equation}
  \frac{\mathd \eta}{\mathd \theta} = \frac{1}{1 + \xi^2}  \frac{\mathd
  \xi}{\mathd \theta} = 1 + \frac{1}{2} \cot \eta \frac{\mathd}{\mathd \theta}
  \log Q^2 .
\end{equation}
Suppose that $\theta' > 0$ initially, so that $\eta > 0$ initially. If $Q^2$
is constant then we see that $\eta$ increases in proportion to $\theta$. This
is simply the equation describing a straight line, and we reproduce the bound
$\Delta \theta < \frac{\pi}{2}$, since $\frac{\pi}{2} - \eta_{\text{init}} <
\frac{\pi}{2}$. If, on the other hand, $Q^2$ is increasing with $\theta$, then
$\eta$ increases \emph{more} quickly. and the bound on $\Delta \theta$
becomes stronger. This is the algebraic form of the geometric argument given
above.

\subsection{A bound on the distance travelled in regions where \alt{$Q^2 (\phi) \geqslant Q_0^2$}{Q²(phi) > Q₀²}}\label{sec:strongerbound}

Before moving on to the multidimensional case, we note that a slightly
stronger bound is possible, which does not require $\partial_{\phi} Q^2
\geqslant 0$ everywhere in the test band.

To derive this bound, first consider the quasiextremal case and suppose that
the fake superpotential $M (\phi)$ describing the hidden attractor is known.
The rate at $M$ decreases per moduli space distance takes a rather simple form
\begin{align}
  {\phi^i}' &= - e^{\chi} G^{i j} \nabla_j M \quad \Longrightarrow \quad (s')^2
  \equiv G_{i j} {\phi^i}' {\phi^j}' = e^{2 \chi} | \nabla M |^2
  \quad \Longrightarrow \nonumber\\ & \phantom{=}\Longrightarrow\frac{\mathd M}{\mathd s} = \frac{M'}{s'} =
  \frac{\nabla_i M}{s'} {\phi^i}' = - | \nabla M | .
\end{align}
Thus, the fake superpotential equation implies that:
\begin{equation}
  \frac{\mathd M}{\mathd s} = - \sqrt{Q^2 - k_N M^2} .
\end{equation}
Let $Q_0$ be the initial value of $Q$ at $\phi = 0$, which implies that $M
\leqslant \frac{Q_0}{\sqrt{k_N}}$ initially. Let us parameterize $M$ as:
\begin{equation}
  M = \frac{Q_0}{\sqrt{k_N}} \cos \vartheta, \qquad \text{for} \qquad
  \vartheta \in \left[ 0, \frac{\pi}{2} \right) .
\end{equation}
Then we obtain:
\begin{equation}
  \frac{\mathd \vartheta}{\mathd s} = \sqrt{k_N}  \frac{1}{\sin \vartheta} 
  \sqrt{\frac{Q^2}{Q_0^2} - \cos^2 \vartheta} .
\end{equation}
We see that $\vartheta$ increases monotonically as $M$ decreases along the
flow. Suppose that the flow passes through a band $\phi \in [\phi_1, \phi_2]$
within which $Q^2 \geqslant Q_0^2$. Then within this band,
\begin{equation}
  \frac{\mathd \vartheta}{\mathd s} \geqslant \sqrt{k_N}  \frac{1}{\sin
  \vartheta}  \sqrt{1 - \cos^2 \vartheta} = \sqrt{k_N}, \qquad \Rightarrow
  \qquad \Delta \vartheta \geqslant \sqrt{k_N}  (\phi_2 - \phi_1) .
\end{equation}
Thus, since $\vartheta \in [ 0, \frac{\pi}{2} )$ to maintain $M >
0$, the distance travelled by the flow in regions where $Q^2 \geqslant Q_0^2$
can be no greater than $\frac{\pi}{2 \sqrt{k_N}}$ \emph{in total},
regardless of whether $Q^2$ is locally increasing or decreasing.

To generalize this bound to non-extremal cases, we define the clock:
\begin{equation}
  \mathfrak{M} \assign k_N^{- 1} e^{- \chi}  \sqrt{(\chi')^2 - (k_N M_0)^2} .
\end{equation}
This coincides with the value of the fake superpotential when one exists.
Moreover,
\begin{align}
  \mathfrak{M}' &= - k_N^{- 1} \chi' e^{- \chi}  \sqrt{(\chi')^2 - (k_N M_0)^2}
  + e^{- \chi}  \frac{\chi' e^{2 \chi} Q^2}{\sqrt{(\chi')^2 - (k_N M_0)^2}}  \nonumber\\
  &=
  \frac{\chi' e^{- \chi}}{\sqrt{(\chi')^2 - (k_N M_0)^2}} G_{i j} {\phi^i}'
  {\phi^j}' \leqslant 0,
\end{align}
so $\mathfrak{M}$ is monotonically decreasing as we approach the horizon.
Eliminating $\chi' = - k_N \sqrt{e^{2 \chi} \mathfrak{M}^2 + M_0^2}$ in favor
of $\mathfrak{M}$ and rewriting $G_{i j} {\phi^i}' {\phi^j}' = (s')^2$, this
becomes:
\begin{equation}
  \mathfrak{M}' = - e^{- \chi}  \frac{\sqrt{\mathfrak{M}^2 + e^{- 2 \chi}
  M_0^2}}{\mathfrak{M}}  (s')^2 .
\end{equation}
Likewise, performing the same substitutions on the constraint equation
\eqref{eqn:BHeqns-cons}, we obtain:
\begin{equation}
  k_N e^{2 \chi} \mathfrak{M}^2 + (s')^2 = e^{2 \chi} Q^2  \qquad \Rightarrow
  \qquad e^{- \chi} s' = \sqrt{Q^2 - k_N \mathfrak{M}^2} . \label{eqn:sprimeM}
\end{equation}
Thus, we obtain:
\begin{equation}
  \frac{\mathd \mathfrak{M}}{\mathd s} = \frac{\mathfrak{M}'}{s'} = - e^{-
  \chi}  \frac{\sqrt{\mathfrak{M}^2 + e^{- 2 \chi} M_0^2}}{\mathfrak{M}} s' =
  - \frac{\sqrt{\mathfrak{M}^2 + e^{- 2 \chi} M_0^2}}{\mathfrak{M}}  \sqrt{Q^2
  - k_N \mathfrak{M}^2} .
\end{equation}
This implies that $\mathfrak{M}$ decreases at a minimum rate:
\begin{equation}
  \frac{\mathd \mathfrak{M}}{\mathd s} \leqslant - \sqrt{Q^2 - k_N
  \mathfrak{M}^2} .
\end{equation}
Therefore, if $\mathfrak{M} \leqslant \frac{Q_0}{\sqrt{k_N}}$ initially, then
parameterizing $\mathfrak{M}= \frac{Q_0}{\sqrt{k_N}} \cos \vartheta$ as before
we conclude that:
\begin{equation}
  \frac{\mathd \vartheta}{\mathd s} \geqslant \sqrt{k_N}  \frac{1}{\sin
  \vartheta}  \sqrt{\frac{Q^2}{Q_0^2} - \cos^2 \vartheta} .
\end{equation}
As before, this implies that $\Delta \vartheta \geqslant \sqrt{k_N} \Delta s$
in regions where $Q^2 \geqslant Q_0^2$. Thus, since $\Delta \vartheta
\leqslant \frac{\pi}{2}$, we conclude that
\begin{tcolorbox}[colback=yellow!20, colframe=black,  boxrule=0.8pt,  arc=0pt,  left=2mm,  right=2mm,  top=1mm,  bottom=1mm]
The total distance $\Delta\phi$ in regions with $Q^2 \ge Q^2_{\text{initial}}$ that can be traversed is bounded,
\begin{equation}
  \Delta\phi < \frac{\pi}{2 \sqrt{k_N}}\,. \label{eqn:totDistBound}
\end{equation}
\end{tcolorbox}
\noindent
Note that the bound on $\Delta \phi$ is strict, since $\Delta \vartheta =
\sqrt{k_N} \Delta s$ requires $Q^2 (\phi) = Q_0^2$ to be constant, but then
$s' \neq 0$ initially (or else we would not move in the moduli space at all)
and therefore $\vartheta_{\text{initial}}$ is strictly positive by
\eqref{eqn:sprimeM}.

\begin{figure}\centering
  \includegraphics[width=0.8\textwidth]{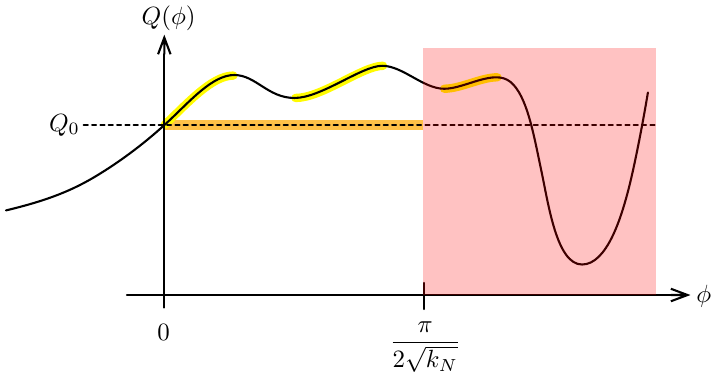}
  \caption{An example where the bound \eqref{eqn:totDistBound} proves that a
  hidden attractor cannot be reached, even though the bound
  \eqref{eqn:monotonicBound} does not. In the orange highlighted region, $Q^2$
  is above its initial value $Q_0^2$. Thus, the red region cannot be reached.
  However, $\frac{\partial}{\partial \phi} Q^2 > 0$ only in the yellow
  highlighted regions, none of which is wide enough to rule out reaching the
  hidden attractor using \eqref{eqn:monotonicBound}.\label{fig:Qosc}}
\end{figure}

\begin{figure}\centering
  \includegraphics[width=0.8\textwidth]{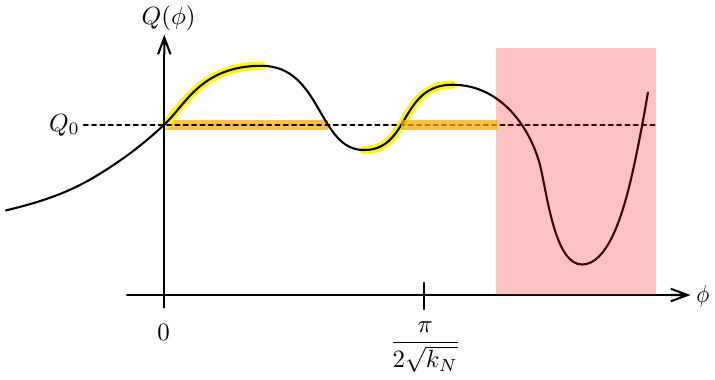}
  \caption{The region where $Q^2 \geqslant Q_0^2$ does not even need to be
  connected. Once the \emph{total} distance in regions where $Q^2
  \geqslant Q_0^2$ exceeds $\frac{\pi}{2 \sqrt{k_N}}$, any further attractors
  become inaccessible. Note that in this case there is an attractor point
  between the two orange regions (in the test band). This attractor point may
  or may not be accessible, depending on the precise choice of $Q^2 (\phi)$,
  but either way the attractor in the red region is
  inaccessible.\label{fig:Qtotdist}}
\end{figure}

\begin{figure}\centering
  \includegraphics[width=0.8\textwidth]{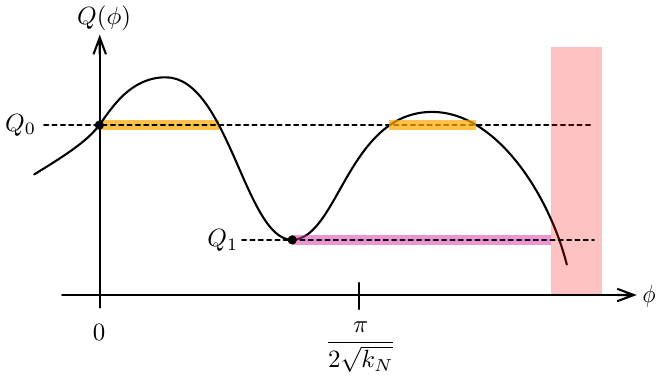}
  \caption{Starting at the first black dot where $Q^2 = Q_0^2$, the total
  distance traversed with $Q^2 \geqslant Q_0^2$ (the orange region) is
  limited. If $Q^2$ decreases below $Q_0^2$ subsequently, e.g., at the second
  black dot where $Q^2 = Q_1^2 < Q_0^2$, then an additional bound can be
  placed on the \emph{subsequent} total distance traversed with $Q^2
  \geqslant Q_1^2$.\label{fig:Qmultiplebounds}}
\end{figure}

The above proof works in a moduli space of arbitrary dimension. In the
one-dimensional case, it implies \eqref{eqn:monotonicBound} but is strictly
stronger than it, as illustrated by the examples shown in figures
\ref{fig:Qosc} and \ref{fig:Qtotdist}. Note that for each choice of threshold
$Q_i^2$, there is a separate bound, as shown in figure
\ref{fig:Qmultiplebounds}.

\begin{figure}\centering
  \includegraphics[width=0.65\textwidth]{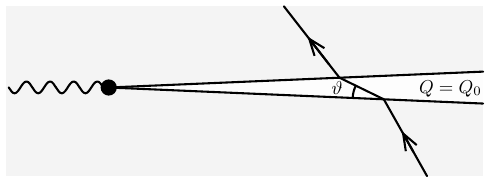}
  \caption{The angle $\vartheta$ can be interpreted in the Maupertuis geometry
  as the angle the path \emph{would} take on the Maupertuis disk if a
  narrow sliver with $Q = Q_0$ were inserted.\label{fig:varthetaMaupertuis}}
\end{figure}

We now comment on the relationship between this strengthened bound and the
Maupertuis geometry. Specializing to the quasiextremal case, one can show
that:
\begin{equation}
  \mathcal{M}= \frac{1}{\sqrt{k_N}} Q \cos \eta,
\end{equation}
where $\eta$ is the angle between the path and the inwards direction $-
\hat{\rho}$, see figure \ref{fig:MaupertuisAngle}. Then Snell's law at an
interface where $Q$ changes abruptly implies that $\mathcal{M}$ is continuous,
whereas straight-line geometry in a region where $Q$ is constant implies that
$\mathcal{M}$ decreases monotonically. Combining these two facts and noting
that any $Q^2 (\phi)$ can be approximated by a piecewise constant function, we
see geometrically that $\mathcal{M}$ decreases monotonically along the path
for any choice of $Q^2 (\phi)$. With $Q^2 = Q_0^2$ initially, we parameterize
\begin{equation}
  \mathcal{M}= \frac{1}{\sqrt{k_N}} Q_0 \cos \vartheta, \qquad \text{where
  $\vartheta = \eta$ initially} .
\end{equation}
Thus, $\vartheta$ increases monotonically along the path. Note that
$\vartheta$ can be interpreted geometrically as the angle between the inwards
direction $- \hat{\rho}$ and the direction that the path \emph{would} take
when traversing a narrow sliver in which $Q = Q_0$, as shown in figure
\ref{fig:varthetaMaupertuis}. This is distinct from $\eta$, which describes
the actual angle of the path, and which is not necessarily monotonically
increasing.

Now consider a sliver with constant $Q \geqslant Q_0$. Within this sliver, the
path is straight and so $\frac{\mathd \eta}{\mathd \theta} = 1$. This implies
that:
\begin{equation}
  \cos \vartheta = \frac{Q}{Q_0} \cos \eta \quad \Rightarrow \quad 
  \frac{\mathd \vartheta}{\mathd \theta} = \frac{Q}{Q_0}  \frac{\sin
  \eta}{\sin \vartheta} = \sqrt{\frac{Q^2 - Q^2 \cos^2 \eta}{Q_0^2 - Q_0^2
  \cos^2 \vartheta}} = \sqrt{1 + \frac{Q^2 - Q_0^2}{Q_0^2 \sin^2 \vartheta}}
  \geqslant 1 .
\end{equation}
Since $\mathcal{M}$ and thus $\vartheta$ does not change at an interface where
$Q$ changes abruptly, and since any $Q^2 (\phi)$ can be approximated by a
piecewise constant function as before, this implies that $\frac{\mathd
\vartheta}{\mathd \theta} \geqslant 1$ in regions where $Q \geqslant Q_0$, and
thus, the total angle $\Delta \theta$ traversed through such regions is at
most $\frac{\pi}{2}$, reproducing the bound \eqref{eqn:totDistBound}.

\section{Multi-dimensional moduli spaces}\label{sec:multidimModuliSpace}

Let us now try to generalize the preceeding argument to a moduli space of
greater than one dimension. We can imagine similarly that the EFT is known in
some region $\mathcal{K}$, and that there is an attractor point inside
$\mathcal{K}$ that generates a family of quasiextremal black hole solutions.
The question, as before, is whether there is some smaller region ``controlled
region'' $\mathcal{C} \subset \mathcal{K}$ in which we can guarantee that
these quasiextremal solutions are indeed extremal.

There are (at least) two ways to do this. The first is a simple corollary of
the bound \eqref{eqn:totDistBound}, so we will consider it first, before
developing a different bound that applies to other situations.

\subsection{The charge threshold bound}\label{sec:basinbound}

Recall the bound \eqref{eqn:totDistBound}, which we now restate as:

\begin{tcolorbox}[colback=yellow!20, colframe=black,  boxrule=0.8pt,  arc=0pt,  left=2mm,  right=2mm,  top=1mm,  bottom=1mm]
  \begin{center}
  \textbf{The charge threshold bound on black hole moduli space excursion} \\[\lineskip]
  \end{center} 
  \noindent If $Q_0^2 = Q^2 (\phi (r_0))$ is the charge of a spherically
  symmetric black hole evaluated in the EFT at $\phi^i = \phi^i (r_0)$, then
  the total moduli space distance $\Delta \phi$ that can be traversed through
  regions with $Q^2 (\phi) \geqslant Q^2_0$ between $r = r_0$ and the horizon
  is bounded:
  \begin{equation}
    \Delta \phi < \frac{\pi}{2 \sqrt{k_N}} . \label{eqn:totDistBound2}
  \end{equation}
\end{tcolorbox}

In a moduli space of dimension larger than one, this bound limits the total
distance that the path $\phi^i (\tau)$ traced by a given black hole solution
can cover in regions where $Q^2 \geqslant Q^2_0$. The implications of this
bound depend of course on the shape of this path, which is not known
\emph{a priori}, however, even without knowing the path we can draw some
universal conclusions as follows.

\begin{figure}
  \centering
  \includegraphics[width=0.55\textwidth]{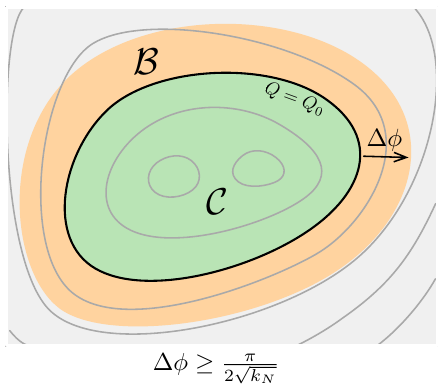}
  \caption{Constraining black hole solutions using the charge threshold bound
  \eqref{eqn:totDistBound2}. The controlled region $\mathcal{C}$ is bounded by
  the surface $Q^2 = Q_0^2$, with $Q^2 \geqslant Q_0^2$ in the surrounding
  test band $\mathcal{B}$ of thickness $\ge \frac{\pi}{2 \sqrt{k_N}}$.
  This ensures that solutions with $\phi_{\infty} \in \mathcal{C}$
  remain within $\mathcal{C} \cup \mathcal{B}$.\label{fig:multidimTestBand}}
\end{figure}

Suppose that there exists a region $\mathcal{C}$ of moduli space bounded by a
$Q^2 = Q_0^2$ surface, such that $Q^2 \geqslant Q_0^2$ in the region
$\mathcal{B}$ directly outside of $\mathcal{C}$. Provided that this test band
$\mathcal{B}$ is at least $\Delta \phi = \frac{\pi}{2 \sqrt{k_N}}$ thick (see
figure \ref{fig:multidimTestBand}) and lies entirely in the known region
$\mathcal{K}$ then the bound \eqref{eqn:totDistBound2} implies that the path
$\phi^i (\tau)$ followed by any black hole with $\phi^i_{\infty} \in
\mathcal{C}$ must stay with $\mathcal{C} \cup \mathcal{B} \subseteq
\mathcal{K}$.

\begin{figure}
  \centering
  \includegraphics[width=0.5\textwidth]{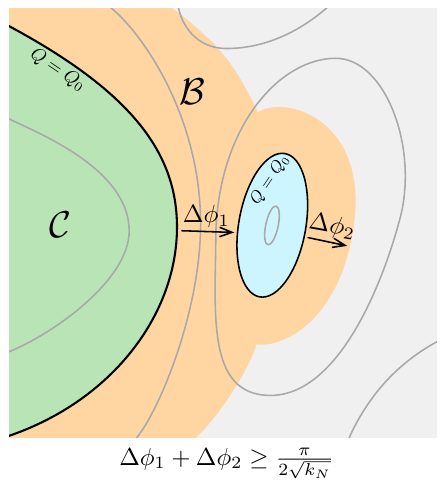}
  \caption{The test band $\mathcal{B}$ can have ``holes'' in it within which $Q^2
  < Q_0^2$. In this case, the ``width'' of $\mathcal{B}$ is the total length
  of the shortest sequence of paths connecting $\mathcal{C}$ to the outer edge
  of $\mathcal{B}$, not counting the distance travelled within the holes.
  Such holes may contain accessible attractor points.\label{fig:TestBandHole}}
\end{figure}

Note that the test band $\mathcal{B}$ can potentially have ``holes'' in it
within which $Q^2 < Q_0^2$, in which case the ``width'' of $\mathcal{B}$ means
the least total length for a sequence of paths from $\mathcal{C}$ to the outer
edge of $\mathcal{B}$ potentially passing through the holes, where the
distance travelled through each hole does not contribute to this ``width'',
see figure~\ref{fig:TestBandHole}. As in
\S\ref{sec:strongerbound}, there may be accessible attractor
points within these holes, so when $\mathcal{B}$ has such holes the attractor
point need not lie within $\mathcal{C}$, but still cannot lie outside the test
band $\mathcal{B}$.

Note that, for any given choice of $\phi^i_{\infty}$, the optimal bound is
obtained by taking $\mathcal{C}$ to be the connected region adjacent to
$\phi^i_{\infty}$ in which $Q^2 < Q^2 (\phi_{\infty})$. However, $\mathcal{C}$
can be chosen to be a larger region if convenient, provided that the above
criteria are satisifed.

\subsection{The directional derivative bound}\label{sec:distFuncBound}

\begin{figure}\centering
  \includegraphics[width=0.5\textwidth]{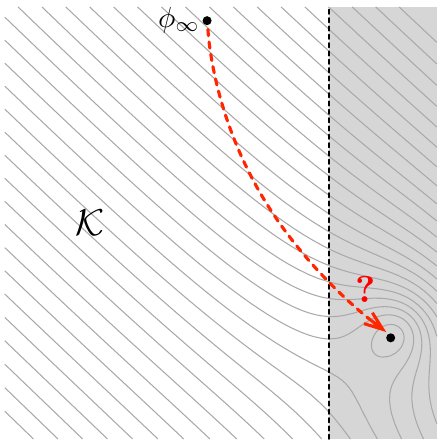}
  \caption{A hidden attractor that is ``downhill'' from the vacuum $\phi_\infty$---i.e., one connected to it by a path along which $Q^2(\phi) \le Q^2(\phi_\infty)$ everywhere---cannot be shown to be inaccessible using the charge threshold bound \eqref{eqn:totDistBound2}. The accessibility of such attractor points is constrained by the directional derivative bound developed below.\label{fig:sidehill}}
\end{figure}

The bound derived above is powerful, but can only be applied if the connected
region where $Q^2 < Q_0^2$ lies entirely within the known region. Otherwise,
there can be trajectories $\phi^i (\tau)$ that approach a hidden attractor
while travelling \emph{downhill} in $Q^2 (\phi)$. However, such
trajectories need not solve the black hole equations, particularly if the
lateral component of the force $\vec{\nabla} Q^2$ is large, see figure
\ref{fig:sidehill}.

To constrain this situation, instead of requiring that $Q^2$ increases above
its initial value (which may not occur), we only require that it is increasing
in a certain direction $\hat{n}^i$, i.e., $\hat{n} \cdot \nabla Q^2 \geqslant
0$, and then seek to bound the distance $\Delta \phi_{(\hat{n})} = \int
\hat{n}_i \mathd \phi^i$ travelled in the specified direction. We will show
that if $\hat{n}^i$ satifies an additional constraint this distance is indeed
bounded. Then with a suitable choice of $\hat{n}^i$ (e.g., pointing towards
the unknown region) we may be able to prove that the black hole solution
cannot reach a hidden attractor.

Let $\hat{n}^i$ be a unit vector field. Note that the extrinsic curvature of a
hypersurface normal to this vector field is:
\begin{equation}
  K_{i j} = \nabla_i \hat{n}_j - \hat{n}_i a_j = \nabla_i \hat{n}_j -
  \hat{n}^k  \hat{n}_i \nabla_k  \hat{n}_j,
\end{equation}
where $a_j \assign \hat{n}^k \nabla_k  \hat{n}_j$ is the acceleration (i.e.,
the failure to be geodesic) along an integral curve of $\hat{n}^i$, and one
can show that $K_{i j} = K_{j i}$ and $\hat{n}^i K_{i j} = \hat{n}^i a_i = 0$.
Analogous to before, we define the clock:
\begin{equation}
  \xi (\tau) \assign \frac{\sqrt{k_N} \phi_{(\hat{n})}'}{\sqrt{(\chi')^2 -
  k_N^2 M_0^2}}, \qquad \text{where} \qquad \phi_{(\hat{n})}' \assign
  \hat{n}_i {\phi^i}',
\end{equation}
which comes from replacing $\phi' \rightarrow \phi_{(\hat{n})}'$ in
\eqref{eqn:xidefn}. Much like before, the horizon condition ensures that the
factor inside the square root is positive, so that $\xi < \xi_{\max} =
\infty$. Assuming that $\phi_{(\hat{n})}' \geqslant 0$ initially, we have
$\xi_{\text{init}} \geqslant \xi_{\min} = 0$, as before.

We now compute:
\begin{align}
  \phi_{(\hat{n})}'' &= \left( \hat{n}_i {\phi^i}' \right)' = \hat{n}_i
  {\phi^i}'' + \partial_i \hat{n}_j {\phi^i}' {\phi^j}' = \hat{n}_i  \left(
  {\phi^i}'' + \Gamma^i_{\; j k} {\phi^j}' {\phi^k}' \right) +
  \nabla_i \hat{n}_j {\phi^i}' {\phi^j}' \nonumber\\
  &= \frac{1}{2} e^{2 \chi}  \hat{n}^i \nabla_i Q^2 + \nabla_i \hat{n}_j
  {\phi^i}' {\phi^j}' = \frac{1}{2 k_N} \chi''  \hat{n}^k \nabla_k \log Q^2 +
  K_{i j} {\phi^i}' {\phi^j}' + \phi_{(\hat{n})}' a_j {\phi^j}' . 
\end{align}
We also have:
\begin{equation}
  \chi'' {= \chi'}^2 - k_N^2 M_0^2 + k_N G_{i j} {\phi^i}' {\phi^j}' {=
  \chi'}^2 - k_N^2 M_0^2 + k_N (\phi_{(\hat{n})}')^2 + k_N \Pi_{i j} {\phi^i}'
  {\phi^j}',
\end{equation}
where $\Pi_{i j} \equiv G_{i j} - \hat{n}_i \hat{n}_j$ is the metric projected
onto the directions orthogonal to $\hat{n}^i$. Now we compute:
\begin{align}
  \xi' &= \frac{\sqrt{k_N} \phi_{(\hat{n})}''}{\sqrt{(\chi')^2 - k_N^2
  M_0^2}} + \frac{\sqrt{k_N} \phi_{(\hat{n})}' (- \chi') \chi''}{[(\chi')^2 -
  k_N^2 M_0^2]^{3 / 2}} \nonumber\\
  &= \left[ {\chi'}^2 - k_N^2 M_0^2 + k_N (\phi_{(\hat{n})}')^2 \right]
  \left[ \frac{\frac{1}{2 \sqrt{k_N}}  \hat{n}^i \nabla_i \log
  Q^2}{\sqrt{(\chi')^2 - k_N^2 M_0^2}} + \frac{\sqrt{k_N} \phi_{(\hat{n})}' (-
  \chi')}{[(\chi')^2 - k_N^2 M_0^2]^{3 / 2}} \right] \nonumber\\
  &\noeq + \left[ \frac{\sqrt{k_N} }{\sqrt{(\chi')^2 - k_N^2 M_0^2}}  \left( K_{i
  j} + \frac{1}{2}  \hat{n}^k \nabla_k \log Q^2 \Pi_{i j} \right) +
  \frac{k_N^{3 / 2} \phi_{(\hat{n})}'  (- \chi')}{[(\chi')^2 - k_N^2 M_0^2]^{3
  / 2}} \Pi_{i j} \right] {\phi^i}' {\phi^j}' 
  \nonumber\\&\noeq + \frac{\sqrt{k_N}
  \phi_{(\hat{n})}' a_j {\phi^j}'}{\sqrt{(\chi')^2 - k_N^2 M_0^2}} . 
  \label{eqn:xiprimeD}
\end{align}
Now, since by the horizon condition $\chi' < - k_N M_0$, we see that each
factor in this equation is non-negative if four sufficient conditions are met:
\vspace{-1em}
\begin{subequations} \label{eqn:DDconds} 
\begin{paracol}{2}
\begin{align}
\phi_{(\hat n)}' &\geqslant 0, \phantom{\frac12}\tag{\theequation a} \label{subeqn:DDcondsA} \\
\hat n^i \nabla_i \log Q^2 &\geqslant 0,\tag{\theequation b} \label{subeqn:DDcondsB}
\end{align}
\switchcolumn
\begin{align}
K_{ij}
+\frac12 \hat n^k \nabla_k \log Q^2\,\Pi_{ij}
&\succcurlyeq 0,\tag{\theequation c} \label{subeqn:DDcondsC}\\
a_i &= 0,\tag{\theequation d} \label{subeqn:DDcondsD}
\end{align}
\end{paracol}
\end{subequations}\noindent
where ``$M_{i j} \succcurlyeq 0$'' means that $M_{i j}$ is a positive
semi-definite matrix. 

The condition \eqref{subeqn:DDcondsB}, equivalent to $\hat{n} \cdot \vec{\nabla} Q^2
\geqslant 0$, expresses the fact that $\hat{n}^i$ points towards a region of
larger $Q^2$, as anticipated. Let us discuss and interpret the other
conditions. Firstly, the condition \eqref{subeqn:DDcondsD} $a_i = 0$ ensures that the last term in
\eqref{eqn:xiprimeD}, which generally has no definite sign, does not
contribute. One can show that this condition, i.e., $\hat{n}^k \nabla_k 
\hat{n}_i = 0$, is equivalent to the requirement that $\hat{n}_i$ is closed,
so (at least locally)\footnote{There could be an obstruction to defining
$\mathfrak{D}$ globally if $\pi_1$ of the moduli space is non-trivial, but one
can always worked on the universal cover to avoid this issue.} there exists a
function $\mathfrak{D}$ such that:
\begin{equation}
  \hat{n}_i = \nabla_i \mathfrak{D}.
\end{equation}
Note that since $\hat{n}_i$ is a unit vector field, $\mathfrak{D}$ is a
distance function (see, e.g.,~\cite{Etheredge:2023zjk}), where $\hat{n}_i$ is normal to surfaces of constant
$\mathfrak{D}$, and the integral curves of $\hat{n}_i$ are the geodesics
determined by $\mathfrak{D}$ gradient flow.

Next, since $K_{i j} = \nabla_i \nabla_j \mathfrak{D}$ is the extrinsic
curvature of a constant $\mathfrak{D}$ surface, the condition $K_{i j} +
\frac{1}{2}  \hat{n}^k \nabla_k \log Q^2 \Pi_{i j} \succcurlyeq 0$ places a
lower bound on this curvature. Since by assumption $\hat{n} \cdot \vec{\nabla}
\log Q^2 \geqslant 0$, it is sufficient for this extrinsic curvature to be
non-negative, but in general if $\hat{n} \cdot \vec{\nabla} \log Q^2$ is
strictly positive then somewhat negative extrinsic curvature is permissible.

Now consider the condition \eqref{subeqn:DDcondsA} $\phi_{(\hat{n})}' \geqslant 0$, which is
equivalent to $\xi \geqslant 0$. So long as this is true initially and the
other assumptions discussed above are satisfied, then $\xi' \geqslant 0$ and
so $\phi_{(\hat{n})}' \propto \xi$ remains non-negative thereafter. Thus, this
condition only constrains the initial data at some $\tau =
\tau_{\text{init}}$, requiring that the black hole solution is moving partly
in the positive $\hat{n}^i$ direction (or at least not in the negative
$\hat{n}^i$ direction), which is indeed the kind of solution that we wish to
constrain.

Finally, note that $\mathfrak{D}' = \phi_{(\hat{n})}'$ and
\begin{equation}
  \mathfrak{D}'' = \phi_{(\hat{n})}'' = \frac{1}{2 k_N}  \bigl[ {\chi'}^2 -
  k_N^2 M_0^2 + k_N (\phi_{(\hat{n})}')^2 \bigr]  \hat{n}^k \nabla_k \log Q^2
  + \biggl[ K_{i j} + \frac{1}{2}  \hat{n}^k \nabla_k \log Q^2 \Pi_{i j}
  \biggr] {\phi^i}' {\phi^j}' .
\end{equation}
If the inequalities \eqref{subeqn:DDcondsB} and \eqref{subeqn:DDcondsC} are saturated then
$\mathfrak{D}'' = 0$, so $\mathfrak{D}'$ must be strictly positive or else the
solution does not move in the direction of increasing $\mathfrak{D}$. Thus,
the three inequalities \eqref{subeqn:DDcondsA}, \eqref{subeqn:DDcondsB}, \eqref{subeqn:DDcondsC} cannot all be saturated together.

Therefore under the above assumptions $\xi$ is a valid clock with maximum
range $0 \leqslant \xi < \infty$. This clock advances at the rate:
\begin{align}
  \frac{\mathd \xi}{\mathd \mathfrak{D}} &= [1 + \xi^2] \left[ \frac{1}{2
  \xi} \nabla^i \mathfrak{D} \nabla_i \log Q^2 + \frac{\sqrt{k_N}}{\sqrt{1 -
  \mu^2 / \mu_h^2}} \right] \nonumber\\
  &\noeq + \left[ \xi \left( \nabla_i \nabla_j \mathfrak{D}+ \frac{1}{2} \nabla^i
  \mathfrak{D} \nabla_i \log Q^2 \Pi_{i j} \right) + \xi^2 
  \frac{\sqrt{k_N}}{\sqrt{1 - \mu^2 / \mu_h^2}} \Pi_{i j} \right] 
  \frac{\mathd \phi^i}{\mathd \mathfrak{D}}  \frac{\mathd \phi^j}{\mathd
  \mathfrak{D}},  \label{eqn:dxidD}
\end{align}
from \eqref{eqn:xiprimeD}, where we reexpress $\tau$ derivatives as
$\mathfrak{D}$ derivatives, $\frac{\mathd}{\mathd \tau} =
\frac{1}{\mathfrak{D}'}  \frac{\mathd}{\mathd \mathfrak{D}}$. Thus, we obtain:
\begin{equation}
  \frac{\mathd \xi}{\mathd \mathfrak{D}} \geqslant \sqrt{k_N}  [1 + \xi^2]
  \qquad \Rightarrow \qquad \Delta \mathfrak{D} \leqslant \frac{1}{\sqrt{k_N}}
  \int_{\xi_{\text{init}}}^{\infty} \frac{\mathd \xi}{1 + \xi^2} \leqslant
  \frac{\pi}{2 \sqrt{k_N}}, \label{eqn:Dineq}
\end{equation}
as before. In fact, the inequality $\Delta \mathfrak{D}< \frac{\pi}{2
\sqrt{k_N}}$ is strict because the inequalities in \eqref{eqn:DDconds} cannot
all be saturated as noted above.

To interpret these results, let $\phi_{\infty}$ lie inside a controlled region
$\mathcal{C}$, with $\mathfrak{D} (\phi)$ the moduli-space distance from the
boundary of $\mathcal{C}$ outwards, i.e., the solution to the equation:
\begin{equation}
  (\nabla \mathfrak{D})^2 = 1 \qquad \text{with the boundary condition} \qquad
  \mathfrak{D} |_{\partial \mathcal{C}} = 0,
\end{equation}
such that $\mathfrak{D}$ is increasing as we move outwards away from
\ensuremath{\mathcal{C}}. Then, if the region $\mathcal{C}$ is surrounded by a test band $\mathcal{B}$
of thickness at least $\frac{\pi}{2 \sqrt{k_N}}$ within which the following
conditions are satisfied
\begin{equation}
  \text{(i)} \quad \nabla^i \mathfrak{D} \nabla_i \log Q^2 \geqslant 0, \qquad
  \text{(ii)} \quad \nabla_i \nabla_j \mathfrak{D}+ \frac{1}{2} \nabla^i
  \mathfrak{D} \nabla_i \log Q^2 \Pi_{i j} \succsim 0, \label{eqn:Dconds}
\end{equation}
then there are no black hole solutions with $\phi^i_{\infty} \in \mathcal{C}$
that venture outside $\mathcal{C} \cup \mathcal{B}$.

We can summarize our conclusions as follows:

\begin{tcolorbox}[colback=yellow!20, colframe=black,  boxrule=0.8pt,  arc=0pt,  left=2mm,  right=2mm,  top=1mm,  bottom=1mm]
  \begin{center}
  \textbf{The directional derivative bound on black hole moduli space excursion} \\[\lineskip]
  \end{center}
 
  \noindent
  If the path $\phi^i (r)$ followed by a charge $Q^2 (\phi)$
  spherically symmetric black hole between $r = r_1$ and $r = r_2$ passes
  through a region of moduli space in which there exists a distance function
  $\mathfrak{D}$ satisfying
  \begin{equation}
    \text{(i)} \quad \nabla^i \mathfrak{D} \nabla_i \log Q^2 \geqslant 0,
    \qquad \text{(ii)} \quad \nabla_i \nabla_j \mathfrak{D}+ \frac{1}{2}
    \nabla^i \mathfrak{D} \nabla_i \log Q^2 \Pi_{i j} \succsim 0,
    \label{eqn:Dboundcons}
  \end{equation}
  then the total increase in $\mathfrak{D}$ along this portion of the
  trajectory is bounded,
  \begin{equation}
    \Delta \mathfrak{D}< \frac{\pi}{2 \sqrt{k_N}} . \label{eqn:DistFuncBound}
  \end{equation}
\end{tcolorbox}

Note that the first condition on $\mathfrak{D}$ is the condition that the
directional derivative of $Q^2 (\phi)$ along $\hat{n}_i = \partial_i
\mathfrak{D}$ is non-negative. This is equivalent to the requirement that
$Q^2$ is nondecreasing along $\mathfrak{D}$ gradient flows (which are
geodesics), or vice versa. The second condition can also be stated as the
requirement that each constant $\mathfrak{D}$ surface has non-negative
extrinsic curvature with respect to the Maupertuis-like metric $\tilde{G}_{i
j} \assign Q^2 G_{i j}$.\footnote{More precisely, this is the pullback of the
quasiextremal ($M_0 = 0$) Maupertuis metric \eqref{eqn:MaupertuisMetric} to a
slice of fixed $\chi$.}

\begin{figure}\centering
  \includegraphics[width=0.5\textwidth]{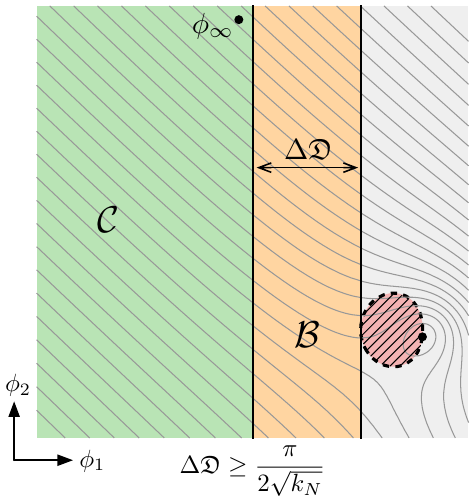}
  \caption{Applying the directional derivative bound to constrain the accessibility of the ``downhill'' hidden attractor shown in figure~\ref{fig:sidehill}. Here we choose $\mathfrak{D} = \phi_1$, assuming a Euclidean moduli space metric for simplicity. Condition (ii) of \eqref{eqn:Dboundcons} is identically satisfied, whereas condition (i) is satisfied everywhere outside the red crosshatched region. Thus, if the test band $\mathcal{B}$ has thickness at least $\frac{\pi}{2\sqrt{k_N}}$ then for $\phi_\infty \in \mathcal{C}$ the hidden attractor point cannot be reached, even when it is ``downhill'' from the vacuum.\label{fig:KDistFunc}}
\end{figure}

Specializing to a one-dimensional moduli space, the directional derivative bound \eqref{eqn:DistFuncBound} reduces to the bound \eqref{eqn:monotonicBound}, which is a weaker consequence of the one-dimensional charge threshold bound~\eqref{eqn:totDistBound}. However, this is not true in moduli spaces of dimension greater than one. For instance, as illustrated in figure \ref{fig:KDistFunc}, the directional derivative bound constrains accessibility of hidden attractor points that are ``downhill'' from the vacuum, whereas the charge threshold bound does not. More generally, the two bounds are complementary to each other.

Similar to \eqref{eqn:QslopeCond}, if we strengthen the first condition to
\begin{equation}
  \nabla^i \mathfrak{D} \nabla_i \log Q^2 \geqslant 2 \alpha \sqrt{k_N},
\end{equation}
then we obtain the stronger bound:
\begin{equation}
  \frac{\mathd \xi}{\mathd \mathfrak{D}} \geqslant \sqrt{k_N} [1 + \xi^2]
  \left[ 1 + \frac{\alpha}{\xi} \right] \quad \Rightarrow \quad \Delta
  \mathfrak{D} \leqslant \frac{1}{\sqrt{k_N}} \int \frac{\mathd \xi}{(1 +
  \xi^2) (1 + \alpha / \xi)} = \frac{\pi + 2 \alpha \log \alpha}{2 \sqrt{k_N} 
  (1 + \alpha^2)}, \label{eqn:refinedDistFuncBound}
\end{equation}
so with this stronger assumption the test band can be thinner.\footnote{To
saturate this bound, we would require $\xi_{\text{init}} = 0$, but then
${\phi^i}'$ could not be parallel to $\hat{n}^i$ and so the second line in
\eqref{eqn:dxidD} would not vanish. Thus, this bound is also strict.}

Note also that the extrinsic curvature condition can be weakened somewhat to:
\begin{equation}
  \nabla_i \nabla_j \mathfrak{D}+ \left( \frac{1}{2} \nabla^i \mathfrak{D}
  \nabla_i \log Q^2 + \xi \sqrt{k_N} \right) \Pi_{i j} \succcurlyeq 0 .
\end{equation}
This is solution dependent, but using $\frac{\mathd \xi}{\mathd \mathfrak{D}}
\geqslant \sqrt{k_N}  [1 + \xi^2]$ we can place the lower bound:
\begin{equation}
  \arctan \xi \geqslant \sqrt{k_N} \mathfrak{D},
\end{equation}
where we chose $\mathfrak{D}= 0$ at the boundary of $\mathcal{C}$. Thus, the
condition
\begin{equation}
  \nabla_i \nabla_j \mathfrak{D}+ \left( \frac{1}{2} \nabla^i \mathfrak{D}
  \nabla_i \log Q^2 \right) \Pi_{i j} \succcurlyeq - \sqrt{k_N} \tan \left(
  \sqrt{k_N} \mathfrak{D} \right) \Pi_{i j}, \label{eqn:DextCond2}
\end{equation}
is sufficient.

\subsection{Geometric interpretation}

Having proven the bound \eqref{eqn:DistFuncBound}, we now provide some
intuition for why it works and why it is necessary to place a lower bound on
the intrinsic curvature of constant $\mathfrak{D}$ surfaces for it to do so.
We first turn to the Maupertuis geometric interpretation of this system. The
Maupertuis metric \eqref{eqn:MaupertuisCone} in the quasiextremal case is:
\begin{equation}
  \widetilde{\mathd s}^2 = \frac{1}{k_N} Q^2 (\phi)  [\mathd \rho^2 + \rho^2
  k_N G_{i j} (\phi) \mathd \phi^i \mathd \phi^j] .
\end{equation}
For convenience, we choose coordinates on the moduli space $\theta, \varphi^a$
where $\theta = \sqrt{k_N} \mathfrak{D}$ is fixed by the distance function,
and the remaining coordinates $\varphi^a$ are fixed along the geodesics
generated by gradient flows of $\mathfrak{D}$. Thus,
\begin{equation}
  \widetilde{\mathd s}^2 = \frac{1}{k_N}  [\mathd \theta^2 + h_{a b} (\theta,
  \varphi) \mathd \varphi^a \mathd \varphi^b], \qquad \text{where} \qquad
  \mathfrak{D}= \theta / \sqrt{k_N} .
\end{equation}
In these coordinates, the extrinsic curvature of a constant $\theta$ surface
works out to be:
\begin{equation}
  K_{a b} = \nabla_a \nabla_b \mathfrak{D}= \frac{1}{2 \sqrt{k_N}}
  \partial_{\theta} h_{a b} .
\end{equation}
Thus, in the quasiextremal case, \eqref{eqn:dxidD} becomes:
\begin{equation}
  \frac{\mathd \xi}{\mathd \theta} = [1 + \xi^2] \left[ 1 + \frac{1}{2 \xi}
  \partial_{\theta} \log Q^2 \right] + \left[ \xi^2 h_{a b} + \frac{\xi}{2} 
  (\partial_{\theta} h_{a b} + \partial_{\theta} \log Q^2 h_{a b}) \right] 
  \frac{\mathd \varphi^a}{\mathd \theta}  \frac{\mathd \varphi^b}{\mathd
  \theta} . \label{eqn:dxidetaHigherDim}
\end{equation}

As before, we focus on the Maupertuis disk with polar coordinates $(\rho,
\theta)$. Let $\eta$ be the angle of trajectory along this disk relative to
the radially inwards direction, see figure \ref{fig:MaupertuisAngle}, so that
\begin{equation}
  \tan \eta = - \frac{\rho \theta'}{\rho'} = \frac{\sqrt{k_N} \mathfrak{D}'}{-
  \chi'} = \xi .
\end{equation}
Expressed in terms of $\eta$, \eqref{eqn:dxidetaHigherDim} becomes:
\begin{equation}
  \frac{\mathd \eta}{\mathd \theta} = 1 + \frac{1}{2} \cot \eta
  \partial_{\theta} \log Q^2 + \sin \eta \left[ \sin \eta h_{a b} +
  \frac{1}{2} \cos \eta (\partial_{\theta} h_{a b} + \partial_{\theta} \log
  Q^2 h_{a b}) \right]  \frac{\mathd \varphi^a}{\mathd \theta}  \frac{\mathd
  \varphi^b}{\mathd \theta} .
\end{equation}
To simplify this further, define $\tilde{h}_{a b} = Q^2 h_{a b}$, so that:
\begin{equation}
  \frac{\mathd \eta}{\mathd \theta} = 1 + \frac{1}{2} \cot \eta
  \partial_{\theta} \log Q^2 + \frac{\sin \eta}{Q^2}  \left[ \sin \eta
  \tilde{h}_{a b} + \frac{1}{2} \cos \eta \partial_{\theta}  \tilde{h}_{a b}
  \right]  \frac{\mathd \varphi^a}{\mathd \theta}  \frac{\mathd
  \varphi^b}{\mathd \theta}, \label{eqn:detadthetaDistFunc}
\end{equation}
where the Maupertuis metric can be written as:
\begin{equation}
  k_N  \widetilde{\mathd s}^2 = Q^2 (\phi)  [\mathd \rho^2 + \rho^2 \mathd
  \theta^2] + \rho^2  \tilde{h}_{a b} (\theta, \varphi) \mathd \varphi^a
  \mathd \varphi^b .
\end{equation}

We can now give a geometric interpretation to each term in
\eqref{eqn:detadthetaDistFunc}. As in the case of a one-dimensional moduli
space, $\frac{\mathd \eta}{\mathd \theta} = 1$ corresponds to straight line
motion along the Maupertuis disk. Thus, each additional term describes
deviations from this motion. The second term is the same refractive effect
that we saw before. The final two terms are new, and only contribute when the
motion is not fully parallel to the Maupertuis disk. Firstly, we observe that
as $\rho$ decreases, an effect analogous to conservation of angular momentum
in the conical Maupertuis geometry will increase the velocity in the moduli
directions in proportion to any initial velocity in these directions. In
particular, if $(\varphi^a)' \neq 0$ then $| (\varphi^a)' |$ will increase at
the expense of $\rho'$, while having no effect on $\theta'$. The net effect of
this is to increase the angle $\eta$, as described by the third term in
\eqref{eqn:detadthetaDistFunc}.

Similarly, if the other angular directions $\varphi^a$ have a size that
depends on $\theta$ (as would occur, e.g., if $\theta$ were the polar angle in
spherical polar coordinates) then there is a similar effect: when
$\partial_{\theta} \tilde{h}_{a b} < 0$ then $| (\varphi^a)' |$ increases at
the expense of $\theta'$, with no direct effect on $\rho'$, leading to a
\emph{decrease} in the angle $\eta$. If, however, $\partial_{\theta}
\tilde{h}_{a b} > 0$, then $\theta'$ increases at the expense of $|
(\varphi^a)' |$ without affecting $\rho'$, leading to an \emph{increase}
in the angle $\eta$. This is accounted for by the last term in
\eqref{eqn:detadthetaDistFunc}.

Since $\partial_{\theta} \tilde{h}_{a b}$ is proportional to the (warped)
extrinsic curvature of constant $\mathfrak{D}$ hypersurfaces, this explains
why this curvature cannot be too negative, since otherwise any initial
velocity in the $\varphi$ directions will grow, leading to decceleration in
the $\theta$ direction, which can enable a large displacement $\Delta \theta$
without violating the horizon condition. Conversely, if the warped extrinsic
curvature $\partial_{\theta} \tilde{h}_{a b}$ is non-negative then any effect
of the extra directions $\varphi$ only serves to hasten the point where the
horizon condition is violated, and we obtain the same bound as in a
one-dimensional moduli space.

\subsection{Examples violating the extrinsic curvature condition}

Since the extrinsic curvature condition \eqref{eqn:Dconds}.ii, or more sharply
\eqref{eqn:DextCond2}, is novel to the multidimensional case, we now give some
examples that demonstrate that violating this condition can lead to a
violation of the bound $\Delta \mathfrak{D}< \frac{\pi}{2 \sqrt{k_N}}$, even
when $\nabla \mathfrak{D} \cdot \nabla Q^2 \geqslant 0$ everywhere.

A simple class of examples occur in 2d moduli spaces with the following form:
\begin{equation}
  \mathd s^2 = \mathd \phi^2 + G_{\psi \psi} (\phi) \mathd \psi^2, \qquad Q^2
  (\phi, \psi) =\mathcal{Q}^2 (\phi) e^{- 2 \gamma \psi}, \qquad \text{for
  some $\gamma > 0$},
\end{equation}
where we are interested in the distance function $\mathfrak{D}= \phi$. We can
then reduce the fake superpotential equation to an ODE using the ansatz $M
(\phi, \psi) =\mathcal{M} (\phi) e^{- \gamma \psi}$, which gives
\begin{equation}
  \left( k_N + \frac{\gamma^2}{G_{\psi \psi}} \right) \mathcal{M}^2 +
  (\mathcal{M}')^2 =\mathcal{Q}^2 .
\end{equation}
This equation can be written more intuitively in the form:
\begin{equation}
  k_N \mathcal{M}^2 + \frac{1}{G_{\phi \phi}^{\text{eff}}}  (\mathcal{M}')^2 =
  Q^2_{\text{eff}}, \qquad \text{where} \qquad G_{\phi \phi}^{\text{eff}} = 1
  + \frac{\gamma^2}{k_N G_{\psi \psi}}, \quad Q^2_{\text{eff}} =
  \frac{1}{G_{\phi \phi}^{\text{eff}}} \mathcal{Q}^2 .
\end{equation}
Thus, the quasiextremal black hole equations reduce to an those of an
effective 1d moduli space with a modified metric and charge function.

\begin{figure}\centering
  \includegraphics[width=0.65\textwidth]{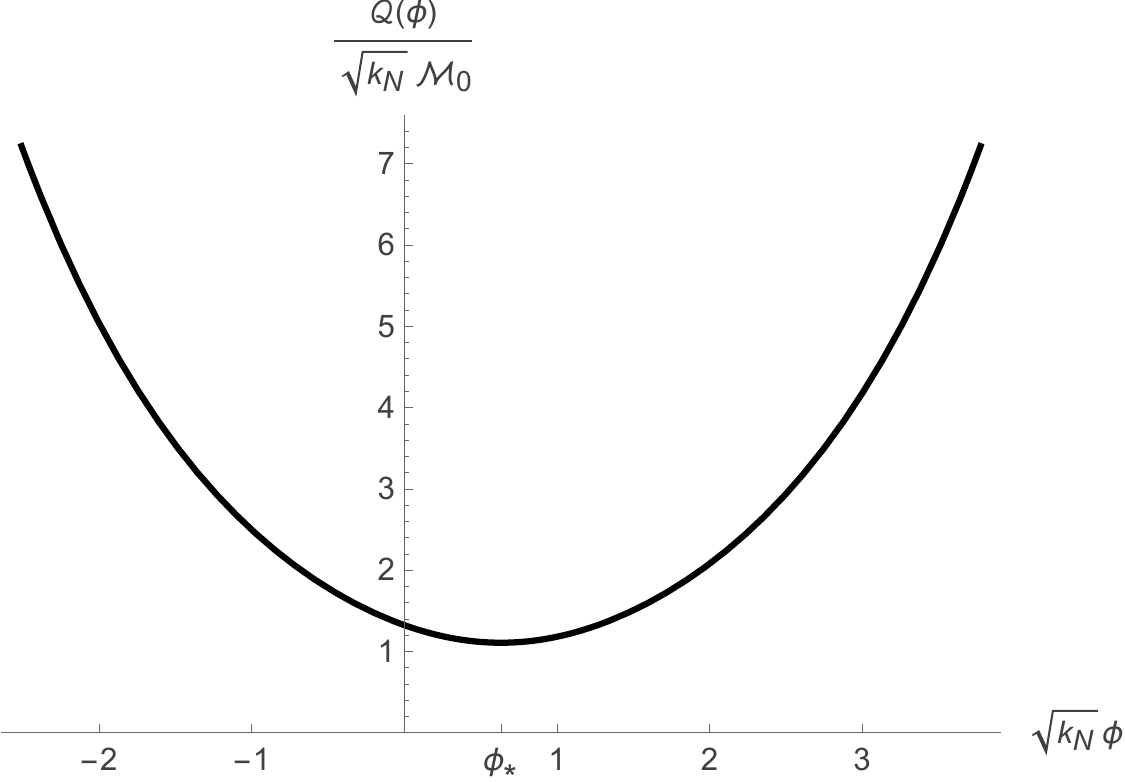}
  \caption{The function $\mathcal{Q}^2 (\phi)$ in
  \eqref{eqn:Q2extrinsicCurvatureViolation}, plotted for the case $\alpha =
  \sqrt{k_N / 2}$, $\beta = \sqrt{2 k_N}$, $\gamma = \sqrt{k_N} /
  2$.\label{fig:calQplot}}
\end{figure}

For simplicity, suppose that $\mathcal{M} (\phi)$ is exponentially decreasing:
\begin{equation}
  \mathcal{M} (\phi) =\mathcal{M}_0 e^{- \alpha \phi}, \qquad \text{for some
  $\alpha > 0$} .
\end{equation}
This implies that
\begin{equation}
  \mathcal{Q}^2 (\phi) = \left[ k_N + \alpha^2 + \frac{\gamma^2}{G_{\psi
  \psi}} \right] \mathcal{M}^2_0 e^{- 2 \alpha \phi} .
\end{equation}
To ensure that ${\mathcal{Q}^2}' > 0$, we require $G_{\psi \psi}$ to be
decreasing sufficiently rapidly, for example
\begin{equation}
  G_{\psi \psi} = e^{- 2 \beta \phi}, \qquad \text{for $\beta > \alpha$} .
\end{equation}
Then we obtain:
\begin{equation}
  \mathcal{Q}^2 (\phi) = [k_N + \alpha^2] \mathcal{M}^2_0 e^{- 2 \alpha \phi}
  + \gamma^2 \mathcal{M}^2_0 e^{2 (\beta - \alpha) \phi},
  \label{eqn:Q2extrinsicCurvatureViolation}
\end{equation}
as plotted in figure \ref{fig:calQplot}. This has a global minimum at
\begin{equation}
  \phi_{\ast} = \frac{1}{2 \beta} \log \left[ \frac{\alpha (k_N +
  \alpha^2)}{(\beta - \alpha) \gamma^2} \right],
\end{equation}
where ${\mathcal{Q}^2}' > 0$ for $\phi > \phi_{\ast}$. Thus, naively for any
choice of $\phi_{\infty} > \phi_{\ast}$, the bound of
\S\ref{sec:distFuncBound} would seem to imply that $\phi_h <
\phi_{\infty} + \frac{\pi}{2 \sqrt{k_N}}$. Indeed, since $Q^2 (\phi, \psi)
=\mathcal{Q}^2 (\phi) e^{- 2 \gamma \psi}$ and $\mathcal{Q}^2 (\phi)$ is
minimized at $\phi = \phi_{\ast}$, one might expect that a quasiextremal black
hole solution would asymptote to a trajectory with $\phi \approx \phi_{\ast}$
and $\psi$ increasing without bound near the horizon.

\begin{figure}\centering
  \includegraphics[width=0.8\textwidth]{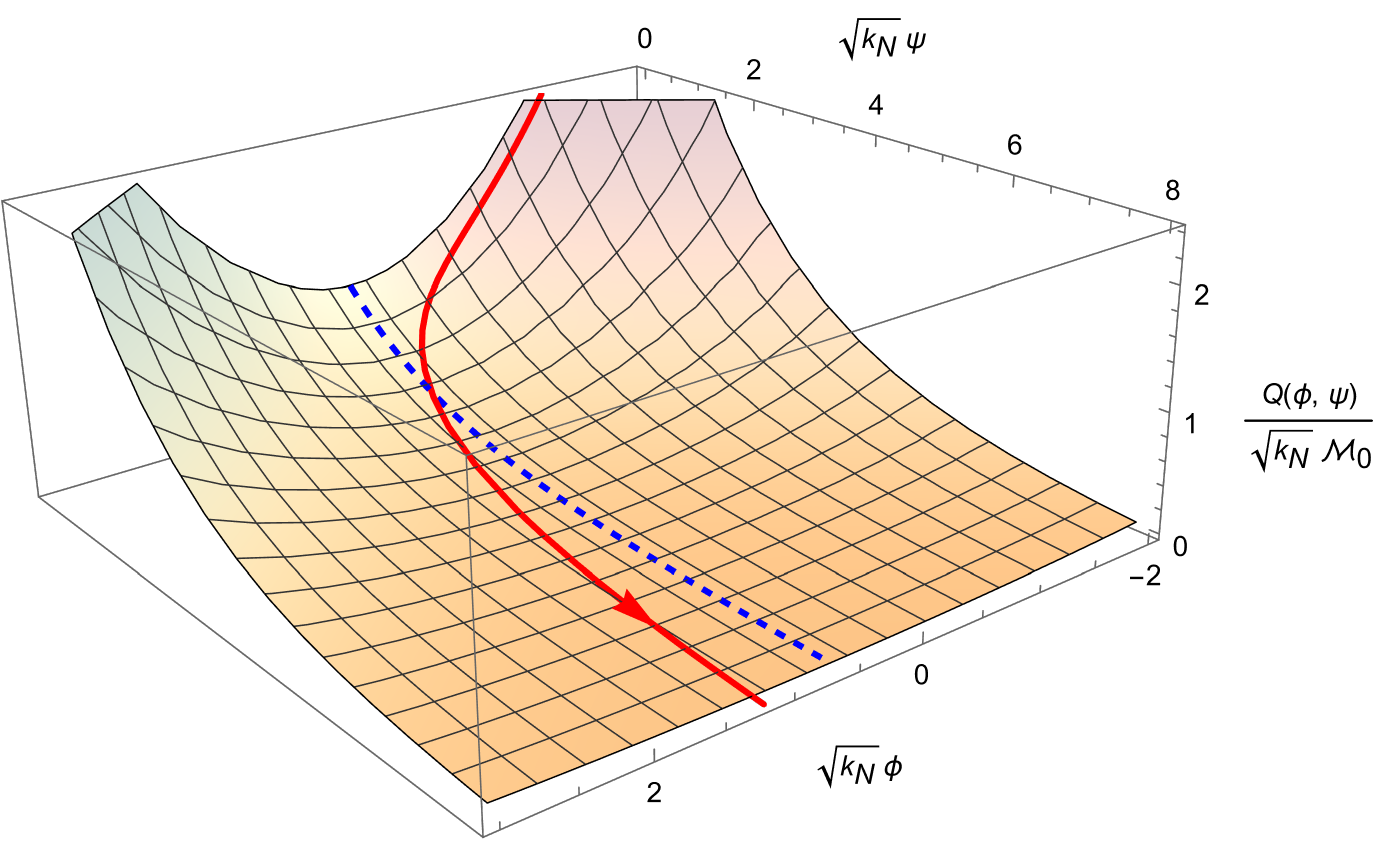}
  \caption{The trajectory (red solid line) of the quasiextremal solution
  \eqref{eqn:phipsisoln}, plotted as a track along a surface of height $Q
  (\phi, \psi)$, for the case $\alpha = \sqrt{k_N / 2}$, $\beta = \sqrt{2
  k_N}$, $\gamma = \sqrt{k_N} / 2$. Note that the trajectory overshoots the
  naive asymptote $\phi = \phi_{\ast}$ (blue dashed line) and continues to
  arbitrarily large values of $\phi$, due to the negative extrinsic curvature
  of each constant $\phi$ surface coupled with the large accumulated velocity
  in the $\psi$ direction.\label{fig:counterexTrajectory}}
\end{figure}

However, this is not the case. Since we know $M (\phi, \psi) =\mathcal{M}_0
e^{- \alpha \phi - \gamma \psi}$ we can construct the entire solution by
solving the gradient flow equations:
\begin{equation}
  \frac{\mathd \chi}{\mathd \tau} = - e^{\chi} M, \qquad \frac{\mathd
  \phi}{\mathd \tau} = - e^{\chi} \partial_{\phi} M, \qquad \frac{\mathd
  \psi}{\mathd \tau} = - e^{\chi}  \frac{1}{G_{\psi \psi}} \partial_{\psi} M .
\end{equation}
Reparameterizing, these are:
\begin{equation}
  \frac{\mathd \phi}{\mathd \chi} = \partial_{\phi} \log M = - \alpha, \qquad
  \frac{\mathd \psi}{\mathd \chi} = \frac{1}{G_{\psi \psi}} \partial_{\psi}
  \log M = - e^{2 \beta \phi} \gamma .
\end{equation}
The solution (plotted in figure \ref{fig:counterexTrajectory}) is:
\begin{equation}
  \phi (\chi) = \phi_0 - \alpha \chi, \qquad \psi (\chi) = \psi_0 +
  \frac{\gamma}{2 \alpha \beta} e^{2 \beta \phi (\chi)} .
  \label{eqn:phipsisoln}
\end{equation}
We see that as we approach the horizon (i.e., as $\chi \rightarrow - \infty$),
$\phi$ increases without bound, hence $\phi_h = \infty$ in contrast with our
naive expectations.

To see how this is consistent with the directional derivative bound, note that
\begin{equation}
  \nabla_{\psi} \nabla_{\psi} \mathfrak{D}= - \Gamma^{\phi}_{\;
  \psi \psi} = \frac{1}{2} G_{\psi \psi}' = - \beta e^{- 2 \beta \phi} .
\end{equation}
Thus,
\begin{multline}
  \nabla_{\psi} \nabla_{\psi} \mathfrak{D}+ \frac{1}{2} \nabla_{\phi} \log Q^2
  G_{\psi \psi} = \left( - \beta + \frac{1}{2} 
  \frac{{\mathcal{Q}^2}'}{\mathcal{Q}^2} \right) e^{- 2 \beta \phi}
  \\
  = - \frac{(\alpha + \beta) [k_N + \alpha^2] \mathcal{M}^2_0 e^{- 2 \alpha
  \phi} + \alpha \gamma^2 \mathcal{M}^2_0 e^{2 (\beta - \alpha) \phi}}{[k_N +
  \alpha^2] \mathcal{M}^2_0 e^{- 2 \alpha \phi} + \gamma^2 \mathcal{M}^2_0
  e^{2 (\beta - \alpha) \phi}} e^{- 2 \beta \phi} < 0, 
\end{multline}
so the condition \eqref{eqn:Dconds}.ii is violated. Indeed, since this is
strictly negative \emph{everywhere}, even the more general condition
\eqref{eqn:DextCond2} is violated, so the directional derivative bound does not
apply.\footnote{Likewise, the charge threshold bound \eqref{eqn:totDistBound2} does not apply,
since due to its $\psi$ dependence, $Q^2 =\mathcal{Q}^2 e^{- 2 \gamma \psi}$
is decreasing towards the horizon in the solution \eqref{eqn:phipsisoln} (except potentially
for one brief period, depending on the parameters).}

To understand why the extremal solution does not flow to the point
$\phi_{\ast}$, despite the fact that this minimizes $Q^2$ for any fixed value
of $\psi$, consider the $\phi$ equation of motion, which in this case is
\begin{equation}
  \ddot{\phi} + \Gamma^{\phi}_{\; \psi \psi}  \dot{\psi}^2 =
  \frac{1}{2} e^{2 (\chi - \gamma \psi)} {\mathcal{Q}^2}' .
\end{equation}
Substituting $- \Gamma^{\phi}_{\; \psi \psi} = \frac{1}{2} G_{\psi
\psi}'$ (which is the extrinsic curvature $K_{\psi \psi}$ of a surface of
constant $\phi$) and rearranging, we obtain:
\begin{equation}
  \ddot{\phi} = \frac{1}{2} e^{2 (\chi - \gamma \psi)} {\mathcal{Q}^2}' +
  \frac{1}{2} G_{\psi \psi}'  \dot{\psi}^2 .
\end{equation}
Thus, the extrinsic curvature $K_{\psi \psi} = \frac{1}{2} G_{\psi \psi}'$
coupled with motion in the $\psi$ direction produces an effective force on
$\phi$. Near the horizon, this force must come into balance with the one
generated by ${\mathcal{Q}^2}'$ in order to have $\dot{\phi} \rightarrow 0$ as
$\tau \rightarrow \infty$, \footnote{The criterion $\dot{\phi} \rightarrow 0$
as $\tau \rightarrow \infty$ is required by the constraint $k_N^{- 1} 
\dot{\chi}^2 + \dot{\phi}^2 + G_{\psi \psi}  \dot{\psi}^2 = e^{2 (\chi -
\gamma \psi)} \mathcal{Q}^2$, where $e^{2 (\chi - \gamma \psi)} \mathcal{Q}^2
= k_N^{- 1}  \ddot{\chi}$ has to vanish as $\tau \rightarrow \infty$ to ensure
that $\dot{\chi}$ remains negative. This ensures that the extremal solution is
a limit of a smooth, subextremal solution (though it does not itself have a
smooth horizon).} which forces $\phi$ to move away from the minimum $\phi =
\phi_{\ast}$ of $\mathcal{Q}^2 (\phi)$.

We can also understand this result by considering the effective
one-dimensional problem, with
\begin{equation}
  G_{\phi \phi}^{\text{eff}} (\phi) = 1 + \frac{\gamma^2}{k_N} e^{2 \beta
  \phi}, \qquad Q^2_{\text{eff}} (\phi) = \frac{1}{G_{\phi \phi}^{\text{eff}}}
  \mathcal{Q}^2 = k_N \mathcal{M}^2_0  \left( 1 + \frac{\alpha^2}{k_N +
  \gamma^2 e^{2 \beta \phi}} \right) e^{- 2 \alpha \phi} .
\end{equation}
Since $Q^2_{\text{eff}} (\phi)$ is monotonically decreasing (due to the rapid
growth of $G_{\phi \phi}^{\text{eff}}$ as $\phi \rightarrow \infty$), in this
formulation it is obvious that $\phi \rightarrow \infty$ at the horizon.

\begin{figure}\centering
  \includegraphics[width=0.8\textwidth]{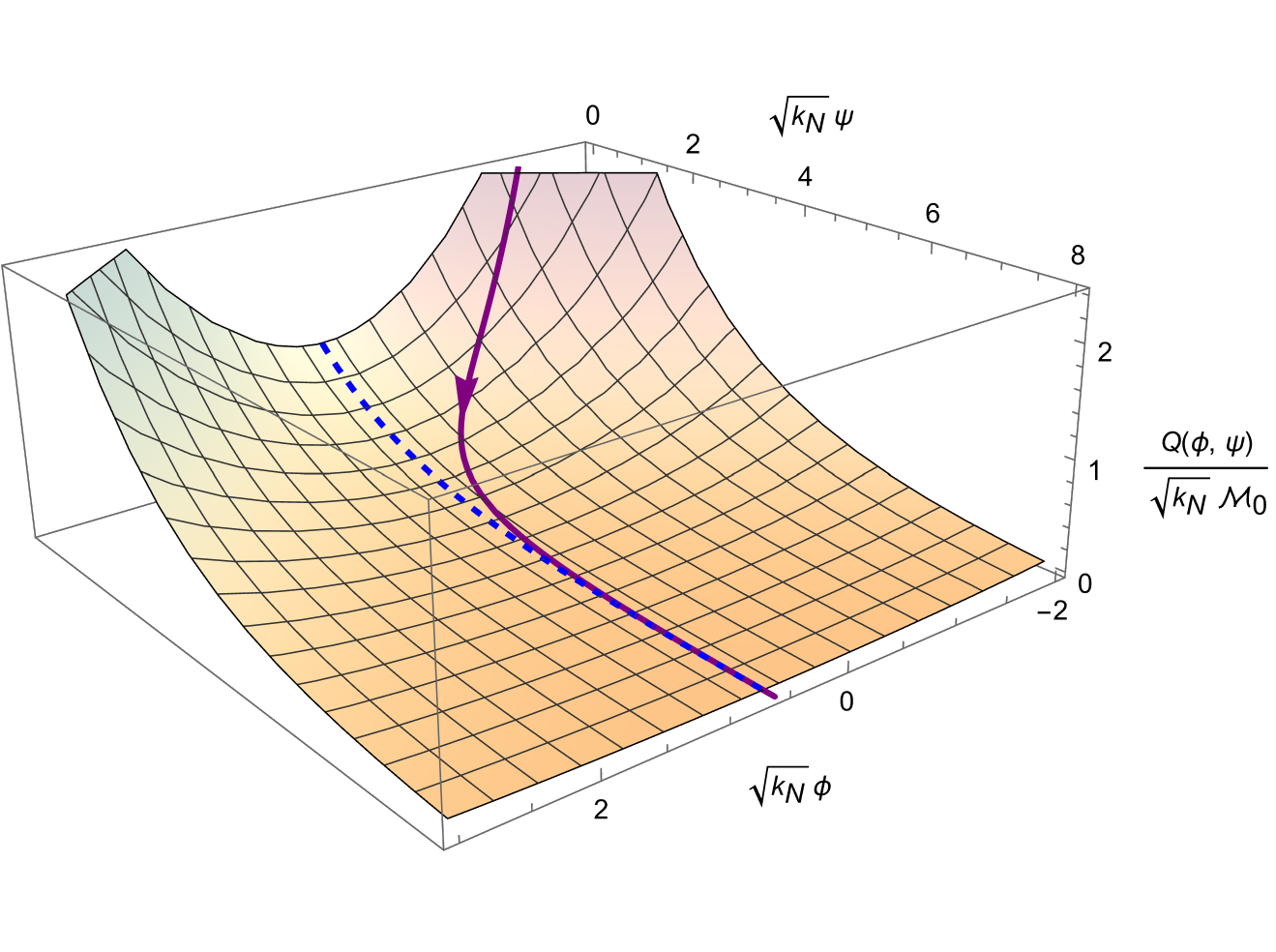}
  \caption{The trajectory (purple solid line) of a quasiextremal solution in
  the case of $G_{\psi \psi} = 1$ (i.e., without extrinsic curvature), plotted
  as a track along a surface of height $Q^2 (\phi, \psi)$, for the case
  $\alpha = \sqrt{k_N / 2}$, $\beta = \sqrt{2 k_N}$, $\gamma = \sqrt{k_N} /
  2$. Note that in this case the trajectory indeed asymptotes to the line
  $\phi = \phi_{\ast}$ (blue dashed line).\label{fig:noncounterexTrajectory}}
\end{figure}

It is also interesting to contrast this behavior with what happens in the
absence of extrinsic curvature. Let us consider the theory:
\begin{equation}
  G_{\psi \psi} = 1, \quad Q^2 (\phi, \psi) = (A e^{- 2 \alpha \phi} + B e^{2
  (\beta - \alpha) \phi}) e^{- 2 \gamma \psi},
\end{equation}
 for $A \assign [k_N + \alpha^2] \mathcal{M}^2_0, B \assign \gamma^2
  \mathcal{M}^2_0$,
i.e., with the same $Q^2 (\phi, \psi)$ but no extrinsic curvature. In this
case, one indeed finds that $\phi \rightarrow \phi_{\ast}$ as $\tau
\rightarrow \infty$. Let us demonstrate this. We consider the following ansatz
for $\mathcal{M} (\phi)$:
\begin{equation}
  \mathcal{M} (\phi) =\mathcal{M}_1 e^{- \alpha \phi} +\mathcal{M}_2 e^{(\beta
  - \alpha) \phi} .
\end{equation}
Thus, we require
\begin{multline}
  (k_N + \gamma^2)  (\mathcal{M}_1 e^{- \alpha \phi} +\mathcal{M}_2 e^{(\beta
  - \alpha) \phi})^2 + (- \alpha \mathcal{M}_1 e^{- \alpha \phi} + (\beta -
  \alpha) \mathcal{M}_2 e^{(\beta - \alpha) \phi})^2 
  \\
  = A e^{- 2 \alpha \phi} +
  B e^{2 (\beta - \alpha) \phi} .
\end{multline}
For this to work, the cross terms on the left-hand side have to cancel, which
requires
\begin{equation}
  k_N + \gamma^2 = \alpha (\beta - \alpha) . \label{eqn:alphabetagammaCond}
\end{equation}
Provided this is satisfied, the solution is:
\begin{align}
  \mathcal{M}_1 &= \sqrt{\frac{A}{k_N + \gamma^2 + \alpha^2 }} =
  \sqrt{\frac{k_N + \alpha^2}{\alpha \beta}} \mathcal{M}_0, \nonumber\\
  \mathcal{M}_2 &= \sqrt{\frac{B}{k_N + \gamma^2 + (\beta - \alpha)^2}} =
  \sqrt{\frac{\gamma^2}{\beta (\beta - \alpha)}} \mathcal{M}_0, 
\end{align}
where we use \eqref{eqn:alphabetagammaCond} to simplify the denominators.
Thus, $\mathcal{M} (\phi)$ has a global minimum at
\begin{equation}
  \phi = \frac{1}{\beta} \log \left[ \frac{\alpha \mathcal{M}_1}{(\beta -
  \alpha) \mathcal{M}_2} \right] = \frac{1}{2 \beta} \log \left[ \frac{\alpha
  [k_N + \alpha^2] }{(\beta - \alpha) \gamma^2 } \right] = \phi_{\ast},
\end{equation}
and indeed we conclude that $\phi \rightarrow \phi_{\ast}$ at the horizon.
This is illustrated in figure \ref{fig:noncounterexTrajectory}.\footnote{The
trajectory in this case is $\psi (\phi) = \psi_0 - \frac{\gamma (2 \alpha -
\beta)}{\alpha (\beta - \alpha)} \phi - \frac{\gamma}{\alpha (\beta - \alpha)}
\log | \mathcal{M}' (\phi) |$, as can be found by solving the gradient flow
equations for $M (\phi, \psi)$.}

\section{Applying the bounds in string theory}\label{sec:applications}

We now discuss how the bounds we have derived can be applied in string theory.

\subsection{Black holes with electric NSNS charge}\label{sec:electricNSNS}

Consider black holes with only electric NSNS charge in a $d$-dimensional
perturbative string vacuum. Let $\Phi_{d, \infty}$ be the asymptotic value of
the $d$-dimensional dilaton,\footnote{Note that $\Phi_d$ controls the
relationship between the string scale $m_{\text{str}} \assign
\frac{1}{\sqrt{\alpha'}}$ and the $d$-dimensional Planck scale $M_d \assign
\kappa_d^{- \frac{2}{d - 2}}$ via:
\begin{equation}
  m_{\text{str}}^{\frac{d - 2}{2}} = e^{\Phi_d} 2^{\frac{d - 4}{2}}
  \pi^{\frac{d - 3}{2}} M_d^{\frac{d - 2}{2}} .
\end{equation}
When the string in question is a compactifcation of some underlying critical
string theory, $\Phi_d$ differs from the dilaton $\Phi$ of this theory, since
$\Phi$ controls the relationship between the string scale and the
\emph{ten-dimensional} Planck scale. Note that $\Phi_d$ is duality
invariant, whereas $\Phi$ is not invariant under, e.g., T dualities.} and
consider the hypersurface in moduli space defined by:
\begin{equation}
  \Phi_d = \Phi_{d, \infty} .
\end{equation}
Let $\mathfrak{D}$ be the distance function that vanishes along this
hypersurface, with $\mathfrak{D}$ positive on the side $\Phi_d > \Phi_{d,
\infty}$. At string tree-level:
\begin{equation}
  G_{\Phi_d \Phi_d}^{\text{tree}} = \frac{4}{\kappa_d^2 (d - 2)}, \qquad
  G_{\Phi_d \varphi}^{\text{tree}} = 0,
\end{equation}
where $\varphi$ is any other modulus (whether in the NSNS or RR sector)
besides the dilaton. Thus, near the hypersurface $\Phi_d = \Phi_{d, \infty}$
we have:
\begin{equation}
  \mathfrak{D} \simeq \frac{2}{\kappa_d  \sqrt{d - 2}}  (\Phi_d - \Phi_{d,
  \infty}), \label{eqn:DPhieqn}
\end{equation}
where ``$\simeq$'' indicates that the expression is true up to string loop
corrections.\footnote{To compute these corrections one has to choose a scheme
for how the dilaton is defined at loop level. While we will not attempt to
show it here, we claim that a scheme exists where \eqref{eqn:DPhieqn} is
\emph{exact}.}

Likewise, at string tree level we find:
\begin{equation}
  \nabla^i \mathfrak{D} \nabla_i \log Q^2 \simeq \frac{2 \kappa_d}{\sqrt{d -
  2}} = 2 \alpha \sqrt{k_N} \qquad \text{for} \qquad \alpha = \frac{1}{\sqrt{d
  - 3}},
\end{equation}
where we use the fact that $Q^2 \propto \frac{1}{\alpha'}$ in Planck units for
a black hole with only electric NSNS charge, together with $\partial_{\Phi_d}
\log \alpha' = - \frac{4}{d - 2}$. Now, consider the quantity:
\begin{equation}
  \tilde{K}_{i j} \assign \nabla_i \nabla_j \mathfrak{D}+ \frac{1}{2} \nabla^i
  \mathfrak{D} \nabla_i \log Q^2 \Pi_{i j},
\end{equation}
where $\Pi_{i j}$ is the projected metric $G_{i j} - \nabla_i \mathfrak{D}
\nabla_j \mathfrak{D}$. For NSNS moduli, the metric on moduli space
$G_{\varphi \varphi'}$ is independent of the dilaton, so we obtain
\begin{equation}
  \nabla_{\varphi} \nabla_{\varphi'} \mathfrak{D} \simeq 0 \qquad \Rightarrow
  \qquad \tilde{K}_{\varphi \varphi'} \simeq \frac{\kappa_d}{\sqrt{d - 2}}
  G_{\varphi \varphi'} .
\end{equation}
Now let $\psi$ be an RR modulus and $\varphi$ be an NSNS modulus. We have:
\begin{equation}
  G^{\text{tree}}_{\Phi_d \psi} = 0, \qquad G_{\varphi \psi}^{\text{tree}} =
  0, \qquad G_{\psi \psi'}^{\text{tree}} = e^{2 \Phi_d} \mathcal{G}_{\psi
  \psi'} (\varphi) .
\end{equation}
In particular, the sphere one-point function of an RR vertex operator vanishes
and the logarithmic dilaton coupling $\frac{\langle \Phi_d \psi \psi'
\rangle}{\langle \mathfrak{g} \psi \psi' \rangle}$ is universal by analogous
arguments to those in \cite{Heidenreich:2024dmr}.\footnote{The
superstring generalization of \cite{Heidenreich:2024dmr} will appear in \cite{HLSuperstring}.} Thus, for RR moduli we obtain:
\begin{equation}
  K_{\psi \psi'} = \nabla_{\psi} \nabla_{\psi'} \mathfrak{D} \simeq \frac{1}{2
  \sqrt{G_{\Phi_d \Phi_d}^{\text{tree}}}} \partial_{\Phi_d} G_{\psi
  \psi'}^{\text{tree}} = \frac{\sqrt{d - 2}}{2} \kappa_d G_{\psi \psi'} \;
  \Rightarrow \quad \tilde{K}_{\psi \psi'} \simeq \frac{d}{2 \sqrt{d - 2}}
  \kappa_d G_{\psi \psi'} .
\end{equation}
Either way, we satisfy the intrinsic curvature constraint with room to spare.

Thus, applying the directional derivative bound \eqref{eqn:DistFuncBound}, we
conclude that a spherically symmetric black hole solution with only electric
NSNS charge cannot reach a dilaton value greater than
\begin{equation}
  \Phi_d^{\text{max}} = \Phi_{d, \infty} + \frac{\pi (d - 2)}{4 \sqrt{d - 3}}
  . \label{eqn:PhiMax}
\end{equation}
Note that although we have only used the string tree level EFT couplings in
determining this bound, since the conditions \eqref{eqn:Dboundcons} are all
satisfied with room to spare, the bound is rigorous provided that string loop
corrections remain small (but need not be \emph{infinitesimal}) at $\Phi_d
= \Phi_{d, \infty}$. For instance, if we assume that the perturbative limit is
$g_d^{\text{pert}} \lesssim 1$ then we only need $g_{d, \infty} \lesssim 0.1$
to ensure that the black hole cannot reach a non-perturbative attractor, since
the second term in \eqref{eqn:PhiMax} is about $1.6 - 2.4$, depending on $d$.

In fact, using the refined bound \eqref{eqn:refinedDistFuncBound}, we get:
\begin{equation}
  \Phi_d^{\text{max}} - \Phi_{d, \infty} \simeq \frac{\sqrt{d - 3} \pi}{4} -
  \frac{1}{2} \log \sqrt{d - 3} .
\end{equation}
Since the right-hand side is a bit smaller than in \eqref{eqn:PhiMax}, we can
further relax the bound on $g_{d, \infty}$. However, the value of $\alpha$ in
\eqref{eqn:refinedDistFuncBound} is subject to string loop corrections, so we
cannot be fully precise about this.

Thus, for small (but not \emph{parametrically} small) of the asymptotic
string coupling $g_{d, \infty}$ we can guarantee that spherically symmetric
black holes with only NSNS electric charge do not explore strongly coupled
regions of the moduli space, and so we can bound the spectrum of these black
holes sharply using only perturbative information. This places the argument of
section 3.4 of [\href{https://arxiv.org/abs/2401.14449}{2401.14449}] on a
rigorous footing.

\subsection{Black holes with dyonic NSNS charges}

In perturbative string theory there are in general both one-form $A_{\mu}^a$
and two-form $B_{\mu \nu}$ NSNS gauge potentials, the latter being the
Kalb-Ramond two form. In $d > 5$, only extended objects can carry magnetic
charge under these gauge fields. However, in $d = 5$ particles can carry
magnetic $B_{\mu \nu}$ charge as well as electric $A_{\mu}^a$ charge, and in
$d = 4$ they can carry both electric and magnetic $A_{\mu}^a$ charge.

Thus, in $d = 4, 5$ we can have black holes with dyonic NSNS charges. We now
show that if these charges and the asymptotic values of the moduli satisfy
certain conditions then we can again guarantee that spherically-symmetric
black holes of this type do not explore strongly-coupled regions of the moduli
space, and thus the spectrum of such black holes can be computed
perturbatively.

For simplicity, we will focus on toroidal compactifications of the heterotic
string, even though we expect that similar reasoning should apply to general
string compactifications. At a generic point in the moduli space,\footnote{At
special points in the moduli space, the $U (1)^{36 - 2 d}$ gauge group is
enhanced to a non-Abelian group of the same rank. However, these enhancements
have no effect on spherically symmetric charged black hole solutions and thus
are not important for our analysis.} the two-derivative string-frame action
for the bosonic fields is:\footnote{There is a well-known four-derivative
gravitational correction to $\mathd \tilde{H}_3$ which is important for
anomaly cancellation, but we do not write it here as it is not part of the
two-derivative effective action.}
\begin{multline}
  S = \frac{1}{2 \hat{\kappa}_d^2}  \int \mathd^d x \sqrt{-
  g} e^{- 2 \Phi_d}  \biggl[ R + 4 (\nabla \Phi_d)^2 - \frac{1}{2}  |
  \tilde{H}_3 |^2 - \frac{\alpha'}{2} \varphi_{A B} F^A \cdot F^B +
  \frac{1}{8} \nabla \varphi_{A B} \cdot \nabla \varphi^{A B} \biggr],
  \\
   \text{where} \qquad \mathd \tilde{H}_3 = - \frac{\alpha'}{2} \eta_{A B}
  F^A \wedge F^B, \qquad 2 \hat{\kappa}^2_d = (2 \pi)^{d - 3}
  {\alpha'}^{\frac{d - 2}{2}},  \label{eqn:hetEffAction}
\end{multline}
$\eta_{A B} = \eta_{B A}$ has signature $(26 - d, 10 - d)$, and $\varphi_{A B}
= \varphi_{B A}$ is a positive definite quadratic form satisfying the
constraint $\varphi^A_{\; B} \varphi^B_{\; C} =
\delta^A_C$ where we raise and lower indices using $\eta_{A B}$. Note that
consequently, $\varphi^{A B} \assign \eta^{A C} \eta^{B D} \varphi_{C D}$ is
also the inverse of $\varphi_{A B}$. This action can be derived via
dimensional reduction, as in \cite{Heidenreich:2024dmr}, appendix C. In Einstein
frame, it becomes:
\begin{align}
  S_E &= \begin{multlined}[t][0.75\textwidth]
  \frac{1}{2 \kappa^2_d}  \int \mathd^d x \sqrt{- g} 
  \Bigg[ R - \frac{4}{d - 2}  (\nabla \Phi_d)^2 - \frac{g_{d,
  \infty}^{\frac{8}{d - 2}}}{2} e^{- \frac{8 \Phi_d}{d - 2}}  | \tilde{H}_3
  |^2 \\
   - \frac{\alpha' g_{d, \infty}^{\frac{4}{d - 2}}}{2} \varphi_{A B} e^{-
  \frac{4 \Phi_d}{d - 2}} F^A \cdot F^B + \frac{1}{8} \nabla \varphi_{A B}
  \cdot \nabla \varphi^{A B} \Bigg],
  \end{multlined} \nonumber \\
  \mathd \tilde{H}_3 &= - \frac{\alpha'}{2} \eta_{A B} F^A \wedge F^B, 
\end{align}
where $\kappa_d = e^{\Phi_{d, \infty}}  \hat{\kappa}_d$ is the gravitational
coupling and $g_{d, \infty} = e^{\Phi_{d, \infty}}$ is the asymptotic value of
the $d$-dimensional string coupling, with $2 \kappa_d^2 = (2 \pi)^{d - 3}
g_{d, \infty}^2 {\alpha'}^{\frac{d - 2}{2}}$. As noted above, the two cases of
interest for dyonic black holes are $d = 4, 5$.

\subsubsection{Dyonic black holes in \alt{$d = 5$}{d = 5}}

In 5d, the Kalb-Ramond two-form gauge field $B_2$ is electromagnetically dual
to a one-form gauge field $\mathcal{B}_1$. To dualize the action, we first
introduce $\mathcal{B}_1$ as a Lagrange multiplier enforcing the $\tilde{H}_3$
Bianchi identity:
\begin{multline}
  S_E = \frac{1}{2 \kappa^2_5}  \int \mathd^5 x \sqrt{- g} 
  \Bigg[ R - \frac{4}{3}  (\nabla \Phi_5)^2 - \frac{g_{5, \infty}^{8 / 3}}{2}
  e^{- \frac{8}{3} \Phi_5}  | \tilde{H}_3 |^2 - \frac{\alpha' g_{5, \infty}^{4
  / 3}}{2} \varphi_{A B} e^{- \frac{4}{3} \Phi_5} F^A \cdot F^B  \\
    + \frac{1}{8}
  \nabla \varphi_{A B} \cdot \nabla \varphi^{A B} \Bigg] 
  + \frac{1}{4 \pi^2 \alpha'} \int \mathcal{B}_1 \wedge \left( \mathd
  \tilde{H}_3 + \frac{\alpha'}{2} \eta_{A B} F^A \wedge F^B \right), 
\end{multline}
where $\tilde{H}_3$ is now an auxilliary field. Integrating it out, we obtain
the dual action:
\begin{multline}
  S_E = \frac{1}{2 \kappa^2_5}  \int \mathd^5 x \sqrt{- g} 
  \left( R - \frac{4}{3}  (\nabla \Phi_5)^2 + \frac{1}{8} \nabla \varphi_{A B}
  \cdot \nabla \varphi^{A B} \right) + \frac{1}{8 \pi^2} \int \eta_{A B}
  \mathcal{B}_1 \wedge F^A \wedge F^B  \\
   - \frac{1}{2 e_5^2}  \int \mathd^5 x \sqrt{- g}  \left[
  e^{- \frac{4}{3} \Phi_5} \varphi_{A B} F^A \cdot F^B + e^{\frac{8}{3}
  \Phi_5}  | \mathcal{H}_2 |^2 \right],
\end{multline}
where $e^2_5 \assign (2 \pi)^{4 / 3}  (2 \kappa_5^2)^{1 / 3}$  
and $\mathcal{H}_2 = \mathd \mathcal{B}_1$.

The gauge kinetic matrix for $A^A, \mathcal{B}$ is thus
\begin{equation}
  \mathfrak{f}_{I J} (\varphi, \Phi_5) = \frac{1}{e^2_5} 
  \begin{pmatrix}
    e^{- \frac{4}{3} \Phi_5} \varphi_{A B} & 0\\
    0 & e^{\frac{8}{3} \Phi_5}
  \end{pmatrix},
\end{equation}
where $I, J$ index the combined space of electric and magnetic NSNS gauge
fields. From this, we find the charge function $\mathcal{Q}^2 (\varphi, \Phi)$
for a black hole with dyonic charge $\mathcal{Q}_I = (Q_A, P)$, where $Q_A$
and $P$ are the electric and magnetic charges, respectively:
\begin{equation}
  \mathcal{Q}^2 (\varphi, \Phi_5) = e^2_5 e^{\frac{4}{3} \Phi_5} \varphi^{A B}
  Q_A Q_B + e_5^2 e^{- \frac{8}{3} \Phi_5} P^2 . \label{eqn:5dQ2dyonic}
\end{equation}
Note that
\begin{equation}
  \varphi^{A B} Q_A Q_B \geqslant | Q \circ Q | = | Q_L^2 - Q_R^2 |,
\end{equation}
where $Q \circ Q' \assign \eta^{A B} Q_A Q_B'$. Thus, so long as the electric
charges $Q_A$ satisfy $Q \circ Q \neq 0$ ($Q_L^2 \neq Q_R^2$), the charge
function has a global minimum:
\begin{multline}
  Q^2 (\varphi, \Phi_5) \geqslant e^2_5 e^{\frac{4}{3} \Phi_5}  | Q \circ Q |
  + e^2_5 e^{- \frac{8}{3} \Phi_5} P^2 \geqslant 3 e_5^2  \biggl[ \frac{| Q
  \circ Q |^2 P^2}{4} \biggr]^{1 / 3} 
  \\=
   6 \pi^{4 / 3} \kappa^{2 / 3}_5  |
  Q_L^2 - Q_R^2 |^{2 / 3} P^{2 / 3},
\end{multline}
at which point the dilaton takes the value
\begin{equation}
  \Phi_5^{\text{attr}} = \frac{1}{4} \log \left[ \frac{2 P^2}{| Q \circ Q |}
  \right] \qquad \Rightarrow \qquad g_5^{\text{attr}} =
  e^{\Phi_5^{\text{attr}}} = \left[ \frac{2 P^2}{| Q \circ Q |} \right]^{1 /
  4} .
\end{equation}
Thus, for an extremal dyonic black hole the moduli flow to this attractor
point at the horizon, $g_{5, h} = g_5^{\text{attr}}$. Provided that
$\frac{1}{2}  | Q \circ Q | \gg P^2$, this solution can be described in
perturbative string theory.\footnote{Note that in our conventions the dyonic
charge lattice is $Q_A, \tilde{Q} \in \mathbb{Z}$.}

In the case of heterotic string theory on a torus, we can still understand
black holes with $g_{5, h} \gg 1$ using supersymmetry and string dualities.
However, in more general string compactifications these tools may not be
available. Thus, it is interesting to ask whether a spherically symmetric
black hole could instead flow to a ``hidden'' attractor point in the
strongly-coupled region of moduli space.

As before, we apply the directional derivative bound \eqref{eqn:DistFuncBound}
with the distance function \eqref{eqn:DPhieqn}. We find:
\begin{equation}
  \partial_{\Phi_5} \mathcal{Q}^2 = \frac{4}{3} e^2_5 e^{\frac{4}{3} \Phi_5}
  \varphi^{A B} Q_A Q_B - \frac{8}{3} e_5^2 e^{- \frac{8}{3} \Phi_5} P^2
  \geqslant \frac{4}{3} e_5^2  \left( e^{\frac{4}{3} \Phi_5}  | Q \circ Q | -
  2 e^{- \frac{8}{3} \Phi_5} P^2 \right) .
\end{equation}
Thus, $\partial_{\Phi_5} \mathcal{Q}^2 > 0$ for all $\Phi_5 >
\Phi_5^{\text{attr}}$. Since $K_{i j} \succcurlyeq 0$ as shown in
\S\ref{sec:electricNSNS}, this implies that so long as $g_{5,
\infty}$ and $g_5^{\text{attr}}$ are both sufficiently small then the black
hole solution cannot reach the strongly-coupled region.

\subsubsection{Dyonic black holes in \alt{$d = 4$}{d = 4}}

In 4d, the Kalb-Ramond two-form gauge field $B_2$ is electromagnetically dual
to an axion $\chi$. As before, we first introduce this as a Lagrange
multiplier enforcing the Bianchi identity for $\tilde{H}_3$:
\begin{multline}
  S_E = \frac{1}{2 \kappa^2_4}  \int \mathd^4 x \sqrt{- g} 
  \biggl[ R - 2 (\nabla \Phi_4)^2 - \frac{g_{4, \infty}^4}{2} e^{- 4 \Phi_4}  |
  \tilde{H}_3 |^2 - \frac{\alpha' g_{4, \infty}^2}{2} \varphi_{A B} e^{- 2
  \Phi_4} F^A \cdot F^B \\ 
  + \frac{1}{8} \nabla \varphi_{A B} \cdot \nabla
  \varphi^{A B} \biggr]
   + \frac{1}{4 \pi^2 \alpha'} \int \chi \biggl( \mathd \tilde{H}_3 +
  \frac{\alpha'}{2} \eta_{A B} F^A \wedge F^B \biggr) . 
\end{multline}
Integrating by parts so that $\tilde{H}_3$ appears without any derivatives and
then integrating it out, we obtain:
\begin{multline}
  S_E = \frac{1}{2 \kappa^2_4}  \int \mathd^4 x \sqrt{- g} 
  \biggl[ R - 2 (\nabla \Phi_4)^2 + \frac{1}{8} \nabla \varphi_{A B} \cdot
  \nabla \varphi^{A B} \biggr] - \frac{1}{4 \pi} \int \mathd^4 x \sqrt{- g}
  e^{- 2 \Phi_4} \varphi_{A B} F^A \cdot F^B \\
   - \frac{f^2_{\chi}}{2} \int e^{4 \Phi_4} \mathd \chi \wedge \star \mathd
  \chi + \frac{1}{8 \pi^2} \int \eta_{A B} \chi F^A \wedge F^B,
  \label{eqn:4ddyonicEffaction} 
\end{multline}
where $f_{\chi} \assign \frac{1}{2 \pi \sqrt{2 \kappa^2_4}}$. Thus, the holomorphic gauge kinetic matrix is
\begin{equation}
  \tau_{A B} = - \frac{\chi}{2 \pi} \eta_{A B} + i e^{- 2 \Phi_4} \varphi_{A  B} .
\end{equation}

To handle dyonically charged objects, it is convenient to reformulate the
gauge fields $F^A$ in an electromagnetically democratic fashion. Following the
conventions of \cite{Heidenreich:2020upe}, appendix A, the resulting electromagnetic gauge
kinetic matrix is
\begin{equation}
  t_{I J} = \begin{pmatrix}
    \tau_2 + \tau_1 \tau_2^{- 1} \tau_1 & - \tau_1 \tau_2^{- 1}\\
    - \tau_2^{- 1} \tau_1 & \tau_2^{- 1}
  \end{pmatrix} \qquad \Rightarrow \qquad t^{I J} =
  \begin{pmatrix}
    \tau_2^{- 1} & \tau_2^{- 1} \tau_1\\
    \tau_1 \tau_2^{- 1} & \tau_2 + \tau_1 \tau_2^{- 1} \tau_1
  \end{pmatrix}\,,
\end{equation}
where $(\tau_1)_{A B} = \re \tau_{A B}$, $(\tau_2)_{A B} = \im
\tau_{A B}$, and the indices $I, J$ cover both electric and magnetic
components. We can then read off the charge function:
\begin{align}
  \mathcal{Q}^2 &= 2 \pi t^{I J} \mathcal{Q}_I
  \mathcal{Q}_J = 2 \pi (\tau_2^{- 1})^{A B} [Q_A + (\tau_1)_{A C} P^C] [Q_B +
  (\tau_1)_{B D} P^D] + 2 \pi (\tau_2)_{A B} P^A P^B \nonumber\\
  &= 2 \pi e^{2 \Phi_4} \varphi^{A B} \Bigl[ Q_A - \frac{\chi}{2 \pi} P_A \Bigr] \Bigl[ Q_B - \frac{\chi}{2 \pi} P_B \Bigr] + 2 \pi e^{- 2 \Phi_4} \varphi^{A B} P_A P_B \,,   \label{eqn:4dchargefunc}
\end{align}
as a function of the moduli $(\varphi_{A B}, \Phi_4, \chi)$,
where $\mathcal{Q}_I = (Q_A, P^A)$ combines the electric and magnetic charges $Q_A$ and $P^A$, respectively, and $P_A \assign \eta_{A B} P^B$.

Unlike the 5d case \eqref{eqn:5dQ2dyonic}, this charge function does
\emph{not} satisfy the constraints \eqref{eqn:Dboundcons} required to
apply the directional derivative bound with $\mathfrak{D}$ proportional to the
dilaton, because $\partial_{\Phi_4} \mathcal{Q}^2$ is always negative
somewhere on every fixed-$\Phi_4$ slice. In particular,
\begin{equation}
  \partial_{\Phi_4} \mathcal{Q}^2 = 4 \pi e^{2 \Phi_4}  \left[ Q -
  \frac{\chi}{2 \pi} P \right] \varphi \left[ Q - \frac{\chi}{2 \pi} P \right]
  - 4 \pi e^{- 2 \Phi_4} P \varphi P, \label{eqn:dPhiQ4ddyonic}
\end{equation}
where $P \varphi Q \assign P_A \varphi^{A B} Q_B$, etc., is a convenient
shorthand. Consider a generic point in the $\varphi$ moduli space, and focus
on the $(1, 1)$ plane generated by $\overrightarrow{P_L}, \vec{P}_R$. Boosting
$\varphi$ within this plane, we can make $\varphi^{A B} P_A P_B = P_L^2 +
P_R^2$ arbitrarily large. This invariably makes $\partial_{\Phi_4}
\mathcal{Q}^2$ negative so long as the first term does not also grow
arbitrarily large. Choosing $\chi = 2 \pi \vec{Q}_L \cdot \vec{P}_L / | P_L
|^2$, we find:
\begin{multline}
  \Bigl[ Q - \frac{\chi}{2 \pi} P \Bigr] \varphi \Bigl[ Q - \frac{\chi}{2
  \pi} P \Bigr] = \nonumber \\ 
   \biggl[ Q_L^2 - \frac{(\vec{Q}_L \cdot \vec{P}_L)^2}{P_L^2}
  \biggr] + \biggl[ Q_R^2 - \frac{(\vec{Q}_R \cdot \vec{P}_R)^2}{P_R^2} \biggr]
  + \frac{[(\vec{Q}_L \cdot \hat{P}_L) P_R - (\vec{Q}_R \cdot \hat{P}_R)
  P_L]^2}{P_L^2} .
\end{multline}
The terms in square brackets are all invariant under the boost in question,
whereas $P_L^2$ increases, so the first term in \eqref{eqn:dPhiQ4ddyonic}
\emph{decreases} in magnitude, and therefore $\partial_{\Phi_4}
\mathcal{Q}^2$ eventually becomes negative.\footnote{Note that this argument
does not apply if the theory has only left-moving (or only right-moving) gauge
bosons in it. This of course is not the case for heterotic string theory on
$T^6$, but is the case, e.g., for heterotic string theory on a Calabi-Yau
manifold. If all gauge fields are left-moving and the charge function is still
of the form \eqref{eqn:4dchargefunc} then the directional derivative bound
\eqref{eqn:DistFuncBound} \emph{can} be applied, much like in the 5d
dyonic case discussed previously.}

Even though the bound \eqref{eqn:DistFuncBound} cannot be applied, we may
still be able to apply the charge threshold bound \eqref{eqn:totDistBound2},
provided that $\mathcal{Q}^2 (\varphi, \Phi, \chi)$ is larger near the
strongly-coupled regions of moduli space than in the vacuum we are
considering. To figure out when this is the case, we first find the minimum
value of $\mathcal{Q}^2 (\varphi, \chi)$ for each fixed value of the string
coupling $g_4 \assign e^{\Phi_4}$. As shown in Appendix
\ref{app:4ddyonicQ2bounds}, $\mathcal{Q}^2 (\varphi, \chi)$ has the following
lower bounds:
\begin{tcolorbox}[colback=white, colframe=black,  boxrule=0.8pt,  arc=0pt,  left=2mm,  right=2mm,  top=1mm,  bottom=1mm]
  \begin{subequations}
  \begin{enumerate}
    \item If $P \circ P \neq 0$ and $(P \circ P) (Q \circ Q) - (P \circ Q)^2
    \geqslant - \frac{1}{g_4^4}  (P \circ P)^2$ then
    \begin{equation}
      \mathcal{Q}^2 \geqslant 2 \pi g_4^2  \left| Q \circ Q - \frac{(Q \circ
      P)^2}{P \circ P} \right| + \frac{2 \pi}{g_4^2}  | P \circ P |,
      \label{eqn:Q2dyonicbound1}
    \end{equation}
    where the inequality is strict if $(P \circ P) (Q \circ Q) - (P \circ Q)^2
    = 0$.
    
    \item If $P \circ P = Q \circ P = 0$ then
    \begin{equation}
      \mathcal{Q}^2 > 2 \pi g_4^2  | Q \circ Q | . \label{eqn:Q2dyonicbound2}
    \end{equation}
    \item If $(P \circ P) (Q \circ Q) - (P \circ Q)^2 < - \frac{1}{g_4^4}  (P
    \circ P)^2$ then
    \begin{equation}
      \mathcal{Q}^2 \geqslant 4 \pi \sqrt{(Q \circ P)^2 - (Q \circ Q) (P \circ
      P) } . \label{eqn:Q2dyonicbound3}
    \end{equation}
  \end{enumerate}
  \end{subequations}
\end{tcolorbox}
\noindent Each of these bounds is saturated somewhere in $(\varphi, \chi)$ moduli space
unless the inequality is strict, in which case the bound is optimum, i.e.,
$\mathcal{Q}^2 (\varphi, \chi) -\mathcal{Q}^2_{\text{bound}}$ has no positive
lower bound.

In order to apply the charge threshold bound \eqref{eqn:totDistBound2} to rule out strongly
coupled black holes, we require that $\mathcal{Q}^2 (\varphi, \chi)
>\mathcal{Q}^2_{\infty}$ everywhere in some band $g_4 \in [g_4^{(1)},
g_4^{(2)}]$ where $g_4^{(2)} / g_4^{(1)} \geqslant e^{\pi / 2}$ (see
\eqref{eqn:PhiMax}) and $g_4^{ (2)} \ll 1$. For this to be possible, it must
at least be the case that there is a band where the minimum value
$\mathcal{Q}^2_{\text{bound}} (g_4)$ is increasing. Inspecting
\eqref{eqn:Q2dyonicbound1}--\eqref{eqn:Q2dyonicbound3}, we see that this is
the case either when (i) $(P \circ P) (Q \circ Q) - (P \circ Q)^2 > 0$ or (ii)
$P \circ P = Q \circ P = 0$ with $Q \circ Q \neq 0$.

In the first case, where $(P \circ P) (Q \circ Q) - (P \circ Q)^2 > 0$, we
have:
\begin{equation}
  \mathcal{Q}^2_{\text{bound}} (g_4) = 2 \pi g_4^2  \left| Q \circ Q -
  \frac{(Q \circ P)^2}{P \circ P} \right| + \frac{2 \pi}{g_4^2}  | P \circ P | .
\end{equation}
This reaches a minimum at
\begin{equation}
  g_4^{\text{attr}} = \left[ \frac{(P \circ P)^2}{(P \circ P) (Q \circ Q) - (P
  \circ Q)^2} \right]^{1 / 4},
\end{equation}
where $\mathcal{Q}^2_{\text{bound}} (g_4)$ is increasing (decreasing) for $g_4
> g_4^{\text{attr}}$ ($g_4 < g_4^{\text{attr}}$).

\begin{figure}\centering
  \includegraphics[width=0.8\textwidth]{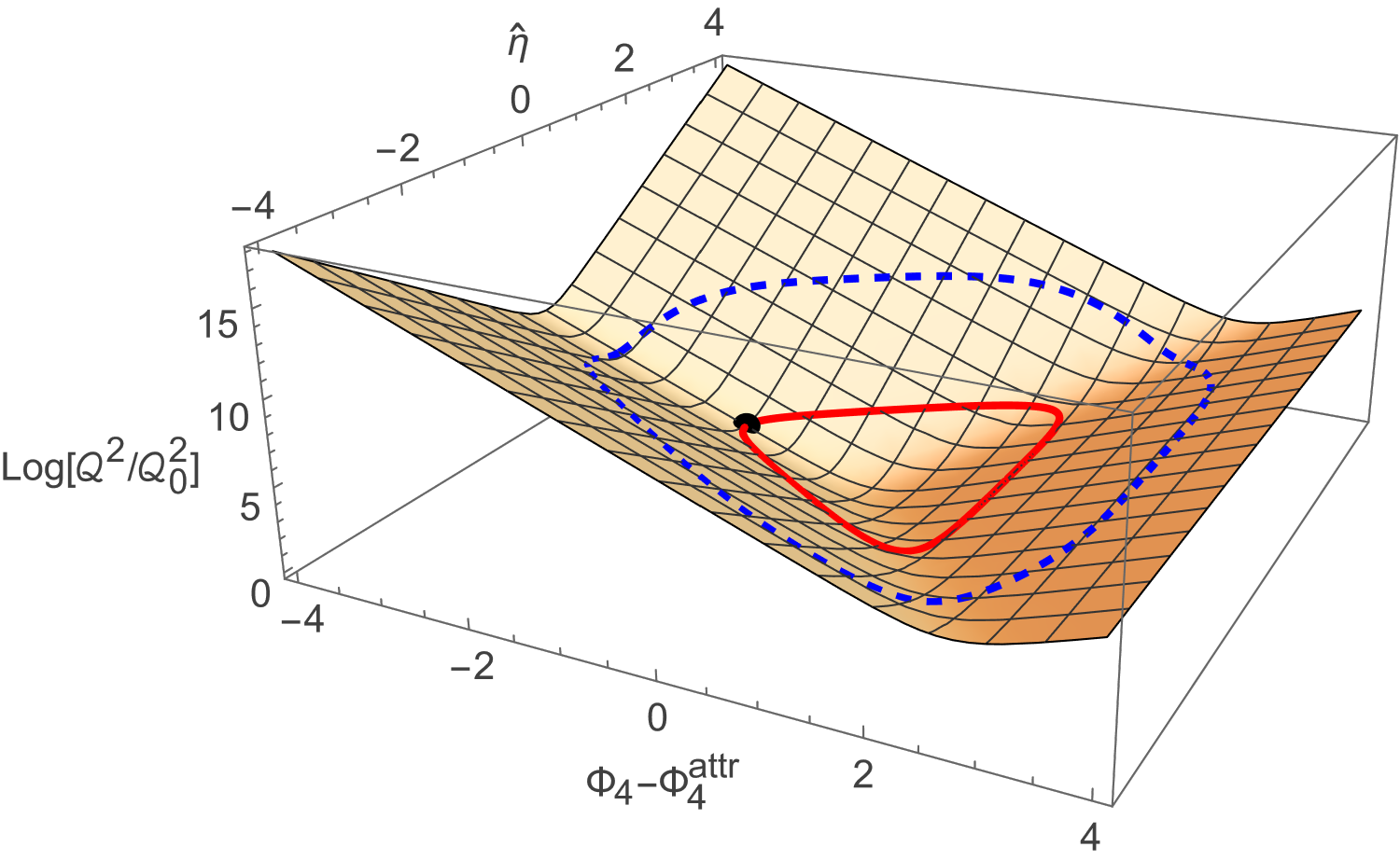}
  \caption{The log of the charge function $\mathcal{Q}^2$ in the
  $\Phi$-$\hat{\eta}$ plane in the $P \circ P, Q \circ Q > 0$, $P \circ Q = 0$
  case. If the black point denotes the asymptotic values of the moduli then
  $\mathcal{Q}^2 \geqslant \mathcal{Q}_{\infty}^2$ everywhere outside the
  solid red line, and thus the charge threshold bound \eqref{eqn:totDistBound2} implies that
  the black hole solution cannot reach regions of moduli space outside the
  blue dashed line.\label{fig:Q2basin}}
\end{figure}

To illustrate this, consider the case where $P \circ Q = 0$ with $P \circ P, Q
\circ Q > 0$. For simplicity, we set $\chi = 0$ and examine a slice of moduli
space where $Q$ is purely left-moving, while $P$ has both left and
right-moving components, so that
\begin{equation}
  \vec{P}_L = | P | \cosh \eta \hat{P}_L, \qquad \vec{P}_R = | P | \sinh \eta
  \hat{P}_R, \qquad \vec{Q}_L = | Q |  \hat{Q}_L, \qquad \vec{Q}_R = \vec{0},
\end{equation}
where $| P | = \sqrt{P \circ P}$, $| Q | = \sqrt{Q \circ Q}$, and $\hat{P}_L,
\hat{Q}_L$ are orthogonal since $P \circ Q = 0$. Here $\eta$ is one of the
moduli in $\varphi_{A B}$. In particular, focusing on the plane spanned by
$\hat{P}_L, \hat{P}_R$ and choosing a fixed basis (orthonormal with respect to
$\eta_{A B}$) in which $P_A = | P |  \bigl(\begin{smallmatrix} 1\\ 0 \end{smallmatrix}\bigr)$, we have
\begin{equation}
  \varphi_{A B} = \begin{pmatrix}
    \cosh 2 \eta & \sinh 2 \eta\\
    \sinh 2 \eta & \cosh 2 \eta
  \end{pmatrix}, \qquad \eta_{A B} = \begin{pmatrix}
    1 & 0\\
    0 & - 1
  \end{pmatrix},
\end{equation}
within this plane. Plugging into \eqref{eqn:4ddyonicEffaction}, we read off
the moduli-space metric $G_{\eta \eta} = \frac{1}{2 k_N}$. It is convenient to
define $\hat{\eta} \assign \eta / \sqrt{2}$, so that $G_{\hat{\eta}
\hat{\eta}} = G_{\Phi_4 \Phi_4} = \frac{1}{k_N}$. In terms of these moduli, we
obtain:
\begin{equation}
  \mathcal{Q}^2 = \frac{\mathcal{Q}^2_0}{2}  \left[ e^{2 \left( \Phi_4 -
  \Phi_4^{\text{attr}} \right)} + e^{- 2 \left( \Phi_4 - \Phi_4^{\text{attr}}
  \right)} \cosh \left( \sqrt{8} \hat{\eta} \right) \right],
\end{equation}
where $\mathcal{Q}^2_0 = 4 \pi | P | | Q |,
  \Phi_4^{\text{attr}} = \frac{1}{2} \log \frac{| P |}{| Q |} $.
This is shown in figure \ref{fig:Q2basin}.

As seen in the figure, for a suitable choice of the charges and the moduli at
infinity, we can indeed ensure that the black hole solution does not reach
strongly coupled regions of moduli space. One can also see why the directional derivative bound with $\mathfrak{D} \propto \Phi_4$ was not successful in this
case: for any fixed value of $\Phi_4$, $\partial_{\Phi_4} \mathcal{Q}^2$
eventually becomes negative for large enough $\hat{\eta}$. However, since this
happens where $\mathcal{Q}^2$ is very large, we can still apply the charge threshold bound
\eqref{eqn:totDistBound2}.\footnote{The directional derivative bound might also be
applicable for a different choice of $\mathfrak{D}$. We leave this question to
future work.}

In the second case, where $P \circ P = Q \circ P = 0$ with $Q \circ Q \neq 0$,
we have:
\begin{equation}
  \mathcal{Q}^2_{\text{bound}} (g_4) = 2 \pi g_4^2  | Q \circ Q |,
\end{equation}
which is increasing for all $g_{\text{eff}}$.

\begin{figure}\centering
  \includegraphics[width=0.8\textwidth]{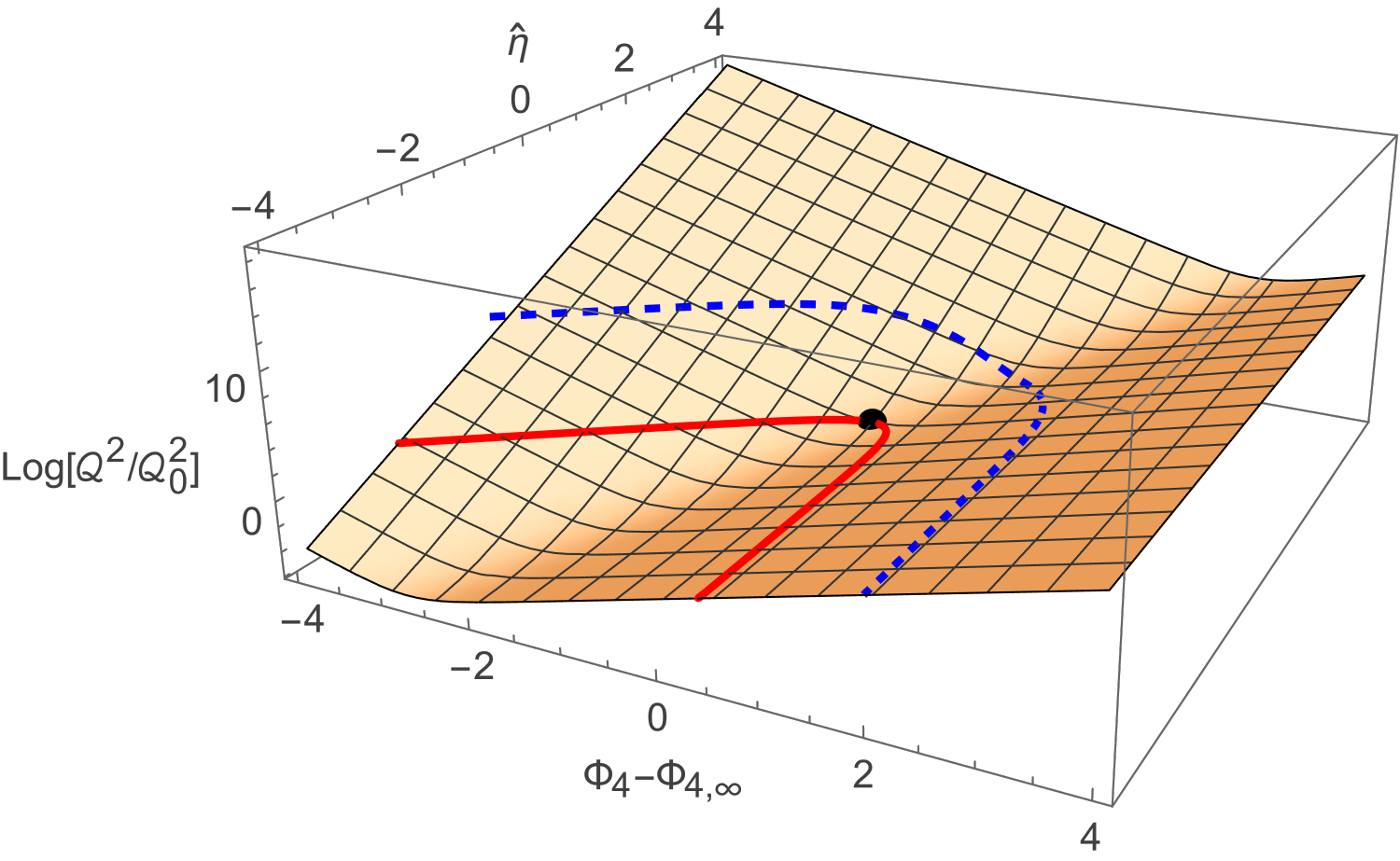}
  \caption{The log of the charge function $\mathcal{Q}^2$ in the
  $\Phi_4$-$\hat{\eta}$ plane in the $P \circ P = P \circ Q = 0$, $Q \circ Q >
  0$ case. This is similar to figure \ref{fig:Q2basin}, except that because $P
  \circ P = 0$, black hole solutions can flow arbitrarily far towards the
  lower left ($\Phi_4, \hat{\eta} \rightarrow -
  \infty$).\label{fig:Q2runaway}}
\end{figure}

To illustrate this, consider the case where $Q \circ Q > 0$. We again consider
a slice of moduli space where $\chi = 0$ and $Q$ is purely left-moving, so
that
\begin{equation}
  \vec{P}_L = P_0 e^{\eta}  \hat{P}_L, \qquad \vec{P}_R = P_0 e^{\eta} 
  \hat{P}_R, \qquad \vec{Q}_L = | Q |  \hat{Q}_L, \qquad \vec{Q}_R = \vec{0},
\end{equation}
As before, $\eta$ is one of the moduli in $\varphi_{A B}$. Choosing a fixed
basis for the $\hat{P}_L, \hat{P}_R$ plane in which $P_A = P_0 
\bigl(\begin{smallmatrix} 1\\ 1 \end{smallmatrix}\bigr)$, we have:
\begin{equation}
  \varphi_{A B} = \begin{pmatrix}
    \cosh 2 \eta & \sinh 2 \eta\\
    \sinh 2 \eta & \cosh 2 \eta
  \end{pmatrix}, \qquad \eta_{A B} = \begin{pmatrix}
    1 & 0\\
    0 & - 1
  \end{pmatrix},
\end{equation}
within this plane, so $G_{\eta \eta} = \frac{1}{2 k_N}$ as above. Note that we
can shift $\eta$ by a constant at the expense of rescaling $P_0$. We will
choose
\begin{equation}
  P_0 = \frac{g_{4, \infty}^2}{\sqrt{2}}  | Q | .
\end{equation}
We then find:
\begin{equation}
  \mathcal{Q}^2 (\Phi_4, \hat{\eta}) = \frac{\mathcal{Q}_0^2}{2}  \left[ e^{2
  (\Phi_4 - \Phi_{4, \infty})} + e^{- 2 (\Phi_4 - \Phi_{4, \infty}) + \sqrt{8}
  \hat{\eta}} \right],
\end{equation}
where $\mathcal{Q}_0^2 = 4 \pi g_{4, \infty}^2 Q^2$,
along this slice of the moduli space. This is shown in figure
\ref{fig:Q2runaway}.

To understand why the charge threshold bound \eqref{eqn:totDistBound2} cannot be applied in
other cases, consider, e.g., the case where $P \circ P \neq 0$ and $(P \circ
P) (Q \circ Q) - (P \circ Q)^2 < 0$. Then we have:
\begin{equation}
  \mathcal{Q}^2_{\text{bound}} (g_4) = \begin{cases}
    2 \pi g_4^2  \Bigl| Q \circ Q - \frac{(Q \circ P)^2}{P \circ P} \Bigr| +
    \frac{2 \pi}{g_4^2}  | P \circ P |, &  g_4 \leqslant
    g_4^{\text{crit}},\\
    4 \pi \sqrt{(Q \circ P)^2 - (Q \circ Q) (P \circ P) }, & g_4 >
    g_4^{\text{crit}},
  \end{cases}
\end{equation}
where
\begin{equation}
  g_4^{\text{(crit)}} \assign \left[ \frac{(P \circ P)^2}{(P \circ Q)^2 - (P
  \circ P) (Q \circ Q)} \right]^{1 / 4} .
\end{equation}
In this case, $\mathcal{Q}^2_{\text{bound}} (g_4)$ still decreases to a
minimum at $g_4^{\text{crit}}$, but for larger $g_4$ it stays flat rather than
increasing.

\begin{figure}\centering
  \includegraphics[width=0.8\textwidth]{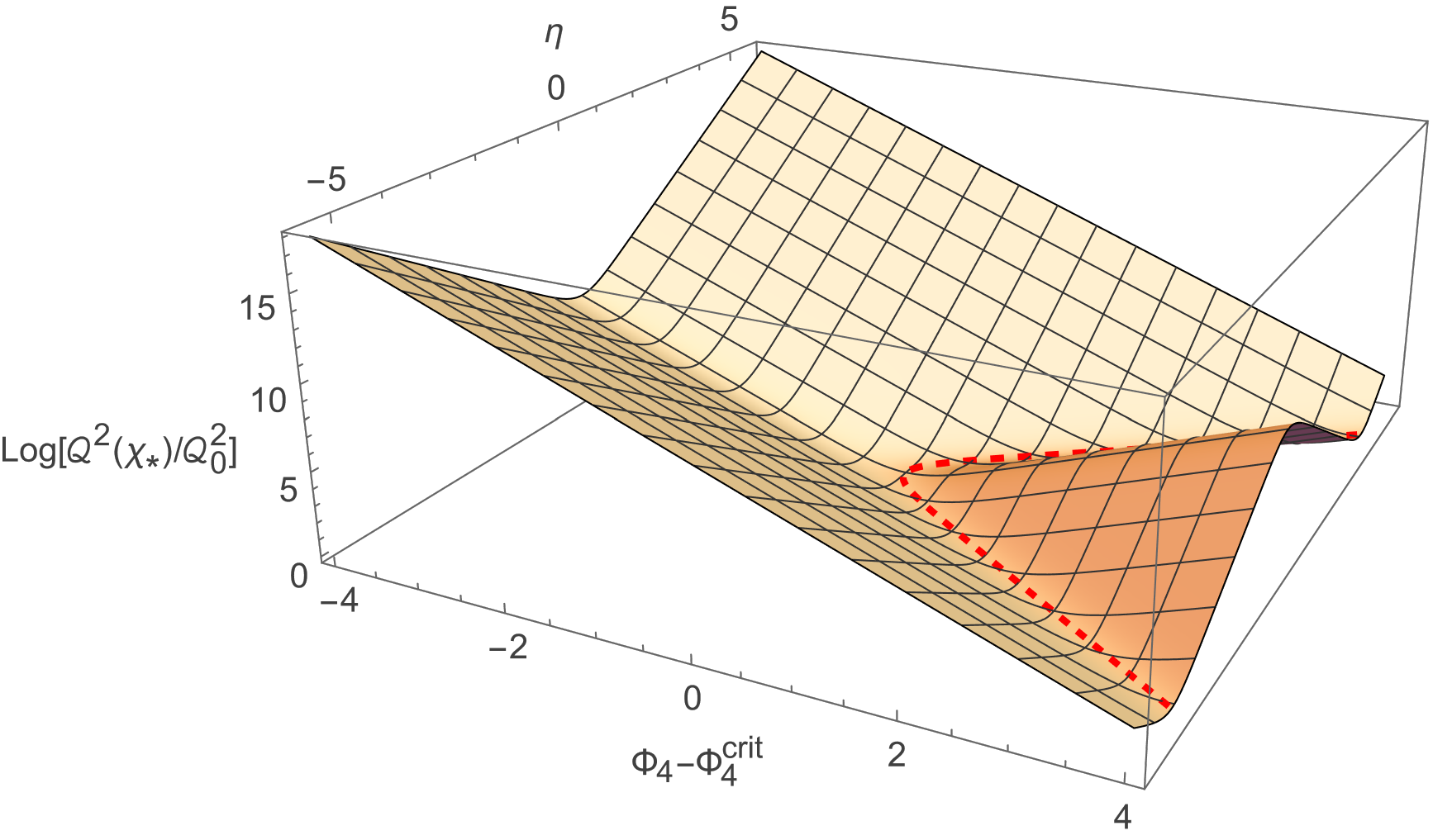}
  \caption{The log of the charge function $\mathcal{Q}^2 (\chi_{\ast})$
  (minimized with respect to the axion $\chi$) plotted in the $\Phi$-$\eta$
  plane in the $P \circ P > 0$, $Q \circ Q < 0$, $P \circ Q = 0$ case. Note
  the flat valley starting at $\Phi_4 = \Phi_4^{\text{crit}}$ and continuing
  into the strongly coupled regime, as highlighted by the red dashed line.
  This prevents us from applying the bound
  \eqref{eqn:totDistBound2}.\label{fig:Q2bifurcate} }
\end{figure}

To illustrate this, let $P \circ P > 0$, $Q \circ Q < 0$ and $P \circ Q = 0$
for definiteness. We consider a slice of moduli space in which $\vec{Q}_L
\propto \vec{P}_L$ and $\vec{Q}_R \propto \vec{P}_R$, so that:
\begin{equation}
\begin{aligned}
  \vec{P}_L &= | P | \cosh \eta \hat{P}_L, & \vec{Q}_L &= | Q | \sinh \eta \hat{P}_L, \\ \vec{P}_R &= | P | \sinh \eta
  \hat{P}_R, & \vec{Q}_R &= | Q | \cosh \eta \hat{P}_R .
\end{aligned}
\end{equation}
Note that in a fixed basis within the $\hat{P}_L, \hat{P}_R$ plane in which
$P_A = | P |  \bigl(\begin{smallmatrix} 1\\ 0 \end{smallmatrix}\bigr)$ and $Q_A = | Q | \bigl(\begin{smallmatrix} 0\\ 1 \end{smallmatrix}\bigr)$, we again have:
\begin{equation}
  \varphi_{A B} = \begin{pmatrix}
    \cosh 2 \eta & \sinh 2 \eta\\
    \sinh 2 \eta & \cosh 2 \eta
  \end{pmatrix}, \qquad \eta_{A B} = \begin{pmatrix}
    1 & 0\\
    0 & - 1
  \end{pmatrix},
\end{equation}
so that $G_{\eta \eta} = \frac{1}{2 k_N}$. We find:
\begin{equation}
  \mathcal{Q}^2 = 2 \pi \left( e^{2 \Phi_4}  \left[ | Q |^2 + \left(
  \frac{\chi}{2 \pi} \right)^2 | P |^2 \right] + e^{- 2 \Phi_4}  | P |^2
  \right) \cosh 2 \eta - 2 e^{2 \Phi_4}  | P |  | Q | \chi \sinh 2 \eta,
\end{equation}
where $\chi$ is the axion. With $\eta \neq 0$, the minimum value of
$\mathcal{Q}^2$ lies at $\chi \neq 0$, which is why we have included it. To
simplify the problem, we note that $\mathcal{Q}^2$ is a quadratic function of
$\chi$, so we start by minimizing with respect to $\chi$ with $\Phi_4, \eta$
held fixed:
\begin{equation}
  \partial_{\chi} \mathcal{Q}^2 = e^{2 \Phi_4}  \frac{\chi}{\pi}  | P |^2
  \cosh 2 \eta - 2 | P |  | Q | e^{2 \Phi_4} \sinh 2 \eta = 0 \qquad
  \Rightarrow \qquad \frac{\chi_{\ast}}{2 \pi} = \frac{| Q |}{| P |} \tanh 2
  \eta .
\end{equation}
Thus,
\begin{equation}
  \mathcal{Q}^2 (\chi) \geqslant \mathcal{Q}^2 (\chi_{\ast}) = 2 \pi \left(
  \frac{e^{2 \Phi_4}}{\cosh 2 \eta}  | Q |^2 + \frac{\cosh 2 \eta}{e^{2
  \Phi_4}}  | P |^2 \right),
\end{equation}
or more succinctly
\begin{equation}
  \mathcal{Q}^2 (\chi_{\ast}) = \frac{\mathcal{Q}_0^2}{2}  \left( \frac{e^{2
  (\Phi_4 - \Phi_4^{\text{crit}})}}{\cosh 2 \eta} + \frac{\cosh 2 \eta}{e^{2
  (\Phi_4 - \Phi_4^{\text{crit}})}} \right),
\end{equation}
 where $\mathcal{Q}_0^2 = 4 \pi | Q | | P |, \Phi^{\text{crit}}_4 =
  \frac{1}{2} \log \frac{| P |}{| Q |}$.
This is plotted in figure \ref{fig:Q2bifurcate}.

In fact, since $\mathcal{Q}^2$ has a locus of global minima extending to
strong coupling, there are genuine attractor flows (depending on the
asymptotic values of the moduli) that reach the strongly coupled
regime.\footnote{However, when $| Q | \gg | P |$, there are also attractors
flow that end at weak coupling, depending on the asymptotic values of the
moduli. It is possible that one can devise a distance function $\mathfrak{D}$
that proves that there are no strongly coupled black holes for these values of
the moduli.} Thus, it is not surprising that we fail to rule out strongly
coupled black holes in this case.

\section{Conclusions}\label{sec:conclusions}

In this paper, we derived sharp bounds on the total moduli space excursion of
a spherically symmetric black hole in situations in which the charge function
$Q^2 (\phi)$ increases above its initial asymptotic value and on the moduli
space excursion in a specified direction in situations in which $Q^2 (\phi)$
has a positive directional derivative in that direction.

We have applied these bounds to rule out an \emph{a priori} possible
loophole in the proof \cite{Heidenreich:2024dmr} of the electric WGC in
perturbative (bosonic) string theory. In particular, since the WGC compares
the exact spectrum of a theory to the spectrum of black hole solutions of the
two-derivative low energy EFT, it is important to know the latter exactly to
properly normalize the bound. If there were an electrically charged black hole
solution with $g_s \gtrsim 1$ somewhere outside its horizon then we would not
be able to determine its properties without first understanding the
strong-coupled string theory in question (or, at least, the low-energy EFT
thereof). Fortunately, our bounds establish that this does \emph{not}
occur, at least for spherically symmetric black holes.

The same methods extend to certain types of dyonic black holes in $d = 4, 5$.
For the same reason, this could become an essential component of a proof of
the WGC for such dyonic charges, provided that the exact spectrum of dyonic
states can be found using perturbative string methods for $g_s \ll 1$. More
generally, our bounds could be applied to any problem where it is important to
know the spectrum of charged black holes with only local knowledge of the
moduli space.

It would be very interesting to see if these bounds extend to non-spherically
symmetric charged black holes. Unfortunately, not much is known about the
\emph{general} spectrum of such black holes in dimension $d > 4$ outside of specific
cases (often supersymmetric ones) where solutions have
been obtained explicitly. However, it may be possible to prove theorems about
such solutions, along the lines of \cite{Gibbons:1982jg}.

Another interesting question is whether our methods can be generalized beyond
the specific criteria that we have used---in either the ``charge threshold''
or ``directional derivative'' variants---to obtain more general bounds that
unify different types of constraints. We leave these topics for the future.

Finally, we note that our bounds bear a strong resemblance to the recently proposed ``Imaginary Distance Bound'' (IDB)~\cite{DiUbaldo:2026rly,Maldacena:2026jqd}, despite the different physical contexts. This seems to be the result of some formal (and apparently accidental) connections between the black hole equations and the equations satisfied by Euclidean wormholes~\cite{Giddings:1987cg,Arkani-Hamed:2007cpn}, as explained in Appendix~\ref{app:wormhole}. It would be interesting to explore these connections further to see if they have some physical origin or consequences.

\section*{Acknowledgments}

We thank the anonymous referee of~\cite{Heidenreich:2024dmr} for motivating us to pursue this topic, and Matthew Reece for noticing the similiarity between our bounds and the IDB. ML would like to acknowledge the hospitality of the Instituto Balseiro, Centro At\'omico Bariloche and ICTP-SAIFR during the later stages of this work.
The work of BH is supported by NSF grants PHY-2112800 and PHY-2412570.
This work of ML is supported by the Ayuda RYC2023-043268-I funded by MICIU/AEI/10.13039/501100011033 and FSE+, as well as through the grants PID2024-156043NB-I00, PID2021-123017NB-I00 and CEX2020-001007-S, funded by MCIN/AEI/10.13039/50110 0011033, and ERDF, EU.

\appendix\section{Maupertuis's principle and geometry}\label{app:Maupertuis}

Maupertuis's principle is an alternative to Hamilton's principle of least
action\footnote{In fact, Maupertuis's principle \emph{predates} the usual
principle of least action.} that exists when the Lagrangian has no explicit
time dependence. The principle reformulates the dynamics as the critical
points of an ``abbreviated action'', which involves an energy-dependent
integral over the path taken by the system, with no appearance of the time
variable at all. Because of this, Maupertuis's principle has a natural
geometric interpretation that will prove useful in interpreting the bounds we
derive in this paper.

Consider a general action of the form:
\begin{equation}
S = \int_{t_1}^{t_2}  \left[ \frac{1}{2} G_{i j} (\phi)  \dot{\phi}^i
   \dot{\phi}^j + A_i (\phi)  \dot{\phi}^i - V (\phi) \right] \mathd t \,.
\end{equation}
Here we assume that the Lagrangian is time-independent and quadratic in
the generalized velocities, but make no further assumptions. The
variation of the action is
\begin{align}
  \delta S &= \int_{t_1}^{t_2}  \biggl[ G_{i j} (\phi) \delta \dot{\phi}^i
  \dot{\phi}^j + \frac{1}{2} G_{i j, k} (\phi) \delta \phi^k  \dot{\phi}^i
  \dot{\phi}^j + A_{i, j} (\phi) \delta \phi^j  \dot{\phi}^i + A_i (\phi)
  \delta \dot{\phi}^i - V_{, i} (\phi) \delta \phi^i \biggr] \mathd t
  \nonumber\\
  &= \int_{t_1}^{t_2}  \biggl[ - \delta \phi^i  \frac{\mathd}{\mathd t} [G_{i
  j} \dot{\phi}^j] + \frac{1}{2} G_{i j, k} (\phi) \delta \phi^k 
  \dot{\phi}^i \dot{\phi}^j + (A_{i, j} - A_{j, i}) \delta
  \phi^j  \dot{\phi}^i - V_{, i}\, \delta \phi^i \biggr] \mathd t, 
\end{align}
where we integrate by parts and drop the boundary terms, which vanish because
$\delta \phi^i (t_1) = \delta \phi^i (t_2) = 0$ in accordance with Hamilton's
principle. Thus, the Euler-Lagrange equations are:
\begin{equation}
  \frac{\mathd}{\mathd t} [G_{i j}  \dot{\phi}^j] - \frac{1}{2} G_{j k, i} 
  \dot{\phi}^j \dot{\phi}^k = - (A_{i, j} - A_{j, i}) \dot{\phi}^j - V_{, i} . 
\end{equation}
Contracting with $\dot{\phi}^i$, and rearranging we get
\begin{equation}
  \frac{\mathd}{\mathd t} \left[ \frac{1}{2} G_{i j} (\phi)  \dot{\phi}^i 
  \dot{\phi}^j + V (\phi) \right] = 0 \qquad \text{so that} \qquad \frac{1}{2}
  G_{i j} (\phi)  \dot{\phi}^i  \dot{\phi}^j + V (\phi) = E,
\end{equation}
where $E$ is the conserved energy. We now rewrite our original action as
follows:
\begin{multline}
  S = \int_{t_1}^{t_2} \left( \sqrt{\frac{1}{2} G_{i j} (\phi)  \dot{\phi}^i
  \dot{\phi}^j} - \sqrt{E - V (\phi)} \right)^2 \mathd t - E (t_2 - t_1) \\
   + \int_{t_1}^{t_2} \sqrt{2 (E - V (\phi)) G_{i j} (\phi)  \dot{\phi}^i 
  \dot{\phi}^j} \mathd t + \int_{t_1}^{t_2} A_i (\phi)  \dot{\phi}^i \mathd t . 
\end{multline}
For solutions of energy $E$, the variation of the terms on the first line is
zero, so the solutions to the Euler-Lagrange equations of this energy are the
paths $\mathcal{P}$ that extremize the \textbf{abbreviated action}:
\begin{equation}
  W = \int_{t_1}^{t_2} \sqrt{2 (E - V (\phi)) G_{i j} (\phi)  \dot{\phi}^i 
  \dot{\phi}^j} \mathd t + \int_{t_1}^{t_2} A_i (\phi)  \dot{\phi}^i \mathd t .
\end{equation}
This can be written in the purely geometric form:
\begin{equation}
  W = \int_{\mathcal{P}} \widetilde{\mathd s} + \int_{\mathcal{P}} A, \qquad
  \text{where} \qquad \widetilde{\mathd s}^2 = 2 (E - V (\phi)) G_{i j} (\phi)
  \mathd \phi^i \mathd \phi^j, \label{eqn:MaupertuisW}
\end{equation}
where the time variable $t$ no longer appears. We call the warped metric
$\widetilde{\mathd s}^2 = 2 (E - V (\phi)) \mathd s^2$ the
\textbf{Maupertuis metric}. While we have carried along the vector potential
$A_i (\phi)$ throughout this calculation to illustrate the wide range of
applicability of Maupertuis's principle, in this paper we will only need the
case $A_i = 0$. In this case, the paths $\mathcal{P}$ that extremize the
abbreviated action are precisely the geodesics of the Maupertuis metric.

\section{Lower bounds on the 4d dyonic charge
function}\label{app:4ddyonicQ2bounds}

In this appendix we prove the optimal lower bounds
\eqref{eqn:Q2dyonicbound1}--\eqref{eqn:Q2dyonicbound3} on the 4d dyonic charge
function \eqref{eqn:4dchargefunc}, reproduced below:
\begin{equation}
  \mathcal{Q}^2 (\varphi, \chi) = 2 \pi g_4^2 \varphi^{A B} \left[ Q_A -
  \frac{\chi}{2 \pi} P_A \right] \left[ Q_B - \frac{\chi}{2 \pi} P_B \right] +
  \frac{2 \pi}{g_4^2} \varphi^{A B} P_A P_B,
\end{equation}
for each fixed value of $g_4 \assign e^{\Phi_4}$. We also derive the criteria
under which these lower bounds can be saturated.

First, we minimize $\mathcal{Q}^2$ with respect to $\chi$:
\begin{equation}
  \partial_{\chi} \mathcal{Q}^2 = - 2 g_4^2 \varphi^{A B} P_A  \left[ Q_B -
  \frac{\chi}{2 \pi} P_B \right] = 0 \quad \Rightarrow \quad 
  \frac{\chi_{\ast}}{2 \pi} = \frac{P_A \varphi^{A B} Q_B}{P_C \varphi^{C D}
  P_D} = \frac{P \varphi Q}{P \varphi P} .
\end{equation}
Thus,
\begin{equation}
  \mathcal{Q}^2 (\varphi, \chi) \geqslant \hat{\mathcal{Q}}^2 (\varphi)
  \assign 2 \pi g_4^2  \left[ Q \varphi Q - \frac{(P \varphi Q)^2}{P \varphi
  P} \right] + \frac{2 \pi}{g_4^2} P \varphi P .
\end{equation}
Note that $\hat{\mathcal{Q}}^2 (\varphi)$ is invariant under shifts $Q
\rightarrow Q + \lambda P$, corresponding to a shift in the critical value of
the axion $\chi_{\ast} \rightarrow \chi_{\ast} + 2 \pi \lambda$.

To place lower bounds on $\hat{\mathcal{Q}}^2$, we begin by considering a few
special cases:

\subsection{Case I: \alt{$P \circ Q = 0$}{P ◦ Q = 0}}

If $P \circ Q = 0$, then under certain assumptions to be specified the optimal
bound is
\begin{equation}
  \hat{\mathcal{Q}}^2 \geqslant \hat{\mathcal{Q}}^2_{\min} = 2 \pi g_4^2  | Q
  \circ Q | + \frac{2 \pi}{g_4^2}  | P \circ P | . \label{eqn:Q2boundcaseI}
\end{equation}
To show this, first suppose that $P \circ P, Q \circ Q \geqslant 0$. We then
find:
\begin{equation}
  \hat{\mathcal{Q}}^2 - \hat{\mathcal{Q}}^2_{\min} = 2 \pi g_4^2  \left[ 2
  Q_R^2 - 4 \frac{(P_R \cdot Q_R)^2}{P_L^2 + P_R^2} \right] + \frac{4
  \pi}{g_4^2} P_R^2 \geqslant 4 \pi g_4^2  \left[ Q_R^2 - \frac{(P_R \cdot
  Q_R)^2}{P_R^2} \right] + \frac{4 \pi}{g_4^2} P_R^2 \geqslant 0,
\end{equation}
where we first use $P_L \cdot Q_L = P_R \cdot Q_R$ (since $P \circ Q = 0$)
followed by $P_L^2 \geqslant P_R^2$ (since $P \circ P \geqslant 0$), finally
applying the triangle inequality $| P_R \cdot Q_R | \leqslant | P_R |  | Q_R
|$.

The case where $P \circ P, Q \circ Q \leqslant 0$ is analogous, so let us now
suppose that $P \circ P > 0$ and $Q \circ Q < 0$:
\begin{align}
  \hat{\mathcal{Q}}^2 - \hat{\mathcal{Q}}^2_{\min} &=  4 \pi g_4^2  \left[
  Q_L^2 - \frac{(P_L \cdot Q_L)^2}{P_L^2 + P_R^2} - \frac{(P_R \cdot
  Q_R)^2}{P_L^2 + P_R^2} \right] + \frac{4 \pi}{g_4^2} P_R^2 \nonumber\\
  &\geqslant 4 \pi
  g_4^2  \left[ Q_L^2 - \frac{P_L^2 Q_L^2}{P_L^2 + P_R^2} - \frac{P_R^2
  Q_R^2}{P_L^2 + P_R^2} \right] + \frac{4 \pi}{g_4^2} P_R^2 \nonumber\\
  &= 4 \pi \left[ g_4^2 Q \circ Q + \frac{1}{g_4^2} P \circ P \right] 
  \frac{P_R^2}{P_L^2 + P_R^2} + \frac{8 \pi}{g_4^2}  \frac{P_R^4}{P_L^2 +
  P_R^2}, 
\end{align}
where we again use $P_L \cdot Q_L = P_R \cdot Q_R$ together with the triangle
inequalities for both $P_L \cdot Q_L$ and $P_R \cdot Q_R$. Thus, the bound
holds so long as $Q \circ Q \geqslant - \frac{1}{g_4^4} P \circ P$. Accounting
for the analogous case $P \circ P < 0$ and $Q \circ Q > 0$, we find in summary
that
\begin{equation}
  \hat{\mathcal{Q}}^2 \geqslant \hat{\mathcal{Q}}^2_{\min} = 2 \pi g_4^2  | Q
  \circ Q | + \frac{2 \pi}{g_4^2}  | P \circ P | \quad \text{provided that}
  \quad (P \circ P) (Q \circ Q) \geqslant - \frac{1}{g_4^4}  (P \circ P)^2 .
\end{equation}

If neither $P$ nor $Q$ is null then there exists a simultaneous ``rest frame''
in which $P, Q$ are each either purely left or purely right moving, and it is
straightforward to verify that the bound is saturated in this frame. If $Q$ is
null but $P$ is not then in a $P$ rest frame, $Q$ lies in an orthogonal plane
to $P$ due to $P \circ Q = 0$, and we can boost in this plane to make $Q_R^2$
as small as desired, bringing $\hat{\mathcal{Q}}^2 -
\hat{\mathcal{Q}}^2_{\min}$ arbitrarily close to zero. If instead $P$ is null,
then since $\hat{\mathcal{Q}}^2$ and the condition $P \circ Q = 0$ are both
invariant under $Q \rightarrow Q + \lambda P$, we can set $P_L \cdot Q_L = P_R
\cdot Q_R = 0$ WLOG. Then we can boost in the separate $P$ and $Q$ planes to
make $\hat{\mathcal{Q}}^2 - \hat{\mathcal{Q}}^2_{\min}$ arbitrarily close to
zero.

Thus, the bound (\ref{eqn:Q2boundcaseI}) is optimal and can be saturated so
long as $(P \circ P) (Q \circ Q) \neq 0$.

\subsection{Case II: generalization to \alt{$P \circ Q \neq 0$}{P ◦ Q ≠ 0}}

Now let us move away from the assumption $P \circ Q = 0$. We observe that
$\hat{\mathcal{Q}}^2$ is invariant under $Q \rightarrow Q + \lambda P$, which
takes $P \circ Q \rightarrow P \circ Q + \lambda P \circ P$. Thus, when $P
\circ P \neq 0$, we can use this shift invariance to relate the general case
to the special case $P \circ Q = 0$ treated above.

Noting that both $P \circ P$ and $(P \circ P) (Q \circ Q) - (P \circ Q)^2$ are
invariant under such shifts, we can use this reasoning to conclude that:
\begin{enumerate}
  \item If $P \circ P \neq 0$ and $(P \circ P) (Q \circ Q) - (P \circ Q)^2
  \geqslant - \frac{1}{g_4^4}  (P \circ P)^2$ then the optimum lower bound is:
  \begin{equation}
    \hat{\mathcal{Q}}^2 \geqslant 2 \pi g_4^2  \left| Q \circ Q - \frac{(Q
    \circ P)^2}{P \circ P} \right| + \frac{2 \pi}{g_4^2}  | P \circ P |,
  \end{equation}
  which can be saturated so long as $(P \circ P) (Q \circ Q) - (P \circ Q)^2
  \neq 0$.
  
  \item If $P \circ P = Q \circ P = 0$ then the optimum lower bound is:
  \begin{equation}
    \hat{\mathcal{Q}}^2 > 2 \pi g_4^2  | Q \circ Q |,
  \end{equation}
  which cannot be saturated.
\end{enumerate}

\subsection{Case III: \alt{$(P \circ P) (Q \circ Q) - (P \circ Q)^2 < -
\frac{1}{g_4^4}  (P \circ P)^2$}{(P◦P)(Q◦Q) - (P◦Q)² < -(1/g₄⁴) (P◦P)²}}

The optimal bound when $(P \circ P) (Q \circ Q) - (P \circ Q)^2 < -
\frac{1}{g_4^4}  (P \circ P)^2$ remains to be determined. We will show that in
this case
\begin{equation}
  \hat{\mathcal{Q}}^2 \geqslant 4 \pi \sqrt{(P \circ Q)^2 - (P \circ P) (Q
  \circ Q)} . \label{eqn:PQbound}
\end{equation}
To so, we first note that
\begin{equation}
  \hat{\mathcal{Q}}^2 = 2 \pi \left[ g_4^2  \left( Q \varphi Q - \frac{(P
  \varphi Q)^2}{P \varphi P} \right) + \frac{1}{g_4^2} P \varphi P \right]
  \geqslant 4 \pi \sqrt{(P \varphi P) (Q \varphi Q) - (P \varphi Q)^2},
  \label{eqn:PQbound0}
\end{equation}
where we use $x + y \geqslant 2 \sqrt{x y}$ for non-negative $x, y$. Now
consider,
\begin{multline}
  (P \varphi P) (Q \varphi Q) - (P \varphi Q)^2 + (P \circ P) (Q \circ Q) - (P
  \circ Q)^2 \\
  = 2 [P_L^2 Q_L^2 - (P_L \cdot Q_L)^2] + 2 [P_R^2 Q_R^2 - (P_R \cdot Q_R)^2] \geqslant 0,
\end{multline}
where we use the triangle inequality for $P_L \cdot Q_L$ and $P_R \cdot Q_R$.
Thus, we conclude that
\begin{equation}
  (P \varphi P) (Q \varphi Q) - (P \varphi Q)^2 \geqslant (P \circ Q)^2 - (P
  \circ P) (Q \circ Q) .
\end{equation}
Since the RHS is positive by assumption, in combination with
\eqref{eqn:PQbound0} this implies the lower bound \eqref{eqn:PQbound}.

To show that the bound \eqref{eqn:PQbound} can be saturated, we divide into
two subcases:

\paragraph{Case A: $P \circ P \neq 0$}

In this case, we can set $P \circ Q = 0$ by shifting $Q \rightarrow Q +
\lambda P$. If $P \circ P > 0$, then the bound we wish to saturate becomes
\begin{equation}
  \hat{\mathcal{Q}}^2 = 4 \pi \sqrt{(P \circ P) (- Q \circ Q)} \qquad
  \text{when} \qquad Q \circ Q < - \frac{1}{g_4^4} P \circ P .
\end{equation}
As before, we can reach a frame where $P$ and $Q$ are purely left and
right-moving, respectively. However, this does not saturate the bound, since
in this frame
\begin{equation}
  \hat{\mathcal{Q}}^2 = 2 \pi g_4^2 Q_R^2 + \frac{2 \pi}{g_4^2} P_L^2 > 4 \pi
  \sqrt{P_L^2 Q_R^2},
\end{equation}
by the inequality $x + y \geqslant 2 \sqrt{x y}$, which cannot be saturated
since $g_4^2 Q_R^2 > \frac{1}{g_4^2} P_L^2$.

To saturate the bound, we perform a further boost in the $P$-$Q$ plane, after
which:
\begin{equation}
  P_L = | P | \cosh \eta, \qquad P_R = | P | \sinh \eta, \qquad Q_L = | Q |
  \sinh \eta, \qquad Q_R = | Q | \cosh \eta,
\end{equation}
where $| P | \assign \sqrt{| P \circ P |}$, $| Q | \assign \sqrt{| Q \circ Q
|}$. Then we find:
\begin{equation}
  \hat{\mathcal{Q}}^2 (\eta) = 2 \pi g_4^2  | Q |^2  \frac{1}{\cosh (2 \eta)}
  + \frac{2 \pi}{g_4^2}  | P |^2 \cosh (2 \eta) .
\end{equation}
The minimum is attained by choosing:
\begin{equation}
  \eta_{\ast} = \frac{1}{2} \cosh^{- 1} \left( \frac{g_4^2  | Q |}{| P |}
  \right) \qquad \Rightarrow \qquad \hat{\mathcal{Q}}^2 (\eta_{\ast}) = 4 \pi
  | Q | | P | = 4 \pi \sqrt{(P \circ P) (- Q \circ Q)},
\end{equation}
which indeed saturates the bound. The case $P \circ P < 0$ is completely
analogous.

\paragraph{Case B: $P \circ P = 0$}

When $P$ is null, the assumption $(P \circ P) (Q \circ Q) - (P \circ Q)^2 < -
\frac{1}{g_4^4}  (P \circ P)^2$ reduces to $P \circ Q \neq 0$. Thus, we wish
to saturate the bound:
\begin{equation}
  \hat{\mathcal{Q}}^2 \geqslant 4 \pi | P \circ Q | \qquad \text{when} \qquad
  P \circ P = 0, \quad P \circ Q \neq 0 .
\end{equation}
To do so, note that by shifting $Q \rightarrow Q + \lambda P$ we can set $Q
\circ Q = 0$. Let us define:
\begin{equation}
  V_{\pm} \assign P \pm g_4^2 Q,
\end{equation}
so that
\begin{equation}
  V_+ \circ V_+ = 2 g_4^2 P \circ Q, \qquad V_- \circ V_- = - 2 g_4^2 P \circ
  Q, \qquad V_+ \circ V_- = 0 .
\end{equation}
If $P \circ Q > 0$ then $V_+$ and $V_-$ are orthogonal and left and
right-like, respectively, so they have a simultaneous rest-frame where one is
purely left-moving and the other is purely right moving. In this frame,
\begin{equation}
  P_L = \frac{1}{2} V_{+ L}, \qquad P_R = \frac{1}{2} V_{- R}, \qquad Q_L =
  \frac{1}{2 g_4^2} V_{+ L}, \qquad Q_R = - \frac{1}{2 g_4^2} V_{- R},
\end{equation}
where $V_{+ L}^2 = V_{- R}^2 = 2 g_4^2 P \circ Q$. Thus,
\begin{equation}
  \hat{\mathcal{Q}}^2 = \frac{2 \pi}{g_4^2} V_{+ L}^2 = 4 \pi P \circ Q,
\end{equation}
which saturates the bound as desired. The case where $P \circ Q < 0$ is
analogous.

\section{Connection to Euclidean Wormholes} \label{app:wormhole}

In this appendix, we explain the mathematical connection between our bounds~\eqref{eqn:totDistBound2}, \eqref{eqn:DistFuncBound} and the properties of Euclidean wormholes~\cite{Giddings:1987cg,Arkani-Hamed:2007cpn}. First, we briefly
review the latter. Consider the Euclidean action in $\tilde{d}$ dimensions:
\begin{equation}
  S_E = \int \mathd^{\tilde{d}} x \sqrt{g}  \left( - \frac{1}{2
  \kappa_{\tilde{d}}^2} \mathcal{R}+ \frac{1}{2} G_{i j} (\phi) \nabla \phi^i
  \cdot \nabla \phi^j \right),
\end{equation}
where the moduli-space metric $G_{i j} (\phi)$ may have indefinite signature
in general. A general spherically symmetric solution takes the form:
\begin{equation}
  \mathd s^2_{\tilde{d}} = e^{\frac{2 (\tilde{d} - 1)}{\tilde{d} - 2} \zeta}
  \mathd \lambda^2 + e^{\frac{2}{\tilde{d} - 2} \zeta} \mathd
  \Omega_{\tilde{d} - 1}^2, \qquad \phi^i = \phi^j (\lambda),
  \label{eqn:WormholeMetric}
\end{equation}
where we choose a convenient gauge for the radial coordinate $\lambda$. The
Euler-Lagrange equations are then:
\begin{subequations} \label{eqn:wormholeeqns}
\begin{align}
  \frac{\mathd^2 \zeta}{\mathd \lambda^2} &= (\tilde{d} - 2)^2 e^{2 \zeta}, \\
   \frac{\mathd^2 \phi^i}{\mathd \lambda^2} + \Gamma^i_{\; j k}  \frac{\mathd \phi^j}{\mathd \lambda}  \frac{\mathd \phi^k}{\mathd \lambda} &= 0, \\
   \left[ \frac{\mathd \zeta}{\mathd \lambda} \right]^2 -
  \tilde{k}_N G_{i j}  \frac{\mathd \phi^i}{\mathd \lambda}  \frac{\mathd
  \phi^j}{\mathd \lambda} &= (\tilde{d} - 2)^2 e^{2 \zeta},
\end{align}
\end{subequations}
where $\tilde{k}_N \assign \frac{\tilde{d} - 2}{\tilde{d} - 1}
\kappa_{\tilde{d}}^2$. The first equation has the general solution:
\begin{equation}
  e^{- \zeta} = \frac{\tilde{d} - 2}{a} \sin [a \lambda], \label{eqn:zetasoln}
\end{equation}
up to a shift in $\lambda$ to eliminate one of the integration constants. For
real, positive $a$ this solution is valid for $0 < \lambda < \frac{\pi}{a}$,
where the $S^{\tilde{d} - 1}$ blows up to infinite volume at each end and
reaches a minimum volume
\begin{equation}
  V_{\min} = \left( \frac{a}{\tilde{d} - 2} \right)^{\frac{\tilde{d} -
  1}{\tilde{d} - 2}} V_{\tilde{d} - 1},
\end{equation}
at $\lambda = \frac{\pi}{2 a}$. This is the Euclidean wormhole geometry.

The remaining equations imply that $\phi^I (\lambda)$ is a geodesic, with
\begin{equation}
  \tilde{k}_N G_{i j}  \frac{\mathd \phi^i}{\mathd \lambda}  \frac{\mathd
  \phi^j}{\mathd \lambda} = - a^2 .
\end{equation}
Thus, the geodesic is ``timelike'' ($\mathd s^2 < 0$), with proper length
\begin{equation}
  \Delta \phi_{\text{full}} = \frac{\pi i}{\sqrt{\tilde{k}_N}},
\end{equation}
between the two ends of the wormhole. Equivalently, the proper length
traversed between the center of the wormhole and one end is:
\begin{equation}
  \Delta \phi_{\text{half}} = \frac{\pi i}{2 \sqrt{\tilde{k}_N}} .
\end{equation}
Notice the similarity between this and the bound \eqref{eqn:totDistBound2}. More precisely, the
two agree numerically (up to the $i$) if we set $\tilde{d} = d - 1$. Note,
however, that the Euclidean wormhole produces an ``imaginary'' (i.e.,
timelike) displacement in the moduli space, whereas \eqref{eqn:totDistBound2}
constrains the real displacement in moduli space (as does \eqref{eqn:DistFuncBound}).

To explain this remarkable coincidence, note that saturating \eqref{eqn:totDistBound2} requires a quasiextremal solution with constant $Q^2$, in which case
the black hole equations reduce to:
\begin{subequations}
\begin{align}
  k_N^{- 1}  \frac{\mathd^2 \chi}{\mathd \tau^2} &= e^{2 \chi} Q^2, \\
  \frac{\mathd^2 \phi^i}{\mathd \tau^2} + \Gamma^i_{\; j k} 
  \frac{\mathd \phi^j}{\mathd \tau}  \frac{\mathd \phi^k}{\mathd \tau} &= 0, \\
  k_N^{- 1}  \left[ \frac{\mathd \chi}{\mathd \tau} \right]^2 + G_{i j}  \frac{\mathd \phi^i}{\mathd \tau}  \frac{\mathd \phi^j}{\mathd \tau}
  &= e^{2 \chi} Q^2 .
\end{align}
\end{subequations}
In fact, these are the \emph{same} equations as \eqref{eqn:wormholeeqns}
up to the replacements:
\begin{equation}
  \tau \rightarrow \lambda, \qquad k_N \rightarrow \tilde{k}_N, \qquad \chi
  \rightarrow \zeta, \qquad Q \rightarrow \frac{\tilde{d} -
  2}{\sqrt{\tilde{k}_N}}, \qquad G_{i j} \rightarrow - G_{i j} .
\end{equation}
Thus, they have analogous solutions, of the form:
\begin{equation}
  e^{- \chi} = \frac{\sqrt{k_N} Q}{a} \sin [a \tau], \qquad \text{with $\phi^i
  (\tau)$ a geodesic satisfying} \qquad k_N G_{i j}  \frac{\mathd
  \phi^i}{\mathd \tau}  \frac{\mathd \phi^j}{\mathd \tau} = a^2 .
\end{equation}
The horizon condition $\chi' < 0$ is satisfied for $0 < \tau < \frac{\pi}{2
a}$, which corresponds to one half of the wormhole solution discussed above.
The geodesic distance traversed in this interval is:
\begin{equation}
  \Delta \phi = \frac{\pi}{2 \sqrt{k_N}},
\end{equation}
which saturates \eqref{eqn:totDistBound2}. However, note that the resulting solution is
\emph{not} a black hole. In particular, the metric is:
\begin{equation}
  \mathd s^2_d = - \frac{a^2}{k_N Q^2 \sin^2 [a \tau]} \mathd t^2 + \left[
  \frac{\sqrt{k_N} Q}{(d - 3) V_{d - 2}} \sinc(a \tau)
  \right]^{\frac{2}{d - 3}}  \left[ \frac{\mathd \tau^2}{(d - 3)^2 \tau^2} +
  \mathd \Omega_{d - 2}^2 \right] .
\end{equation}
As $\tau \rightarrow 0$, the geometry asymptotically approaches AdS$_2 \times
S^{d - 2}$. However, at $\tau \rightarrow \pi / a$ there is a naked
singularity. Indeed, this is the expected end result of violating the horizon
condition $\chi' < 0$ at $\tau = \pi / 2 a$, and agrees with our previous
observation that the bounds \eqref{eqn:totDistBound2}, \eqref{eqn:DistFuncBound} cannot be saturated.

This connection is intriguing, but hard to interpret physically. We can take
it a little farther by observing that the timelike reduction of a spherically
symmetric Lorentzian solution in $d$ dimensions is a Euclidean solution in
$\tilde{d} = d - 1$ dimensions, where
\begin{equation}
  \mathd s^2_d = - e^{2 \chi} \mathd t^2 + e^{- \frac{2}{d - 3} \chi} \mathd
  s_{\tilde{d}}^2,
\end{equation}
and the warp factor $\chi$ can be thought of as the ``radion'' associated to
the timelike reduction. Combining this with \eqref{eqn:WormholeMetric}, we
obtain:
\begin{equation}
  \mathd s^2_d = - e^{2 \chi} \mathd t^2 + e^{\frac{2}{d - 3} \zeta -
  \frac{2}{d - 3} \chi}  [e^{2 \zeta} V_{d - 2}^2 \mathd \tau^2 + \mathd
  \Omega_{d - 2}^2], \qquad \phi^i = \phi^i (\tau),
\end{equation}
where we introduce a rescaling factor $\lambda = V_{d - 2} \tau$ for future
convenience. The black hole equations work out to
\vspace{-1em}
\begin{subequations}
\columnratio{0.35}
\begin{paracol}{2}
\begin{align}
  \vphantom{\frac{1}{2}}
  \chi'' &= k_N e^{2 \chi} Q^2 , \tag{\theequation a} \\
  \vphantom{\left( {\chi'}^2 {- \zeta'}^2 \right)}
  \zeta'' &= k_N e^{2 \zeta}  \tilde{Q}^2 , \tag{\theequation b} 
\end{align}
\switchcolumn
\begin{align}
  {\phi^i}'' + \Gamma^i_{\; j k} {\phi^j}' {\phi^k}' &=
  \frac{1}{2} e^{2 \chi} G^{i j} Q^2_{, j} , \tag{\theequation c} \\
  k_N^{- 1}  \left( {\chi'}^2 {- \zeta'}^2 \right) + G_{i j} {\phi^i}'
  {\phi^j}' &= e^{2 \chi} Q^2 - e^{2 \zeta}  \tilde{Q}^2 , \tag{\theequation d} 
\end{align}
\end{paracol}
\end{subequations}\noindent
where $\tilde{Q}^2 \assign k_N^{- 1} (d - 3)^2 V_{d - 2}^2$.\footnote{Note
that substituting $e^{- \zeta} = \frac{\tilde{Q}}{\sqrt{k_N} M_0} \sinh [k_N
M_0 \tau]$ reproduces \eqref{eqn:BHeqnstau}.} Now we observe that for constant $Q$ these
equations have the symmetry:
\begin{equation}
  \chi \leftrightarrow \zeta, \qquad Q \leftrightarrow \tilde{Q}, \qquad G_{i
  j} \rightarrow - G_{i j} .
\end{equation}
As seen above, this exchanges Euclidean wormhole solutions with solutions that
saturate our bounds. It would be interesting to explore this further to see if
there is some physical duality here that explains this symmetry, or if it is
just a mathematical accident.

\bibliographystyle{JHEP}
\bibliography{refs}

\end{document}